\documentclass{article}

     \usepackage[final, nonatbib]{neurips_2020}

\usepackage[utf8]{inputenc} % allow utf-8 input
\usepackage[T1]{fontenc}    % use 8-bit T1 fonts
\usepackage[hidelinks]{hyperref}       % hyperlinks
\usepackage{url}            % simple URL typesetting
\usepackage{booktabs}       % professional-quality tables
\usepackage{amsfonts}       % blackboard math symbols
\usepackage{nicefrac}       % compact symbols for 1/2, etc.
\usepackage{microtype}      % microtypography

\usepackage{subcaption}
\usepackage{glossaries}
\usepackage{xcolor}
\usepackage{amsmath}
\usepackage{amsfonts}
\usepackage{amssymb}
\usepackage{pgfplots}
\usepackage{booktabs}
\usepackage{graphicx}
\usepackage{subcaption}
\usepackage{glossaries}
\usepackage{diagbox}
\usepackage{multirow}
\usepackage{textcomp}
\usepackage{gensymb}
\usepackage{cleveref}
\usepackage{rotating, graphicx}
\usepackage{caption}
\usepackage{subcaption}

\usepackage{xspace}
\newcommand\ie{i.\,e.\xspace}
\newcommand\eg{e.\,g.\xspace}

\newcommand{\CC}[1]{\cellcolor{gray!#1}}

\usepackage{xcolor,colortbl}
\definecolor{Gray}{gray}{0.85}
\definecolor{LightCyan}{rgb}{0.88,1,1}

\DeclareRobustCommand{\bbone}{\text{\usefont{U}{bbold}{m}{n}1}}

\newcolumntype{C}[1]{>{\centering\let\newline\\\arraybackslash\hspace{0pt}}m{#1}}

\newcolumntype{x}{>{\columncolor{Gray}}C}
\newcolumntype{y}{>{\columncolor{white}}C}

\newacronym{emos}{EMOS}{Ensemble Model Output Statistics}
\newacronym{ecmwf}{ECMWF}{European Centre for Medium-Range Weather Forecasts}
\newacronym{eps}{EPS}{Ensemble Prediction System}
\newacronym{nwp}{NWP}{Numerical Weather Prediction}
\newacronym{crps}{CRPS}{Continuous Ranked Probability Score}
\newacronym{crpss}{CRPSS}{CRPS Skill Score}
\newacronym{tigge}{TIGGE}{The International Grand Global Ensemble}

\title{Evaluating Ensemble Post-Processing for Wind Power Forecasts}

\author{%
  Kaleb Phipps\\
  Institute for Automation and Applied Informatics\\
  Karlsruhe Institute of Technology\\
  \texttt{kaleb.phipps@kit.edu} \\
  \And
  Sebastian Lerch \\
  Department of Economics \\
  Karlsruhe Institute of Technology\\
  \texttt{sebastian.lerch@kit.edu} \\
  \And
  Maria Andersson\\
  Department of Management and Engineering,  \\
  Division of Energy Systems \\
  Linköping University \\
  \texttt{maria.h.andersson@liu.se} \\
  \And
  Ralf Mikut\\
  Institute for Automation and Applied Informatics,  \\ 
  Karlsruhe Institute of Technology\\
  \texttt{ralf.mikut@kit.edu} \\
  \And
  Veit Hagenmeyer\\
  Institute for Automation and Applied Informatics,  \\ 
  Karlsruhe Institute of Technology\\
  \texttt{veit.hagenmeyer@kit.edu} \\
  \And
  Nicole Ludwig\\
  Cluster of Excellence Machine Learning,  \\
  University of Tübingen \\
  \texttt{nicole.ludwig@uni-tuebingen.de} \\
}

\begin{document}

\maketitle

\begin{abstract}
Capturing the uncertainty in probabilistic wind power forecasts is challenging, especially when uncertain input variables, such as the weather, play a role. Since ensemble weather predictions aim to capture the uncertainty in the weather system, they can be used to propagate this uncertainty through to subsequent wind power forecasting models. However, as weather ensemble systems are known to be biased and underdispersed, meteorologists post-process the ensembles. This post-processing can successfully correct the biases in the weather variables but has not been evaluated thoroughly in the context of subsequent forecasts, such as wind power generation forecasts.

The present paper evaluates multiple strategies for applying ensemble post-processing to probabilistic wind power forecasts. We use Ensemble Model Output Statistics (EMOS) as the post-processing method and evaluate four possible strategies: only using the raw ensembles without post-processing, a one-step strategy where only the weather ensembles are post-processed, a one-step strategy where we only post-process the power ensembles, and a two-step strategy where we post-process both the weather and power ensembles. Results show that post-processing the final wind power ensemble improves forecast performance regarding both calibration and sharpness, whilst only post-processing the weather ensembles does not necessarily lead to increased forecast performance.
\end{abstract}
\section{Introduction} \label{sec:intro}

Forecasting, especially probabilistic forecasting, is essential to allow decision makers in power systems to optimally operate and maintain the grid~\cite{Ordiano2017, Appino2018}. With the push towards energy systems with high shares of renewable energy sources, forecasting renewable generation, such as wind power, becomes increasingly important. Forecasting wind power is challenging, as the generation depends on weather variables, such as wind speed and temperature, in a non-linear and bounded fashion ~\cite{Pinson.2018}. Additionally, these weather variables are difficult to forecast and are usually described by \gls{nwp} models, which model the physical relationships of the atmosphere to a certain extent.

When forecasting wind power with the help of weather predictions, two models thus play a role: the \gls{nwp} model whose output aims to describe the uncertainty in the weather variables and the wind power forecasting model, whose output describes the uncertainty in the wind power given the weather variables. Within the first model, ensemble predictions which are different runs of the \gls{nwp} model, capture the inherent uncertainty in the weather variables. These ensemble predictions are known to be biased and underdispersed thus post-processing the output of these models is done frequently ~\cite{Thorarinsdottir2010, Schefzik2013, Schefzik2017, gneiting2014calibration, Gneiting2005, Gilbert.2020, Sweeney.2011, Bossavy.2013}. Post-processing is used to alleviate systematic biases in the model and calibrate the forecasts to past observations and there is a large body of work discussing optimal post-processing methods for weather variables (see \eg the review by Vannitsem~et al.~\cite{Vannitsem2021}). Within the second model, \ie the wind power forecast model, the uncertain input as well as the model uncertainty (\ie  epistemic uncertainty) contribute to the final probabilistic forecast. While an optimal forecasting model could theoretically handle all systematic biases, post-processing might be useful when this is not the case.

In the wind power ensemble prediction setting, there are thus three ways in which systematic model biases can be present making post-processing useful: in the first stage concerning the weather output from NWP models, in the second stage concerning the wind power forecasting model, or in both stages. Pinson and Messner~\cite{Pinson.2018} explain the concept behind post-processing for wind power applications, both before the weather is used as input to the forecasting model and afterwards. However, they do not evaluate or compare these different approaches. Although the literature on wind power forecasting is vast (see \eg the review by Zhang~et al.~\cite{Zhang.2014}), only a few examples dealing with post-processing exist, for example Taylor et al.~\cite{Taylor2009}, Pinson and Henrik ~\cite{Pinson.2009}, Messner et al. ~\cite{Messner.2014}, and Worsnop et al. ~\cite{Worsnop.2018}. However, each of these papers only uses post-processing for the weather ensemble data, thus before the weather is used as input to the wind power forecasting model. To the best of our knowledge no paper exists which evaluates whether post-processing at other stages benefits the probabilistic wind power forecast more.

In the present paper, we compare different post-processing strategies to determine at which stage in the wind power forecasting process the post-processing is most beneficial.  We analyse whether one post-processing step can account for all previous biases by only post-processing the wind power ensembles and compare this one-step strategy to post-processing only the weather ensembles, \ie  assuming the biases from the wind power model are negligible. Lastly, we also compare one strategy where we post-process both weather and power ensembles. We use publicly available synthetic benchmark data and wind power data from two bidding zones in Sweden to evaluate these post-processing strategies. The remainder of the present paper is structured as follows. \Cref{sec:methodology} introduces the theoretical background on the \gls{nwp} model. We then describe the post-processing strategies in detail in \Cref{sec:post_processing}, before evaluating these strategies in \Cref{sec:eval}. \Cref{sec:discussion} discusses our approach before \Cref{sec:conclusion} concludes the paper.

\section{Numerical Weather Prediction Models} \label{sec:methodology}

\begin{figure}
	\centering
	\includegraphics[width=0.85\textwidth]{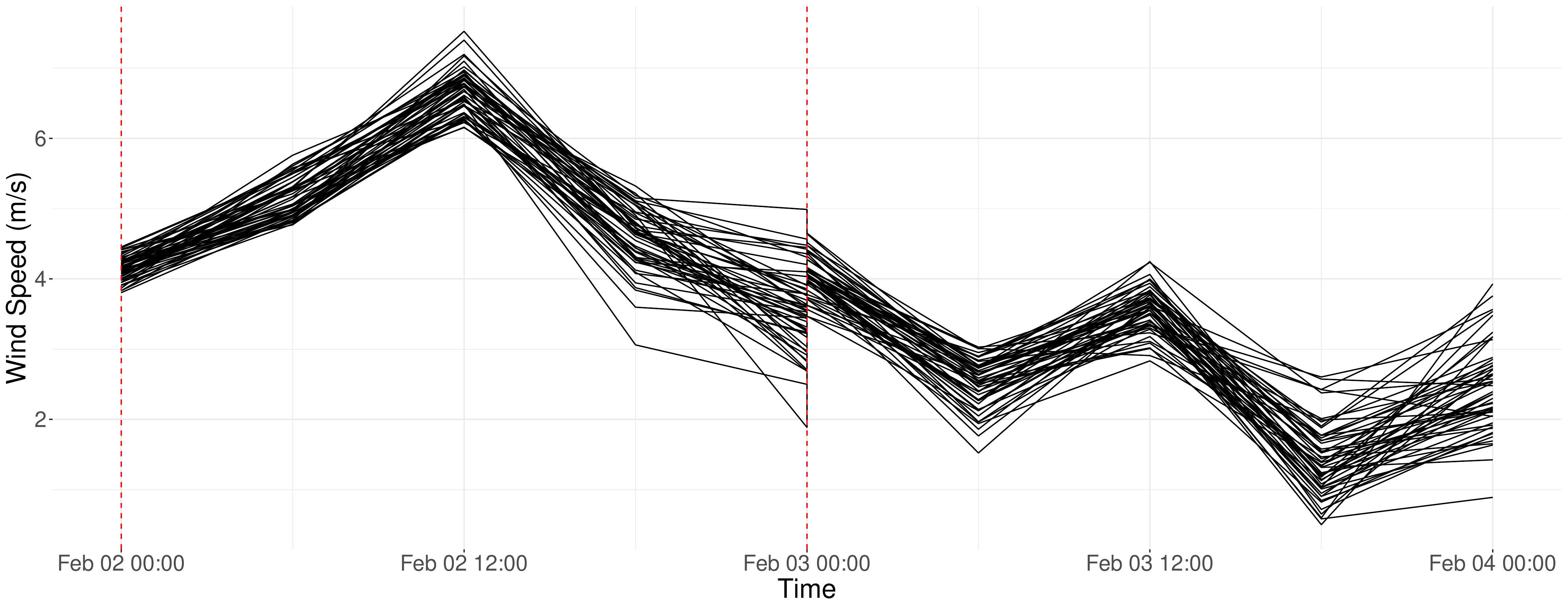}
	\caption{An illustration of two ensemble forecast trajectories over a 24h forecast horizon. The red dotted lines indicate the forecast origins.}
	\label{fig:ensemble_example}
\end{figure}

Atmospheric behaviour is chaotic and considered an unstable system which has finite, state dependent limits of predictability ~\cite{Lorenz.1963, Bauer.2015}. \glsfirst{nwp} models describe and forecast this atmospheric behaviour, and with it the weather, through solving a system of non-linear differential equations starting with the current observed atmospheric conditions.  As this current state of the atmosphere cannot be fully observed at any given point in time and space, there remains some uncertainty with regard to the initial conditions of the \gls{nwp} models.  However, forecasts of non-linear numerical models are highly sensitive to the given initial conditions, and initial errors grow during the forecast ~\cite{Bauer.2015, Poincare.2003}. Accounting for the uncertainty in the initial conditions is therefore crucial and nowadays quantified with the help of \glspl{eps}.  \glspl{eps} generate ensemble forecast by running the \gls{nwp} model several times with slightly different initial conditions, \eg adding perturbations to the initial state. Hence, today's weather forecasts provide an inherent probabilistic uncertainty estimate in the form of ensembles of \glspl{nwp}. \Cref{fig:ensemble_example} exemplarily shows the ensemble forecast trajectories over time for two consecutive forecast origins and a forecast horizon of one day. At each forecast origin, the \gls{eps} generates ensemble predictions for the specified forecast horizon. The forecast horizon describes the number of future time steps into which the weather is predicted.  With increasing forecast horizon the trajectories diverge, resulting in an increased uncertainty associated with the chaotic behaviour of the non-linear weather system. For more information on \gls{nwp} see \eg Bauer~et al.~\cite{Bauer.2015}.

\section{Ensemble Post-Processing Strategies} \label{sec:post_processing}

The focus of the present paper is on evaluating and comparing different post-processing strategies for probabilistic wind power forecasts.  Post-processing is performed to calibrate the forecasts to past observations such that systematic model biases can be alleviated. As the weather ensembles from the \gls{eps} are already known to be biased and underdispersed  ~\cite{Thorarinsdottir2010}, different approaches exist in the meteorological literature to calibrate ensembles~\cite{Vannitsem2021}. Two of the most common methods are \gls{emos} developed by Gneiting~et al.~\cite{Gneiting2005} and Bayesian Model Averaging (BMA) introduced by Raftery~et al.~\cite{Raftery2005}. Both of these approaches have been successful in post-processing various weather ensembles (see for example Javanshiri~et al.~\cite{javanshiri2021} and Han~et al.~\cite{Han2018}), and the difference in performance between the two models has been shown to be negligible~\cite{Han2018, Baran2013, Schulz2021}. However, \gls{emos} is computationally far simpler than BMA~\cite{vannitsem2018statistical,Schulz2021} and currently operational implementations of post-processing at weather services focus almost exclusively on \gls{emos}~\cite{gneiting2014calibration,Vannitsem2021}. For these reasons, the present paper focuses on \gls{emos}~\footnote{We also implement BMA and compare its performance to \gls{emos} when calibrating individual weather ensembles. Through this experiment, we confirm the higher computational cost for BMA and observe that the calibration performance for individual weather ensembles was inferior to \gls{emos}. Based on these results we only consider the better performing and widely used \gls{emos} method when comparing different post-processing strategies. We discuss this decision further in Appendix~\ref{sec:appendix:bma}.} when comparing post-processing strategies. In the following we describe the \gls{emos} post-processing method in detail, before discussing how it is used in our different post-processing strategies.

\subsection{Ensemble Model Output Statistics} \label{sec:emos}

The \gls{emos} method for ensemble post-processing, developed by Gneiting~et al.~\cite{Gneiting2005}, is based on non-homogenous regressions. The standard \gls{emos} approach is designed for ensemble members that are individually distinguishable, however, the present paper uses ensemble members from the \gls{ecmwf}. These \gls{ecmwf} ensemble members are classified as \emph{singular vector synoptic ensembles} and therefore exchangeable ~\cite{Molteni1996,Fraley2010}. Exchangeable ensembles represent equally likely future scenarios and have no distinguishing features or ordering. Thus, they are ensembles with a joint distribution function that is invariant under permutation of the arguments~\cite{Broecker2011}. This exchangeability implies that, for example, the ensemble labelled as first ensemble member $x_1$ at forecast origin $t$ is not related to the ensemble member with the same label at forecast origin $t+1$. Given exchangeable ensembles, \gls{emos} expresses a univariate weather quantity $Y$ in terms of a multiple linear regression on the $M$ ensemble members, with equal weights for each exchangeable ensemble member ~\cite{gneiting2014calibration}, i.e.~
\begin{equation}
	Y = a + b \cdot \sum_{i=1}^{M}  x_i + \epsilon.
\end{equation}
Hereby $x_1, \dots, x_M$ are exchangeable ensemble forecasts,  $a, b$ are regression coefficients, and $\epsilon$ is an error term with a mean of zero.  Given this deterministic forecast, we can create a probability density function or probabilistic forecast by assuming a distribution. The parameters of the distribution can then be modelled given the mean and variance of the individual ensemble members. For example, assuming a normal distribution for the corresponding weather variable,  we use the regression on $y$ as an approximation for the mean $\mu$ and approximate the variance $\sigma^2$ as a linear function of the ensemble spread
\begin{equation}
	\sigma^2 = c + d \cdot S^2
\end{equation}
where $S^2$ is the ensemble variance and $c$ and $d$ are non-negative coefficients. The resulting parametric model is then given by
\begin{equation}
	Y | x_1, \dots x_M \sim  \mathcal{N} \left( a + b \cdot \sum_{i=1}^{M}  x_i, c + d \cdot S^2 \right).
\end{equation}
Different variables require a different choice of distribution, for example, wind speeds are restricted to positive values and exhibit a skewed distribution, and \gls{emos} can be easily extended to these other distributions ~\cite{Gneiting2005}.

In order to estimate the EMOS coefficients Gneiting~et al.~\cite{Gneiting2005} use a minimum \gls{crps} estimation which is based on the minimum contrast estimation approach.  The \gls{crps} is a measure for calibration and sharpness of a predictive cumulative distribution function $F$ and is given by
\begin{equation}
	\mathrm{CRPS}(F, y) = \int_\mathbb{R} \left( F(y) - \bbone\{y \leq y \} \right) ^2 dy,
	\label{eq:crps}
\end{equation}
with $y$ the verifying observation and $\bbone$ denoting an indicator function.  Gneiting~et al.~\cite{Gneiting2005} show that the CRPS can be expressed as an analytical function and the EMOS coefficients that minimise the CRPS can be found through the Broyden-Fletcher-Goldfarb-Shanno~(BFGS) algorithm.

%The standard \gls{emos} approach is designed for ensemble members that are individually distinguishable. Ensemble members from the \gls{ecmwf} are however classified as \emph{singular vector synoptic ensembles} and therefore exchangeable ~\cite{Molteni1996,Fraley2010}. Exchangeable ensembles represent equally likely future scenarios and have no distinguishing features or ordering. Thus, they are ensembles with a joint distribution function that is invariant under permutation of the arguments~\cite{Broecker2011}. This exchangeability implies that for example the ensemble labelled as first ensemble member $x_1$ at forecast origin $t$ is not related to the ensemble member with the same label at forecast origin $t+1$.  Given exchangeable ensembles, the weights $b_1,\dots, b_M$ for each ensemble member should be equal, and we cannot directly apply the EMOS method proposed by Gneiting~et al.~\cite{Gneiting2005}. As a result the EMOS model considered in the present paper is the slightly modified variation from Gneiting and Katzfuss~\cite{gneiting2014calibration} that is adjusted for exchangeable ensembles.

\begin{figure*}
	\centering
	\includegraphics[width=0.7\linewidth]{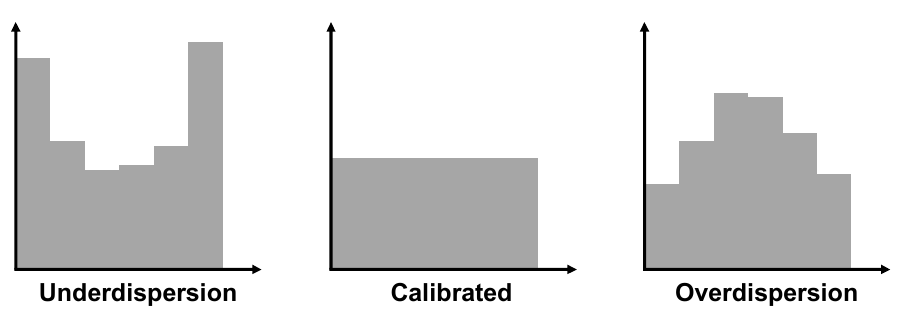}
	\caption{A sketch illustrating how to interpret PIT/ verification rank histograms. The ensembles are underdispersed if there is more mass in the outer bins (left diagram) and overdispersed if there is more mass in the central bins (right diagram). Well-calibrated ensembles are uniformly distributed (middle diagram).}
	\label{fig:pit_intro}
\end{figure*}

The general aim of post-processing and using \gls{emos} is to obtain calibrated forecasts. To check whether a forecast is calibrated, we look at the \emph{probability integral transform}~(PIT)~\cite{pit}. If $F$ denotes a fixed, non-random predictive cumulative distribution function~(CDF) for an observation $Y$, the PIT is the random variable $Z_F = F(Y)$. It is known that if $F$ is continuous and $Y \sim F$ then $Z_F$ is standard uniform. Thus, ideally the PIT from the given forecast is uniform.  In the discrete case, where we do not have a CDF but instead multiple ensemble members, the PIT can be described by the verification rank histogram~\cite{Hamill2001}. The verification rank histogram contains multiple bins formed from two ordered neighbouring ensemble members. Since in an ideal ensemble system the verifying ensembles are equally likely to fall within any of these bins, a rank histogram is also ideally uniformly distributed. Whilst based on slightly different principles, we interpret both PIT histograms and verification rank histograms in the same way~\cite{Hamill2001}. \Cref{fig:pit_intro} sketches the key information present in these histograms. If the post-processing is successful and the forecasts are calibrated then we observe a uniformly distributed histogram. If the forecasts are underdispersed (\ie they underestimate the true spread), then there are more observations in the outlying bins and less in the middle. When there is more mass in the middle of the histogram however, this is a sign of overdispersion (\ie the forecast overestimates the true spread).

%In the case where samples, \ie in our case ensemble members, are given instead of complete CDFs, we can use the empirical cumulative distribution function to calculate the \gls{crps} using
%\begin{equation}
%\hat{F}^{\mathrm{ECDF}}_m(z)= \frac{1}{m} \sum\limits_{i = 1}^m{\bbone \{ X_i \leq z  \}},
%\end{equation}
%which then results in
%\begin{equation}
%\mathrm{CRPS}(\hat{F}_m, y) = \frac{1}{m} \sum\limits_{i=1}^m{ |X_i - y| } - \frac{1}{2m^2} \sum\limits_{i=1}^m \sum\limits_{j=1}^m{|X_i-X_j|}
%\end{equation}
%as implementations of the above equation are rather inefficient, we use the improved formulation based in the order statistics $X_{(i)}, \ldots X_{(m)}$ and the generalized quantile function
%\begin{equation}
%\mathrm{CRPS}(\hat{F}_m, y) = \frac{2}{m^2} \sum\limits_{i=1}^m{(X_{(i)} - y)} \left( m \bbone \{ y < X_{(i)} \} - i + \frac{1}{2} \right).
%\end{equation}

\subsection{Strategies} \label{sec:strategy}

\begin{figure}
	\centering
	\includegraphics[width=\textwidth]{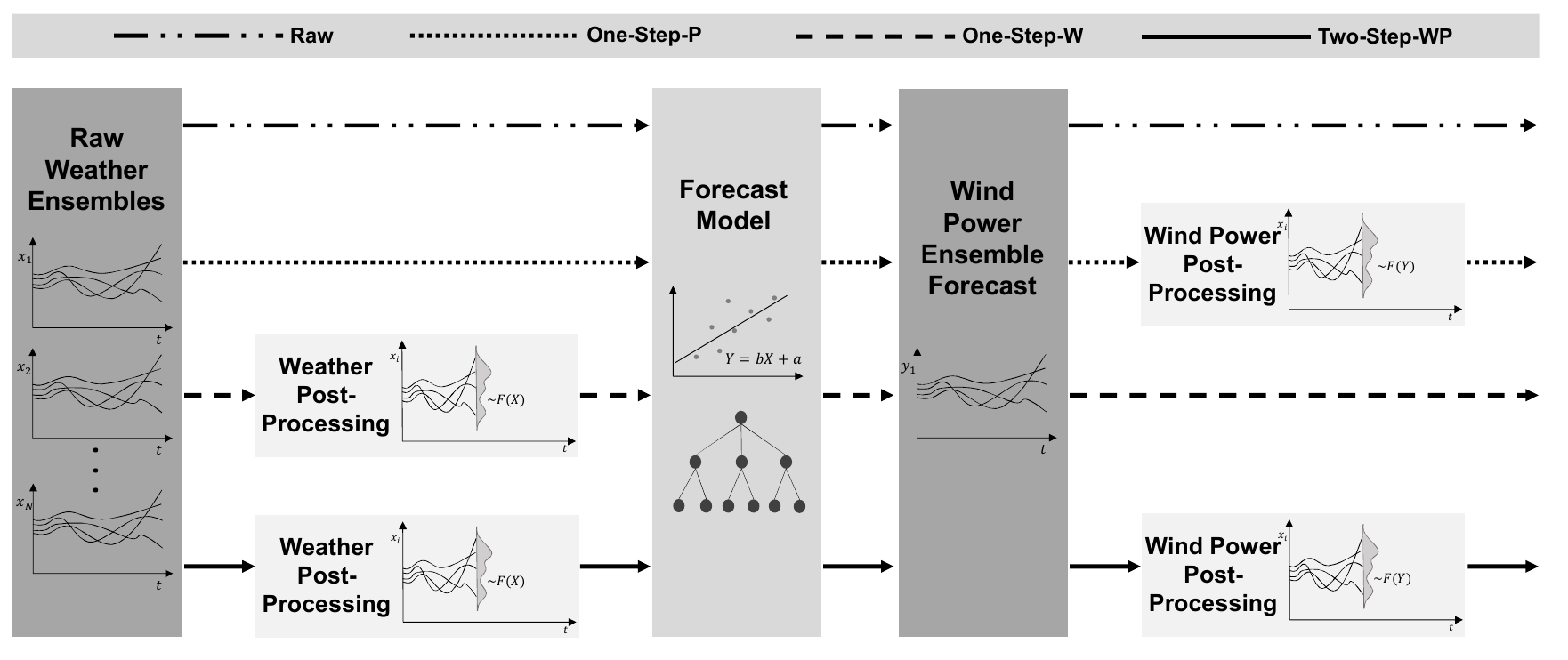}
	\caption{An overview of the post-processing possibilities available. We identify four strategies: No post-processing (Raw), post-processing only the power ensembles (\emph{One-Step-P}), post-processing only the weather ensembles (One-Step-W), and post-processing both weather and power ensembles (\emph{Two-Step-WP}).}
	\label{fig:approach_overview}
\end{figure}

The focus of the present paper is the evaluation and comparison of different post-processing strategies for probabilistic wind power forecasts.  Each strategy we evaluate starts with the raw weather ensembles, includes a forecasting model and has some form of probabilistic wind power forecast as output, which is then evaluated using the \gls{crps} (see \Cref{eq:crps}). The strategies differ in their use of the above introduced \gls{emos} post-processing.  \Cref{fig:approach_overview} provides an overview of the forecasting process given the four different strategies and we introduce each strategy in detail in the following.

The first strategy we introduce is called \emph{Raw}.  This strategy serves as our benchmark and does not include any form of post-processing. All available $M$ ensemble members from the \gls{eps} are fed through the same forecasting model individually, resulting in an ensemble of wind power forecasts with $M$ members.  As the result of this strategy is not an analytical form of a distribution function, we cannot use the above \gls{crps} formula to assess the quality of the forecast. Instead we use the sample-based \gls{crps} which relies on the empirical cumulative distribution function
\begin{equation}
	\hat{F}_M(y)= \frac{1}{M} \sum\limits_{i = 1}^M{\bbone \{ X_i \leq y  \}},
\end{equation}
and is calculated using
\begin{equation}
	\mathrm{CRPS}(\hat{F}_M, y) = \frac{1}{M} \sum\limits_{i=1}^M{ |X_i - y| } - \frac{1}{2M^2} \sum\limits_{i=1}^M \sum\limits_{j=1}^M{|X_i-X_j|}.
	\label{eq:sample_crps}
\end{equation}

In the second strategy, \emph{One-Step-P}, post-processing is applied once for the final wind power forecasting model.  This strategy relies on the assumption that post-processing the wind power ensembles also accounts for the biases in the weather ensembles. We again start with the $M$ raw ensemble members and feed them individually through the forecasting model. In contrast to the \emph{Raw} strategy, the resulting ensemble of wind power forecasts is post-processed. Thus, we fit the ensemble of wind power forecasts to the historical wind power generation using a rolling \gls{emos} approach, where the parameters are estimated every day on the basis of the past $40$ days.  Following Gneiting and Katzfuss~\cite{Gneiting2014}, we assume a truncated normal distribution for the wind power using $\mathcal{N}^+_{[0, \infty)}(\mu, \sigma^2)$, with location $\mu = a + b\bar{x}$ and scale $\sigma^2 = c+dS^2$ as an affine function of the ensemble variance $S^2 = \frac{1}{M} \sum_{i = 1}^M{(x_i - \bar{x})^2}$. The resulting wind power forecast can then be evaluated using the \gls{crps} for truncated normal distributions
\begin{equation}
\begin{split}
	\mathrm{CRPS}(F_{\mathrm{TN}}, y) = & \sigma \Big[ \frac{y - \mu}{\sigma} \Phi (\mu/\sigma) (2\Phi ((x-\mu)/\sigma) + \Phi (\mu/ \sigma) -2) \\
	& + 2\varphi ((y-\mu)/\sigma)\Phi(\mu/\sigma) - \frac{1}{\sqrt{\pi}} \Phi \left( \sqrt{2} \mu/\sigma \right) \Big] [\Phi (\mu/ \sigma)]^{-2}.
\end{split}
\end{equation}
As an alternative, we also consider a gamma distribution for the wind power. For simplicity, the main paper focuses on the results using the truncated normal distribution, with the almost identical gamma distribution results presented in the \Cref{sec:appendix:robust}.

% $\mathrm{Gamma}(\alpha, \beta) = F_{\alpha, \beta}$, with $\alpha\beta = a + b_1x_1 + \ldots b_Mx_M$ and $\alpha\beta^2 = c + d m_{\mathrm{ENS}}$ with $m_{\mathrm{ENS}} = a + b \sum_{m=1}^M{x_m}$.  The resulting wind power forecast is then has a gamma distribution based cumulative density function and can thus be evaluated with the \gls{crps} for gamma distributions
% \begin{equation}
	% \mathrm{CRPS}(F_{\alpha, \beta}, y) = y(2F_{\alpha, \beta}(y)-1) - \frac{\alpha}{\beta} (2F_{\alpha + 1, \beta}(y) -1) - \frac{1}{\beta B (\frac{1}{2}, \alpha)}.
	% \end{equation}

In the third strategy, \emph{One-Step-W}, we post-process only the weather ensembles. Thus, instead of using the raw ensembles, as for the two previous strategies, we post-process the ensembles and then draw $M$ independent samples from each of the resulting calibrated weather distributions. Whilst independent sampling does ignores possible dependency structures that exist between weather variables, Phipps et al.~\cite{phipps2020potential} show that restoring these structures with methods such as Ensemble Copula Coupling (ECC) has no noticeable effect on the results. Therefore we only consider a simple independent sampling method for this paper. Furthermore, we select $M$ samples to replicate the number of raw ensembles available. Overall, the \emph{One-Step-W} strategy relies on the assumption that all biases in the wind power forecast can be eliminated by accounting for the biases in the weather ensembles. For the post-processing, each weather variable is considered separately and for each weather variable we use EMOS with a rolling training window to estimate the parameters for the appropriate distributions. We use the distributions as suggested by Gneiting~\cite{gneiting2014calibration}, thus a normal distribution for temperature, a normal distribution for the u- and v-components of wind, and a truncated normal distribution or a gamma distribution for wind speed~\footnote{Again, the gamma and truncated normal distribution delivered similar performance. For simplicity we focus on the results from the truncated normal distribution in the main paper and report full results in \Cref{sec:appendix:robust}.}. Given these distributions, we draw $M$ independent random samples from each distribution to form weather calibrated ensemble members. We then feed these sampled ensemble members through the forecasting model, resulting in an ensemble of wind power forecasts whose accuracy can be assessed with the sample-based CRPS from \Cref{eq:sample_crps}.

In the fourth and last strategy, \emph{Two-Step-WP}, both the weather ensembles and the ensemble of wind power forecasts are post-processed. This strategy relies on the assumption that neither of the one-step approaches can sufficiently account for all biases in the models and data and the forecasts should be post-processed at all stages in the forecasting process. Thus, the two one-step strategies are coupled together, where we first apply the procedure described for \emph{One-Step-W} to the weather ensembles and then \emph{One-Step-P} to the ensemble of wind power forecasts. The distribution choices, parameters, and CRPS calculations remain the same as for the individual strategies.

\subsection{Forecasting Model} \label{sec:forecastModels}

We aim to evaluate ensemble post-processing at different stages in the forecasting process,  thus, we also need an appropriate wind power forecasting model. This forecasting model should represent the relationship between the different weather variables and the wind power and in contrast to a manufacturers wind power curve model, can be estimated regardless of the turbine type and for aggregated wind power values (\eg the sum of wind power from a large region with an unknown number of wind turbines). Since we focus on the comparison of different post-processing strategies and not raw forecast accuracy, we want the models to be simple, common in the literature, and lightweight with respect to parameter optimisation and computation time. Research has shown, that post-processing generally improves regression results, even when deep learning methods are applied, see \eg~Kuleschov et al.~\cite{pmlr-v80-kuleshov18a}, and therefore we expect the post-processing performance to be similar independent of the model chosen. We therefore choose one linear model, namely a linear regression~\cite{Wu2007}, as a simple benchmark model and one more complex non-linear model, namely a random forest~\cite{Lydia2010}. Furthermore, to show the robustness of our results we additionally implement an alternative linear regression model based on Zhang and Wang~\cite{ZHANG20161074} and a simple artificial neural network as an alternative non-linear forecasting model. In the following, we describe the linear regression and random forest models as well as how we measure the forecasting accuracy. The additional linear regression and neural network models are presented in \Cref{sec:appendix:robust}.

The linear regression model which we use to forecast the wind power given the weather input can be described with
\begin{equation}
	y_{t+h} = \beta_0 + \alpha y_{t+h-24} + \sum\limits_{k = 1}^K{\beta_k W^k_{t+h}} + \sum\limits_{j = 1}^J{\gamma_j D^j_{t+h}} + \varepsilon_{t+h},
\end{equation}

where $y_t$ is the dependent variable, which in this case is the wind power, $y_{t-24}$ is the actual wind power a day before, $W^k$ are weather time series, such as wind speed, the u-component of wind, the v-component of wind, surface pressure, and temperature, and $D^j$ are dummy variables, such as the season, the month, and the year. The models are fitted for each forecast horizon $h = {h_1,  \ldots,  h_H}$ with $h_H \leq 24$, using actual historical weather data in order to describe the real relationship among the variables and remove any bias fitting on historical weather forecasts or ensembles could introduce. Each ensemble $x_1 \ldots x_M$ from the \gls{eps} is then used in a separate prediction run for each forecast horizon to generate an ensemble of wind power predictions with the previously fitted regression coefficients
\begin{equation}
	\hat{y}_{t+h}(x_1,  \ldots,  x_M) = \hat{\beta}_0 + \hat{\alpha} y_{t+h-24} + \sum\limits_{k = 1}^K{\hat{\beta}_k \widehat{W}^{k}_{t+h} (x_1,  \ldots, x_M)} + \sum\limits_{j = 1}^J{\hat{\gamma}_j D^j_{t+h}} + \hat{\varepsilon}_t.
\end{equation}

In order to better forecast non-linear dependencies, we also implement a random forest. 
Random forests are a statistical learning method and are additionally to their ability to model non-linearity also robust to errors when unnecessary predictors are included.  Random forests create several de-correlated regression trees to form a collection of solutions. These de-correlated trees are fitted by selecting a subset of features at each candidate split.  The final prediction is then the average over all regression trees. We use a random forest with $500$ trees and the same input variables as for the linear regression to predict the wind power. Giving the random forest the full set of input variables available, specifically multiple weather features, allows it to automatically identify the most important features and hereby generate accurate forecasts. The parameters of the random forest are, equivalently to the linear regression, estimated using the historical values, while the probabilistic ensemble forecasts is generated by running the forest on each ensemble member.

%We tested multiple neural network configurations before selecting a configuration with two hidden layers of 10 and 7 neurons respectively and trained it with the resilient backpropagation algorithm. This network architecture was selected because it is the simplest we found that still returns accurate forecasts. The chosen activation function is a hyperbolic tangent and the input features remain the same as for the linear regression model explained above. Again, the parameters (\ie weights) are fitted using the actual historical weather data and each ensemble member is passed through the network to get an ensemble of wind power forecasts.

The forecasting strategy is the same, regardless of the specific forecasting model. Given the forecast origin, we use weather forecasts $\widehat{W}^k_{t+h}$ from the \gls{eps} to predict the wind power $\hat{y}_{t+h}$ for each of the forecasts horizons $h$ (between 3h and 24h and determined by the temporal resolution of the EPS). Additionally to the weather variables, we also include historical wind power generation from 24h before the prediction time and dummy features $D^j_{t+h}$, such as the time of day or month. Due to the fact that the historical weather observations are implicitly included in the \gls{nwp} model and the calibration process relies on these observations, we do not include historical weather information as a specific input for our prediction models. All forecasting models are implemented in R, and for the random forests we use the \emph{randomForest} package~\footnote{\url{https://cran.r-project.org/web/packages/randomForest/randomForest.pdf}}. We apply the forecast strategy explained above for every post-processing strategy shown in \Cref{sec:strategy}, such that the forecasting strategy does not influence our comparison. To assess the performance, we calculate the average \gls{crps} over all time steps $t = 1, \ldots, N$ in the test set
\begin{equation}
	\mathrm{CRPS} = \frac{1}{N} \sum\limits_{t = 1}^{N}{\mathrm{CRPS}(F_t, y_t)}.
\end{equation}
As we are mainly interested in a fair comparison of our post-processing strategies, we also calculate the \gls{crpss}. This score measures the improvement in \gls{crps} when compared to a given benchmark model. In our case the \gls{crpss} is calculated for all post-processing strategies with respect to the  \emph{Raw} strategy as the benchmark. We report the \gls{crpss} as a percentage, where positive results indicate an improvement.

\section{Evaluation} \label{sec:eval}

Given the four ensemble post-processing strategies introduced in the previous section, we want to evaluate their performance by comparing the accuracy, sharpness, and calibration of the resulting wind power forecasts. Before we report the results, we introduce the two data sets we use: an open source benchmark data set (\Cref{sec:dataBench}) and a data set with real wind power generation data from two bidding zones in Sweden (\Cref{sec:dataSweden}). In \Cref{sec:results} we present the results of different post-processing strategies using only the truncated normal distribution, whilst the results of further experiments with different distributions are reported in \Cref{sec:appendix:robust}.

%In our case the available ensemble predictions determine our forecast horizon. In the case of the benchmark data set we have forecast horizons from 6h-24h in steps of 6h, whilst the Swedish data set has forecast horizons from 3h-24h in 3h steps. Since all of these forecasts are made based on the weather information available at the forecast origin (i.e.~$t=0$), the larger the forecast horizon the greater the uncertainty associated with the forecast.

\subsection{Benchmark Data} \label{sec:dataBench}

Due to the lack of open source wind power data for specific wind parks, we generate benchmark wind power using the \emph{Renewables.ninja} API~\footnote{\url{www.renewables.ninja}.}. Staffell and Pfenninger~\cite{Staffell2016} verify that the simulation and bias-corrections implemented in the Renewables.ninja API are capable of reproducing accurate wind power time series. We replicate two real German wind power parks, one onshore and one offshore.  \Cref{tab:benchmark_wind_config} shows the parameters which we use, where we selected turbines with similar characteristics than those installed using the wind turbine database~\footnote{\url{https://en.wind-turbine-models.com/turbines}}. 

\begin{table}
	\centering
	\footnotesize
	\caption{Configuration parameters used to generate the wind power time series with the \emph{Renewable.ninja} API.}
	\label{tab:benchmark_wind_config}
	\begin{tabular}{p{3cm}|p{6cm}p{4cm}}
		\toprule
		& Onshore Benchmark & Offshore Benchmark \\
		\midrule
		Coordinates &  51.0\textdegree N, 10.5\textdegree E & 54.5\textdegree N, 6.0\textdegree E \\
		Time Span & 01-02-2017 -- 31-08-2018 & 01-02-2017 -- 31-08-2018 \\
		Capacity & 130 MW & 400 MW \\
		Turbine Height & 95m & 90m \\
		Turbine Type & Vestas V90 2000 & Gamesa G128 5000 \\
		Similar Real Windpark & Windfeld Wangenheim-Hochheim-Ballst\"adt-Westhausen & BARD Offshore I \\
		\bottomrule
	\end{tabular}
\end{table}

We use \gls{tigge} archive~\footnote{\url{https://apps.ecmwf.int/datasets/data/tigge/}} to access open source ensemble weather data. \gls{tigge} archive is a result of \emph{The Observing System Research and Predictability Experiment} which aimed to combine ensemble forecasts from leading forecast centres to improve probabilistic forecasting capabilities ~\cite{Swinbank2016}. The archive includes a limited sample of the \gls{ecmwf} ensembles from October 2006 until the present. The limitations are placed on the available forecast horizons (only in steps of 6h instead of 3h in the licensed version of the \gls{eps}), the number of weather variables (e.g.~wind speed and wind components are only available at a height of 10m and not also on 100m) and a reduced spatial resolution compared to the operational \gls{eps}. \gls{tigge} archive is however publicly accessible and data can be downloaded through the MARS API. We use weather data for the same locations as the synthetically generated wind power data (see \Cref{tab:benchmark_wind_config}). We include the parameters temperature at two meter above ground, surface pressure, u-Component and v-Component of wind  at 10m, and wind speed, for the period from February 2017 until August 2018. The limited time span of data available is due to damaged tapes in \gls{tigge} archive which affected all ensemble data, and the aforementioned period is the longest available with continuous weather data at the time of writing. For the ground truth historical weather data, we use the ERA5 reanalysis data ~\cite{ERA5_cite}. This data is accessed via the Copernicus Climate Data Store (CDS) API~\footnote{\url{https://cds.climate.copernicus.eu/home}}. Here, the same locations and identical parameters are included. When working with the benchmark dataset, we use the entire  year 2017 for training and the remainder of the data (01.2018-08.2018) for the evaluation.

In order to allow replication of the results and further development of the methods presented in the current paper, the ensemble weather data and the synthetic wind power time series are made available through GitHub: \url{https://github.com/KIT-IAI/EvaluatingEnsemblePostProcessing}. Due to licensing constraints, the ERA5 reanalysis data must be downloaded separately. We provide instructions for downloading and formatting the reanalysis data in the repositories documentation.

\subsection{Swedish Data Set} \label{sec:dataSweden}

\begin{figure*}
	\centering
	\includegraphics[width=0.7\linewidth]{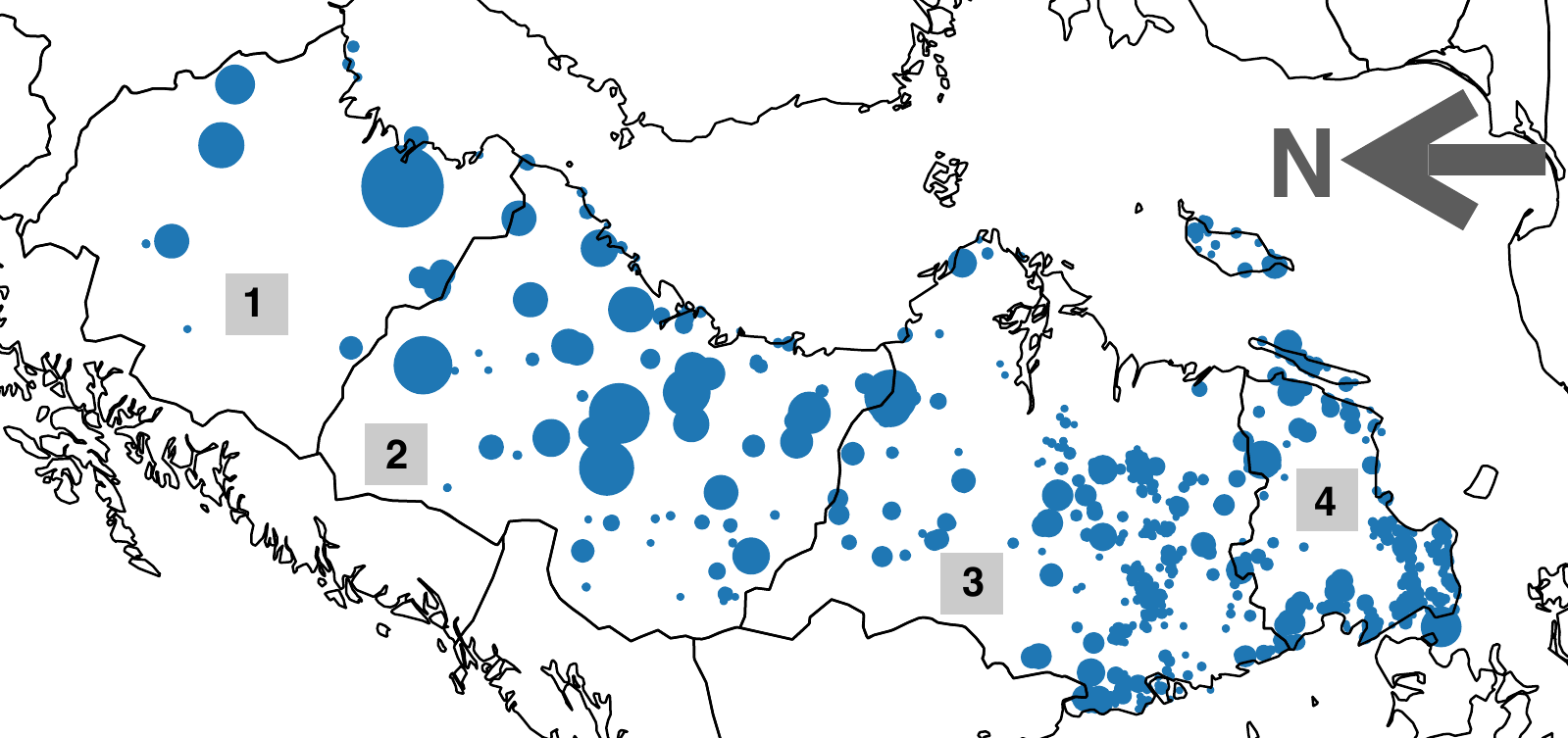}
	\caption{A map of Sweden with the four bidding zones shown through the border lines. The distribution of wind turbines is indicated by the blue circles. The figure is adapted from Olauson and Bergkvist~\cite{Olauson2015}}
	\label{fig:sweden_map}
\end{figure*}

The second data set we use contains data from the Swedish electricity system. The Swedish electricity system is divided into four sub-areas or bidding zones ~\cite{maria7}. The division of the electricity system into sub areas has several purposes ~\cite{maria8}. One purpose is to create regional price differences between the sub areas. This is an incentive for cost-effective further development of the electricity system, as new power plants will be built where there is a shortage of electricity. Another purpose is that state-owned Svenska Kraftnät, which operates the national grid, should receive indications about where the national grid needs to be strengthened to be able to transfer enough electricity to the other sub areas. A third purpose is to comply with EU legislation and thereby facilitate the continued integration of the Swedish electricity system with the European electricity market.

In the present paper, we focus on the area contained in bidding zones 3 and 4, for two reasons. Firstly, the large majority of wind power generation occurs in these two bidding zones (see \Cref{fig:sweden_map}). Secondly, these are the two bidding zones in Sweden that are sometimes faced with a lack of electricity ~\cite{maria8}. Whilst in northern Sweden the supply of electricity is normally greater than the demand, transmission capacity between north and south of Sweden is not always sufficient to transfer the demanded electricity and this can lead to bottlenecks ~\cite{maria8}.

\begin{table}
	\centering
	\footnotesize
	\caption{A summary of the key characteristics of the weather data available for the use case in Sweden.}
	\label{tab:weather_summary}
	\begin{tabular}{p{3cm}|p{10cm}}
		\toprule
		Characteristic & Notes \\
		\midrule
		Temporal Dimension & 05.01.2015 -- 31.08.2019 \\
		Spatial Dimension & 11\textdegree E -- 19.5\textdegree E \newline 54\textdegree N -- 62.5\textdegree N \\
		Spatial Resolution & Grid resolution of 0.25\textdegree $\times$ 0.25\textdegree \\
		Forecast Time & Forecasts for up to 24h ahead made at 00:00:00 \\
		Forecast Horizon & One step ahead forecasts for 3h, 6h, 9h, 12h, 15h, 21h and 24h \\
		Weather Variables & 100m u-component of wind, 100m v-component of wind, 100m wind speed, 2m temperature, surface pressure\\
		\bottomrule
	\end{tabular}
\end{table}

Weather data for bidding zones 3 and 4 consists of the \gls{ecmwf} \gls{eps} (Molteni~et al.~\cite{Molteni1996}) and also the ERA5 reanalysis data C3S~\cite{ERA5_cite}. We use the \gls{eps} as the foundation for probabilistic forecasting methods, and again use the ERA5 reanalysis data as the ground truth for the post-processing. Table~\ref{tab:weather_summary} summarises the key aspects of the data. The weather data available is in a grid-based format. This means that atmospheric models are used to create a \gls{nwp} for certain geographical grid-points on earth. Since we are not considering a single wind park but looking at the aggregated wind power generation for each bidding zone, we considered all of these grid points in the form of a weighted average. The weighted average is calculated as follows; Firstly each of the data points is sorted into the appropriate bidding zone based on the geographical specifications (see Figure~\ref{fig:sweden_map}), then a weighted average of every point is calculated. Due to a lack of accurate location data for various wind turbines in Sweden a rudimentary weighted average method is used; areas with a high concentration of wind turbines are given double weighting, whilst those areas with a lower concentration only receive a standard weight. As seen in Figure ~\ref{fig:sweden_map}, the doubly weighted areas include a central area in bidding zone 3 and the coastal areas in bidding zone 4.

\begin{table}
	\centering
	\footnotesize
	\caption{A statistical summary of the wind power generation data for both benchmark data sets and for bidding zones 3 and 4 in Sweden.}
	\label{tab:wind_summary}
	\begin{tabular}{p{2cm}|p{3cm}p{3cm}p{2cm}p{2cm}}
		\toprule
		& Onshore Benchmark & Offshore Benchmark & Bidding Zone 3 & Bidding Zone 4 \\
		\midrule
		Minimum & 0.13 MW & 1.2 MW & 3 MW & 2 MW \\
		1st Quartile & 11.31 MW & 125.6 MW & 278 MW & 167 MW \\
		Median & 23.53 MW & 236.8 MW & 568 MW & 352 MW \\
		Mean & 29.12 MW & 223.9 MW & 670.96 MW & 447.55 MW \\
		3rd Quartile & 40.04 MW & 333.2 MW & 967 MW & 664 MW\\
		Maximum & 128.83 MW & 385.6 MW & 2246 MW & 1463 MW \\
		\bottomrule
	\end{tabular}
\end{table}

The wind power generation data is available through the open source transparency platform which is operated by the European Network of Transmission System Operators (ENTSO-E) ~\cite{entso_e}. The transparency platform provides aggregated onshore wind power generation data at an hourly resolution from the 05.01.2009 until present. This data is aggregated for each bidding zone in Sweden (zone 1-4), where we consider bidding zones 3 and 4 in the present paper. We use data from 2015-2017 for the training of our models and then from 01.2018-08.2019 for the evaluation. \Cref{tab:wind_summary} provides a statistical summary of the wind power generation data collected. The data collected is the raw wind power generation in Megawatt and therefore affected by structural changes such as an increase in capacity, outages due to maintenance, and upgrades to wind turbines. 

% Figure~\ref{fig:wind_years} plots the wind power generated in each bidding against the weighted mean of the wind speed for the same bidding zone. The data points are colour coded according to the year the data was collected. It can be seen, that whilst there are small differences between the years, there is no clear separation that signifies a structural change. This indicates that calendrical dummy variables that include the year of the forecast should be sufficient to account for inconsistencies across the years and we therefore do not introduce any further correction factors.

%\begin{figure}[t]
%	\centering
%	\includegraphics[width=\textwidth]{figs/new_wind_vs_power_sweden.pdf}
%	\caption{We see the weighted mean of the wind speed plotted against the power generation for both bidding zone 3 (left) and bidding zone 4 (right) in Sweden. The data points are colour coded according to their year. We see, that there is no clear separation between the years, with the colours mostly overlapping. The small trend seen in bidding zone 3 can be accounted for by including the year as an input in the forecast model.}
%	\label{fig:wind_years}
%\end{figure}

\subsection{Results}\label{sec:results}

%\stress{Take means and use MAPE (but this is not probabilistic)?}
In this section we present the evaluation results given the different post-processing strategies and the forecasting models introduced in \Cref{sec:post_processing}.  We start with the results for the benchmark data set (\Cref{sec:evalBench}) before reporting the results for the Swedish data set (\Cref{sec:evalSweden}).

\subsubsection{Benchmark Data} \label{sec:evalBench}

\begin{figure}
	\centering
	\includegraphics[width=\textwidth]{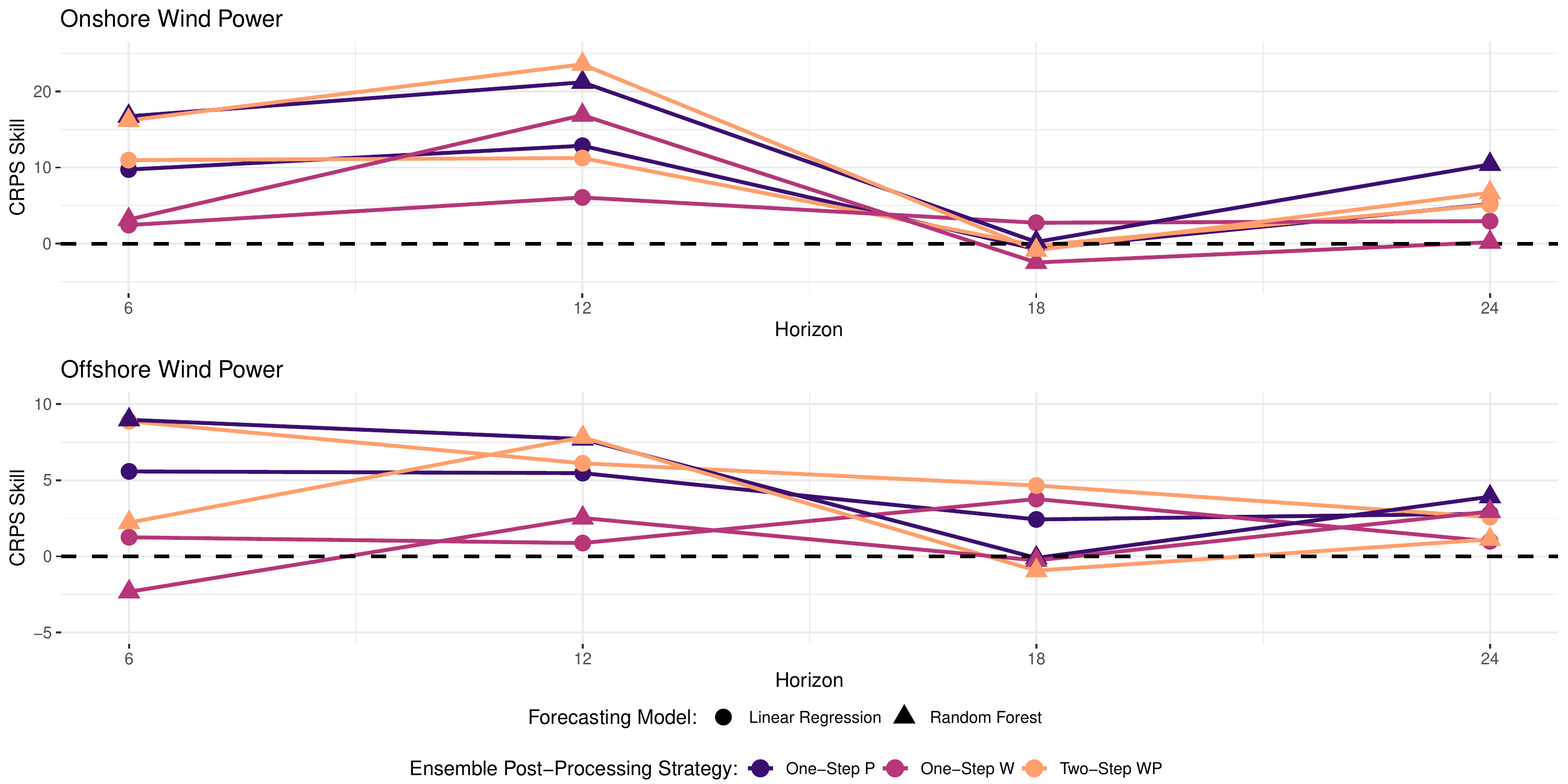}
	\caption{The CRPS skill score plotted against the forecast horizon on the test data for the onshore benchmark (top figure) and the offshore benchmark (bottom figure). Positive values indicate an improvement over the \emph{Raw} strategy in percent. In general post-processing leads to an improvement in forecasting performance, although only post-processing the weather has little impact or performs slightly worse.}
	\label{fig:crps_benchmark}
\end{figure}

We first perform the different post-processing strategies on the benchmark data set and evaluate the calibration and sharpness based on the above introduced \gls{crps} and PIT/ verification rank histogram for both forecasting models with varying forecast horizon. \Cref{tab:crpsResultsBenchmark} summarises the mean \gls{crps} for the onshore and offshore benchmark data set with the best value highlighted in bold. We can see that for almost every model and every forecast horizon either the \emph{One-Step-P} or \emph{Two-Step-WP} post-processing strategy perform the best. \Cref{fig:crps_benchmark} plots the \gls{crpss} for each post-processing strategy across all forecast horizons. We see that post-processing almost always leads to an improvement in \gls{crps} when compared to the \emph{Raw} strategy. The improvement depends on the forecast horizon and the data set, but is typically between 5\% and 15\%.

\begin{table}[t]
	\centering
	\footnotesize
	\caption{Summary of mean CRPS on the test data for the benchmark data sets and for all forecast horizons. The best prediction for each strategy, forecast horizon, and model is highlighted in bold.}
	\label{tab:crpsResultsBenchmark}
	\begin{tabular}{c|rrrrr}
		\toprule
		Data Set & & 6h & 12h & 18h & 24h \\
		\midrule
		& Linear Raw &   3.91 & 5.60 & 5.67 & 6.08\\
		& Linear \emph{One-Step-P} &   3.53 & \textbf{4.88} & 5.71 & \textbf{5.76}\\
		& Linear One-Step-W & 3.81 & 5.26 & \textbf{5.51} & 5.90 \\
		Onshore & Linear \emph{Two-Step-WP}  &   \textbf{3.48} & 4.97 & 5.68 & 5.77 \\
		Benchmark &\CC{20}Random Forest Raw &\CC{20}  3.98 &\CC{20} 5.34 &\CC{20} 5.68 &\CC{20} 5.84 \\
		&\CC{20}Random Forest \emph{One-Step-P}  &\CC{20} \textbf{3.61} &\CC{20} 4.63 &\CC{20} \textbf{5.84} &\CC{20} \textbf{5.38} \\
		&\CC{20}Random Forest One-Step-W  &\CC{20} 3.97 & \CC{20} 4.53 & \CC{20} 5.78 & \CC{20} 5.83  \\
		&\CC{20}Random Forest \emph{Two-Step-WP}  &\CC{20}3.67 &\CC{20} \textbf{4.34} &\CC{20} \textbf{5.84} &\CC{20} 5.70 \\
		\midrule
		& Linear Raw &  18.98 & 23.51 & 25.46 & 24.11 \\
		& Linear \emph{One-Step-P} &  17.92 & 22.23 & 24.85 & \textbf{23.44}  \\
		& Linear One-Step-W  &  18.74 & 23.31 & 24.51 & 23.88 \\
		Offshore & Linear \emph{Two-Step-WP} &  \textbf{17.30}& \textbf{22.08} & \textbf{24.28} & 23.49 \\
		Benchmark &\CC{20}Random Forest Raw  & \CC{20}20.77 & \CC{20}23.80 & \CC{20}25.00 &\CC{20} 25.54  \\
		&\CC{20}Random Forest \emph{One-Step-P} &\CC{20} \textbf{19.47} &\CC{20} \textbf{22.15} &\CC{20}\textbf{24.24} &\CC{20} \textbf{23.83} \\
		&\CC{20}Random Forest One-Step-W  &\CC{20}21.45 &\CC{20} 23.22 &\CC{20} 25.41 &\CC{20} 24.93 \\
		&\CC{20}Random Forest \emph{Two-Step-WP}  &\CC{20}20.47 &\CC{20} 22.43 &\CC{20} 25.56 &\CC{20} 24.08 \\
		\bottomrule
	\end{tabular}
\end{table}

\begin{sidewaysfigure}
	\centering
	\includegraphics[width=\textwidth]{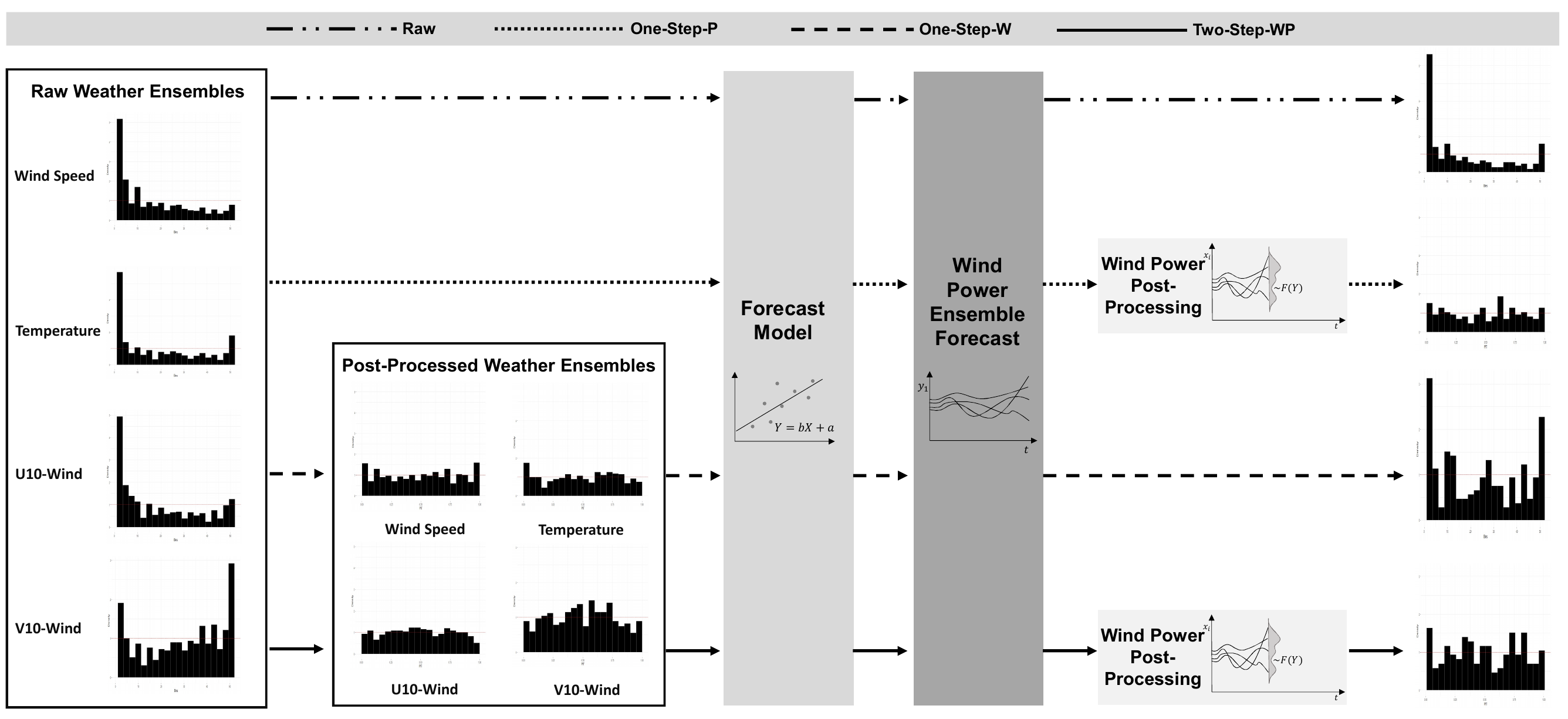}
	\caption{An overview of the post-processing results for a forecast horizon of 12h on the onshore benchmark data set. Although each individual weather variable is calibrated before the wind power forecast, this does not result in a calibrated wind power ensemble. In order to achieve a calibrated wind power ensemble, post-processing after the wind power forecast is required.}
	\label{fig:onshore_calibration_results}
\end{sidewaysfigure}

We next take a closer look at the calibration,  \ie PIT/ verification rank histograms. For brevity's sake, we focus in the following on the onshore data set and pick a forecast horizon of 12h (note that the other data set and forecast horizons behave similarly and are therefore not presented in detail). The effect of the various post-processing strategies on the calibration is shown in \Cref{fig:onshore_calibration_results}, where we plot the different verification rank histograms and PITs at each forecasting step. The raw weather ensembles for wind speed, temperature, and the u-component of wind measured at 10m show a clear positive bias and are slightly underdispersed. The v-component of wind measured at 10m does not show the same positive bias, but is also underdispersed. After the weather ensembles are post-processed the resulting PIT histograms appear almost uniform. Interestingly, the v-component of wind now appears to be slightly overdispersed which indicates an overcorrection in the post-processing. The wind power ensemble forecasts generated with the raw weather ensembles are underdispersed, but there is almost no improvement when the post-processed weather ensembles are used. The ensemble forecast from \emph{\emph{One-Step-P}} appears uniform, whilst the PIT histogram for \emph{\emph{Two-Step-WP}} also indicates a slight overcorrection with overdispersion present. 

\subsubsection{Swedish Data} \label{sec:evalSweden}

\begin{figure}[t]
	\centering
	\includegraphics[width=\textwidth]{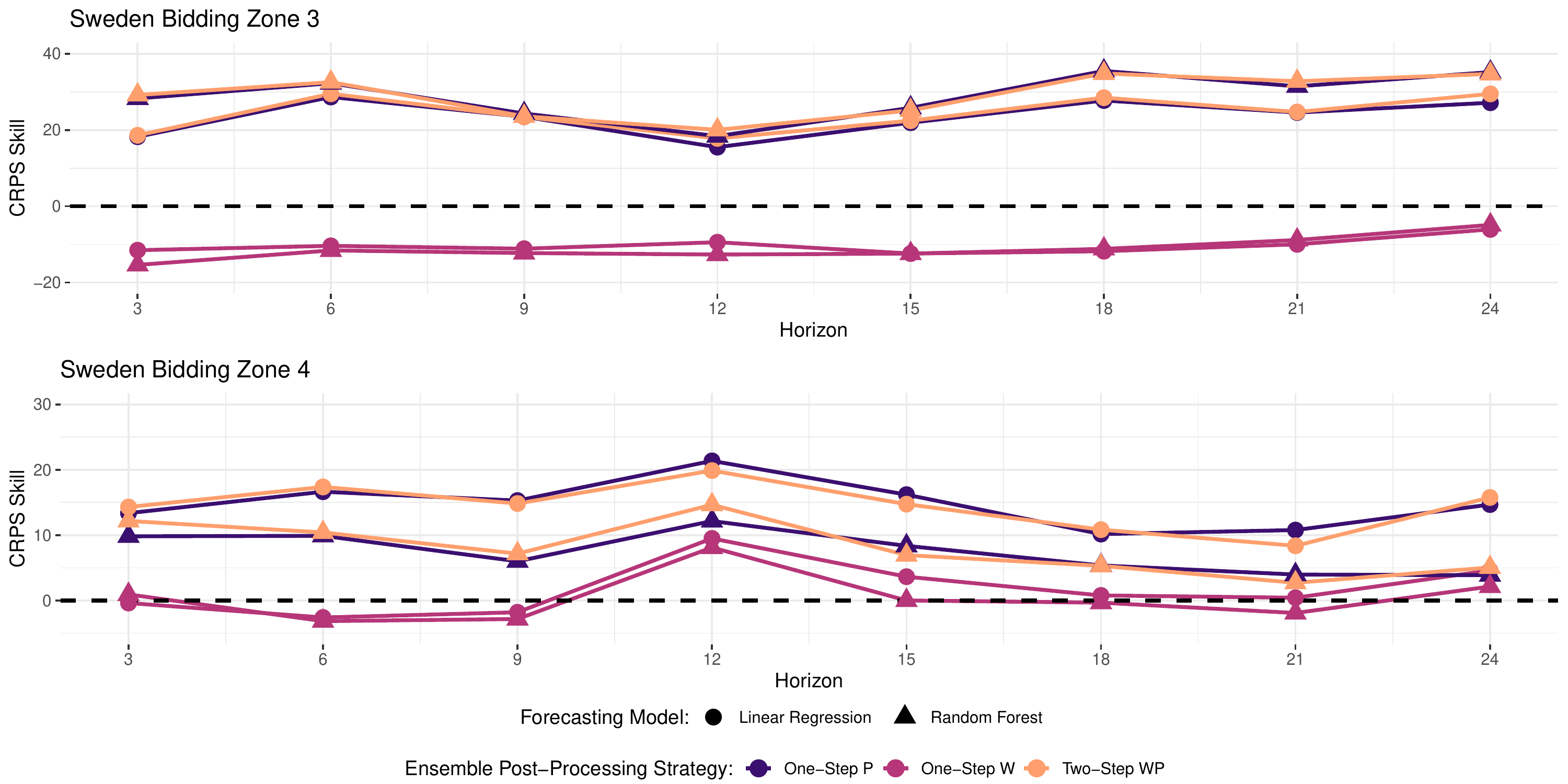}
	\caption{The CRPS skill score plotted against the forecast horizon on the test data for both bidding zone 3 (top figure) and bidding zone 4 (bottom figure) in Sweden. Positive values indicate an improvement over the \emph{Raw} strategy in percent. Post-processing the final power ensemble, either directly or as part of a two-step process clearly improves the forecast, whilst only post-processing the weather ensemble leads to worse or similar forecast performance.}
	\label{fig:crps_sweden}
\end{figure}

This section focuses on the evaluation of the Swedish data set. We again evaluate the mean \gls{crps} (a summary of these values is shown in \Cref{tab:crpsResultsSweden}) for each model and each forecast horizon. We also consider the \gls{crpss} in \Cref{fig:crps_sweden}, where the improvement in \gls{crps} compared to the \emph{Raw} strategy is plotted for each bidding zone in Sweden against the forecast horizon. Considering bidding zone 3, we see that processing the final power ensemble (\emph{One-Step-P} or \emph{Two-Step-WP}), leads to vast improvements in \gls{crps}, with a \gls{crpss} ranging between 20\% and 40\%. However, the \emph{One-Step-W} strategy, which only processes the weather ensembles, leads to a decrease in performance of approximately 10\%. For bidding zone 4, the results are similar. Calibrating the power ensemble again results in clear \gls{crps} improvements, whilst only calibrating the weather ensembles leads to slightly worse performance, or performance very similar to the \emph{Raw} strategy (excluding a forecast horizon of 12h). Furthermore, we observe that although the mean \gls{crps} for different forecasting models (linear regression and random forests) are noticeably different, the \gls{crpss} results are very similar. This can be observed on both the bidding zone 3 and bidding zone 4 data set. 

\begin{table}[t]
	\centering
	\scriptsize
	\caption{Summary of mean CRPS on the test data for the use case in Sweden for all forecast horizons. The best prediction for each strategy and each forecast model is highlighted in bold.}
	\label{tab:crpsResultsSweden}
	\begin{tabular}{c|rrrrrrrrr}
		\toprule
		Data Set & & 3h & 6h & 9h & 12h & 15h & 18h & 21h & 24h \\
		\midrule
		& Linear Raw &  57.84 & 78.46 & 81.03 & 79.75 & 86.20 & 89.05 & 89.06 & 93.33 \\
		& Linear \emph{One-Step-P}& 47.30 & 55.99 & \textbf{62.01} & 67.39 & 67.31 & 64.36 & 67.17 & 68.00 \\
		& Linear One-Step-W & 64.51 & 86.61 & 90.05 & 87.27 & 96.89 & 99.55 & 97.95 & 98.99 \\
		Bidding & Linear \emph{Two-Step-WP} & \textbf{47.08} & \textbf{55.32} & 62.04 & \textbf{65.56} & \textbf{66.86} & \textbf{63.70} & \textbf{67.05} & \textbf{65.83}  \\
		Zone 3 &\CC{20}Random Forest Raw  & \CC{20}60.71 & \CC{20}71.31 & \CC{20}69.72 & \CC{20}67.74 & \CC{20}73.39 & \CC{20}82.01 & \CC{20}83.67 &\CC{20} 88.09 \\  
		&\CC{20}Random Forest \emph{One-Step-P} &\CC{20}46.08 &\CC{20} 50.28 &\CC{20} \textbf{55.09} &\CC{20} 55.96 &\CC{20} \textbf{55.17} &\CC{20} \textbf{53.44} &\CC{20} 57.83 &\CC{20} 58.86 \\
		&\CC{20}Random Forest One-Step-W &\CC{20}69.96 &\CC{20} 79.95 &\CC{20} 78.21 &\CC{20} 76.50 &\CC{20} 82.77 &\CC{20} 90.38 &\CC{20} 91.47 &\CC{20} 91.15\\
		&\CC{20}Random Forest \emph{Two-Step-WP} &\CC{20}\textbf{45.20} &\CC{20} \textbf{49.92} &\CC{20} 55.48 &\CC{20} \textbf{54.27} &\CC{20} 55.36 &\CC{20} 53.95 &\CC{20} \textbf{57.34} &\CC{20} \textbf{57.24} \\ 
		\midrule
		& Linear Raw &  41.75 & 49.91 & 50.93 & 63.17 & 63.85 & 53.27 & 51.46 & 51.21 \\
		& Linear \emph{One-Step-P} &  36.17 & 41.60 & \textbf{43.15} & \textbf{49.68} & \textbf{53.51} & 47.86 & \textbf{45.92} & 43.69 \\
		& Linear One-Step-W &  41.91 & 51.20 & 51.85 & 57.17 & 61.52 & 52.86 & 51.24 & 48.86 \\
		Bidding &  Linear \emph{Two-Step-WP} & \textbf{35.78} & \textbf{41.24} & 43.38 & 50.61 & 54.43 & \textbf{47.50} & 47.15 & \textbf{43.14} \\
		Zone 4 & \CC{20}Random Forest Raw & \CC{20}36.97 & \CC{20}41.33 &\CC{20} 40.74 &\CC{20} 50.67 &\CC{20} 51.60 &\CC{20} 45.54 &\CC{20} 42.26 &\CC{20} 39.41 \\
		&\CC{20}Random Forest \emph{One-Step-P} &\CC{20} 34.33 &\CC{20} \textbf{38.25} &\CC{20} 39.30 &\CC{20} 44.60 &\CC{20} \textbf{48.02} &\CC{20} \textbf{44.28} &\CC{20} \textbf{41.72} &\CC{20} 38.18\\
		&\CC{20}Random Forest One-Step-W &\CC{20} 36.81 &\CC{20} 42.91 &\CC{20} 41.94 &\CC{20} 46.51 &\CC{20} 51.16 &\CC{20} 46.24 &\CC{20} 43.16 &\CC{20} 38.31 \\
		&\CC{20}Random Forest \emph{Two-Step-WP} &\CC{20} \textbf{33.52} &\CC{20} \textbf{38.25} &\CC{20} \textbf{39.21} &\CC{20} \textbf{43.95} &\CC{20} 48.08 &\CC{20} 44.74 &\CC{20} 42.04 &\CC{20} \textbf{37.33}\\
		\bottomrule
	\end{tabular}
\end{table}

\begin{figure}[t]
	\centering
	\subfloat[Linear Model]{
		\includegraphics[width=0.48\textwidth]{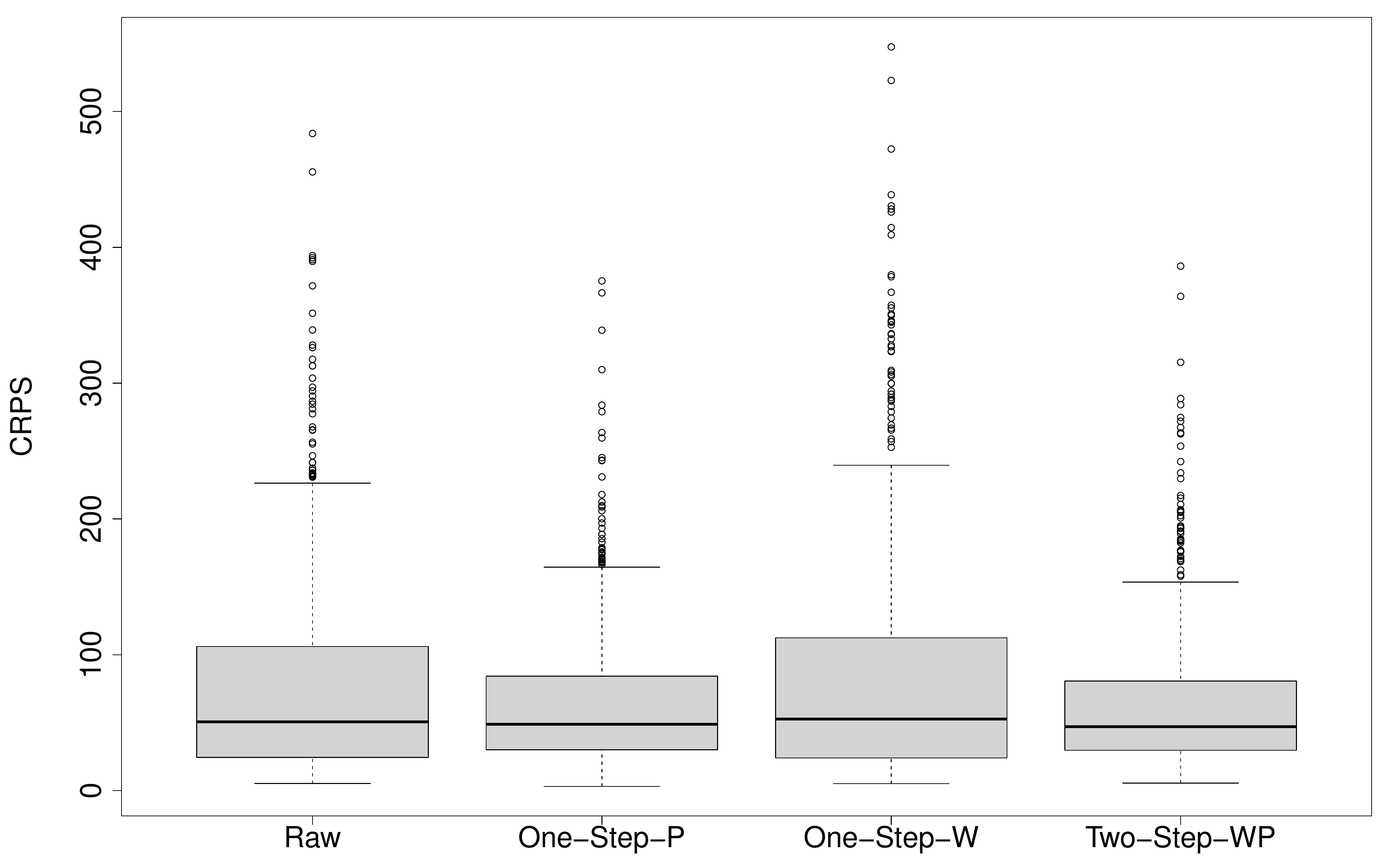}
		\label{fig:hist_Sweden_lin}	} %
	\subfloat[Random Forest]{
		\includegraphics[width=0.48\textwidth]{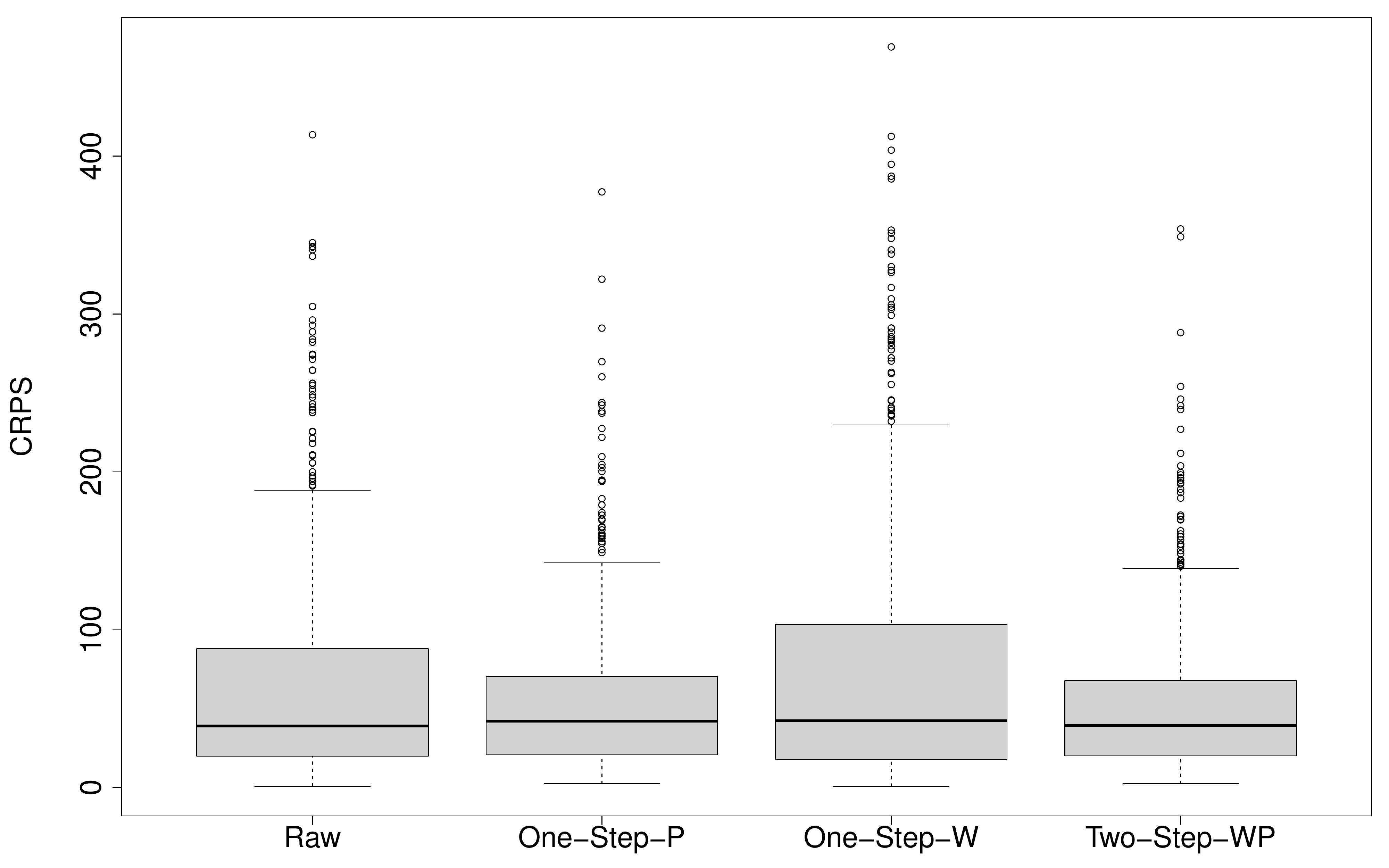}
		\label{fig:hist_Sweden_rf}} %
	\caption{A boxplot of all CRPS values for a forecast horizon of 12h for bidding zone 3 in Sweden. In (a) the forecast using a linear model is shown and in (b) the forecast using a random forest. Although the post-processing does not show large improvement in the median, it does reduce the spread.
	}
	\label{fig:hist_Sweden}
\end{figure}

%\stress{I'm missing here why we're suddenly interested in outliers and why only for the Swedish data. }
Additionally to the mean \gls{crps}, we also consider box plots of the \gls{crps} to help assess the effect of post-processing on the spread of the forecast, \ie the sharpness. \Cref{fig:hist_Sweden} shows these box plots of all \gls{crps} for a forecast horizon of 12h, exemplary for bidding zone 3 in Sweden. A smaller box indicates a sharper forecast. \Cref{fig:hist_Sweden_lin} details the performance of the linear model and it can be seen that both \emph{One-Step-P} and \emph{Two-Step-WP} improve the spread of the \gls{crps} quite noticeably. \emph{One-Step-W} has little effect on the median \gls{crps} but does lead to a larger spread, suggesting that this model struggles with values that are difficult to predict.

For the random forests (\Cref{fig:hist_Sweden_rf}), the \gls{crps} is smaller (note the scale of the y-axis). Apart from an overall improvement in \gls{crps}, the sharpness results are similar to the linear models. There is no clear improvement in median, but both \emph{One-Step-P} and \emph{Two-Step-WP} strategies show a noticeably smaller spread in \gls{crps} values. This smaller spread is an indication that the post-processing results in a better approximation of those values that are normally difficult to predict (\ie result in a large \gls{crps} value), even if the median accuracy is not noticeably different.

\section{Discussion} \label{sec:discussion}

The results above show that post-processing the ensembles leads to more accurate wind power forecasts. As the two strategies which include post-processing the final wind power ensemble (\emph{One-Step-P and Two-Step-WP}) outperform the others (\emph{One-Step-W} and \emph{Raw}), this final post-processing seems crucial.  Only post-processing the weather (\emph{One-Step-W}) does not improve the forecast performance regarding both calibration and sharpness, in some cases it even leads to a decrease in performance. A possible explanation for this behaviour is that post-processing only the weather neglects other bias sources for the wind power model. We train the models on historical data such that they learn the true relationship among the variables, however, this makes it harder for the forecasting model to properly propagate the uncertainty from the set of ensembles through the model, as they are not designed to do that. Thus, despite the weather ensembles being well-calibrated after post-processing, this information seems to not be accurately used by the forecasting model. 

It is also worth noting that to enable comparisons across all strategies, we only consider a sample size of 51, i.e. the number of raw ensembles, when drawing from the calibrated weather ensembles. Although it is possible to vary the number of independently drawn samples, there are two reasons why we conclude this will not have a noticeable effect on the results. Firstly, the considered sample size of 51 is large enough to accurately approximate the estimated distribution and, secondly, increasing the number of samples will also increase computational complexity and could lead to poorer performance in subsequent calibration steps. However, a systematic analysis of the effect of sample size could be interesting in future work.

When we compare \emph{One-Step-P} and \emph{Two-Step-WP}, both strategies perform well and in most cases there is no noticeable difference between them. The key factor that differentiates the two strategies is the number of post-processing steps required. When we apply the \emph{Two-Step-WP} strategy,  we need to post-process all weather inputs (in the case of the current paper this is four, but in more complicated models this number may increase). Once this post-processing is complete, we generate a wind power forecast and then are again required to process the resulting ensembles. The \emph{One-Step-P} strategy involves only one post-processing step, irrespective of how many weather inputs the model includes. Although \gls{emos}' computational complexity is low and multiple post-processing steps are still feasible, \emph{One-Step-P} is a more computationally efficient strategy that achieves similar forecasting accuracy.  Overall, the post-processing of wind power forecasts does increase forecasting accuracy, independent of the forecasting model, with the last post-processing step (in this case the wind power) having the biggest effect.

\section{Conclusion} \label{sec:conclusion}

The present paper evaluates at which stage ensemble post-processing is the most useful for wind power forecasts. Although Pinson and Messner~\cite{Pinson.2018} introduced ensemble post-processing for wind power applications, they did not evaluate which post-processing strategy is most effective. We, therefore, identify four post-processing strategies that can be applied; (1) no post-processing (\emph{Raw}), (2) a one-step strategy with post-processing of the resulting wind power ensemble (\emph{One-Step-P}), (3) a one-step strategy with post-processing of the weather ensembles (\emph{One-Step-W}), and (4) a two-step strategy with both post-processing of the weather ensembles and the resulting wind power ensembles (\emph{Two-Step-WP}). These strategies are evaluated on two data sets with a linear regression and a random forest used as forecasting models. Results show that post-processing generally improves performance, specifically when the wind power ensemble is post-processed. \emph{One-Step-P} and \emph{Two-Step-WP} deliver similar results in terms of \gls{crps} and \gls{crpss} performance, but since it requires significantly fewer post-processing steps the \emph{One-Step-P} strategy is preferred.

Whilst initial results show that post-processing the final wind power ensemble leads to improvements in forecasting accuracy, there are numerous aspects that should be investigated further. 
%Firstly, when \gls{emos} is applied, possible dependency structures between weather variables are lost. Since no dependency between the weather variables is considered in the present paper, Ensemble Copula Coupling (ECC) should be applied after the \gls{emos} weather post-processing in order to restore possible dependencies in future work. 
Firstly, EMOS is limited in that it assumes a parametric distribution in the calibration process and wind power generation and other weather variables do not always perfectly follow such a distribution. Therefore, non-parametric post-processing methods that do not rely on the assumption of a parametric probability distribution should also be considered. Furthermore, the current paper focuses on a one-step-ahead forecast for either a single location or a broad region. This removes the need to account for spatial and temporal dependencies. Ensemble Copula Coupling (ECC) could be used to account for these dependencies and future work should focus on multi-step ahead and multi-location forecasts. It could also be interesting to systematically analyse the effect of sample size on forecasting performance in future work. Finally, the current methods use traditional forecasting methods which do not appear to be able to learn the uncertainty in the data. Therefore, approaches that learn this uncertainty in the weather ensembles and directly propagate it through to a wind power forecast should be investigated. 
%The present paper has focused on evaluating post-processing strategies for wind power forecasts. Other renewable energy sources, such as solar, are also heavily linked with the weather. It would therefore be interesting to evaluate post-processing strategies for other renewable energy forecasts to determine whether results remain consistent.

\section*{Data Availability Statement}
The ensemble weather data for the onshore and offshore benchmarks that support the findings of this study are openly available via GitHub \url{https://github.com/KIT-IAI/EvaluatingEnsemblePostProcessing} and were adapted from The International Grand Global Ensemble (TIGGE) archive \url{https://apps.ecmwf.int/datasets/data/tigge/}. The wind power generation data for the onshore and offshore benchmarks are generated with the Renewables.ninja API (\url{https://www.renewables.ninja/}) and also openly available via GitHub \url{https://github.com/KIT-IAI/EvaluatingEnsemblePostProcessing}.

The ensemble weather data for Sweden that further supports the findings of this study are generated from the Ensemble Prediction System (EPS) from European Centre for Medium-Range Weather Forecasts (ECMWF). This data was downloaded with a license from ECMWF and is provided with the Creative Commons Attribution 4.0 International (CC BY 4.0) license. We adapt the data from the ECMWF archive and make it openly available thanks to ECMWF via GitHub \url{https://github.com/KIT-IAI/EvaluatingEnsemblePostProcessing} in line with the CC BY 4.0 licence. 
%Restrictions apply to the availability of these data, which were used under license for this study. The full ECMWF EPS, is available with a research license from ECMWF \url{https://www.ecmwf.int/}. 
The wind power generation data for Sweden are openly available via the open source transparency platform, operated by the European Network of Transmission System Operators (ENTSO-E) \cite{entso_e}.

The observational weather data used in this study are ERA5 reanalysis data and openly available via the Copernicus Climate Data Store (CDS) \url{https://cds.climate.copernicus.eu/home}.

\section*{Acknowledgements}

This work is funded by the German Research Foundation (DFG) as part of the Research Training Group 2153 ``Energy Status Data -- Informatics Methods for its Collection, Analysis and Exploitation'' and by the Helmholtz Association's Initiative and Networking Fund through Helmholtz AI, the Helmholtz Association under the Program ``Energy System Design'' and  the Joint Initiative ``Energy System Design - A Contribution of the Research Field Energy''. Nicole Ludwig acknowledges financial support by the DFG under Germany’s Excellence Strategy – EXC number 2064/1 – Project number 390727645. 
Sebastian Lerch acknowledges support by the DFG through SFB/TRR 165 "Waves to Weather", and by the Vector Stiftung through the Young Investigator Group "Artificial Intelligence for Probabilistic Weather Forecasting". The research detailed in the current paper was based on data from the ECMWF obtained through an academic licence for research purposes. The authors would like to thank Jon Olauson for his help creating \Cref{fig:sweden_map} and Marian Turowski for the useful discussions and his continued personal support throughout the development of this paper.

\newpage

\section{Appendix} \label{sec:appendix}

In this appendix we firstly discuss the performance of Bayesian Model Averaging (BMA) when post-processing weather ensembles in Section~\ref{sec:appendix:bma}. In Section ~\ref{sec:appendix:robust}, we then report further results of experiments not included in the main paper that serve to demonstrate the robustness of our results. These experiments include using an alternative linear regression and a neural network as forecasting models, considering alternative distributions for wind speed and power, and showing that the selected distributions are a reasonable fit for the data.

\subsection{Bayesian Model Averaging as an Alternative Post-Processing Method}\label{sec:appendix:bma}
As discussed in Section~\ref{sec:post_processing}, one common alternative to \gls{emos} is BMA, introduced by Raftery~et al.~\cite{Raftery2005}. Although BMA is computationally more expensive than \gls{emos}~\cite{vannitsem2018statistical,Schulz2021}, we initially implement both post-processing methods and compare their performance regarding the calibration of weather ensembles. We implement BMA with the R package \emph{ensembleBMA}~\cite{fraley2007ensemblebma}, selecting the same probability distributions for weather variables as chosen for \gls{emos}.

\begin{figure}
	\centering
	\begin{subfigure}[b]{0.49\textwidth}
		\includegraphics[width=\textwidth]{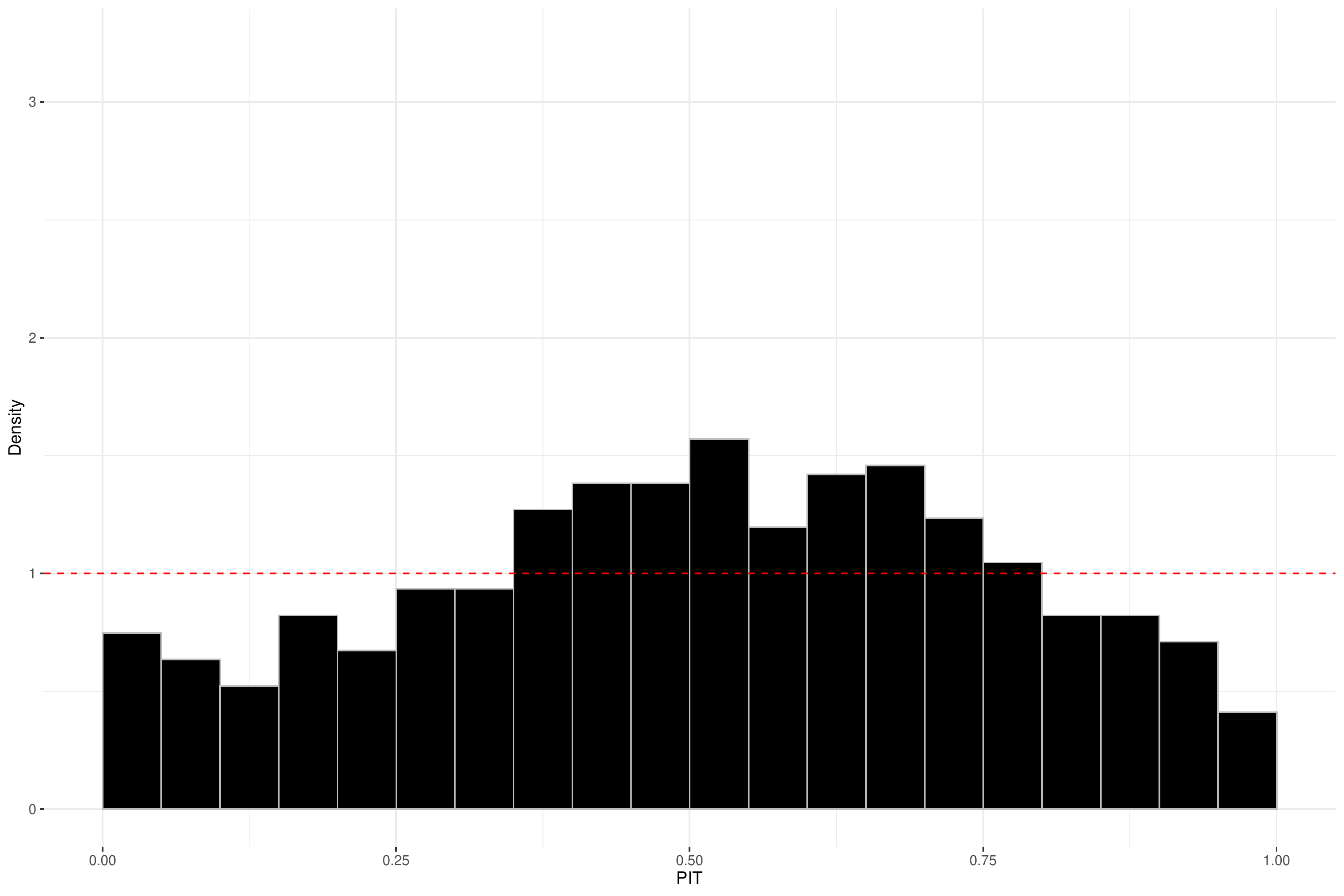}
		\caption{Wind Speed}
	\end{subfigure}
	\begin{subfigure}[b]{0.49\textwidth}
		\includegraphics[width=\textwidth]{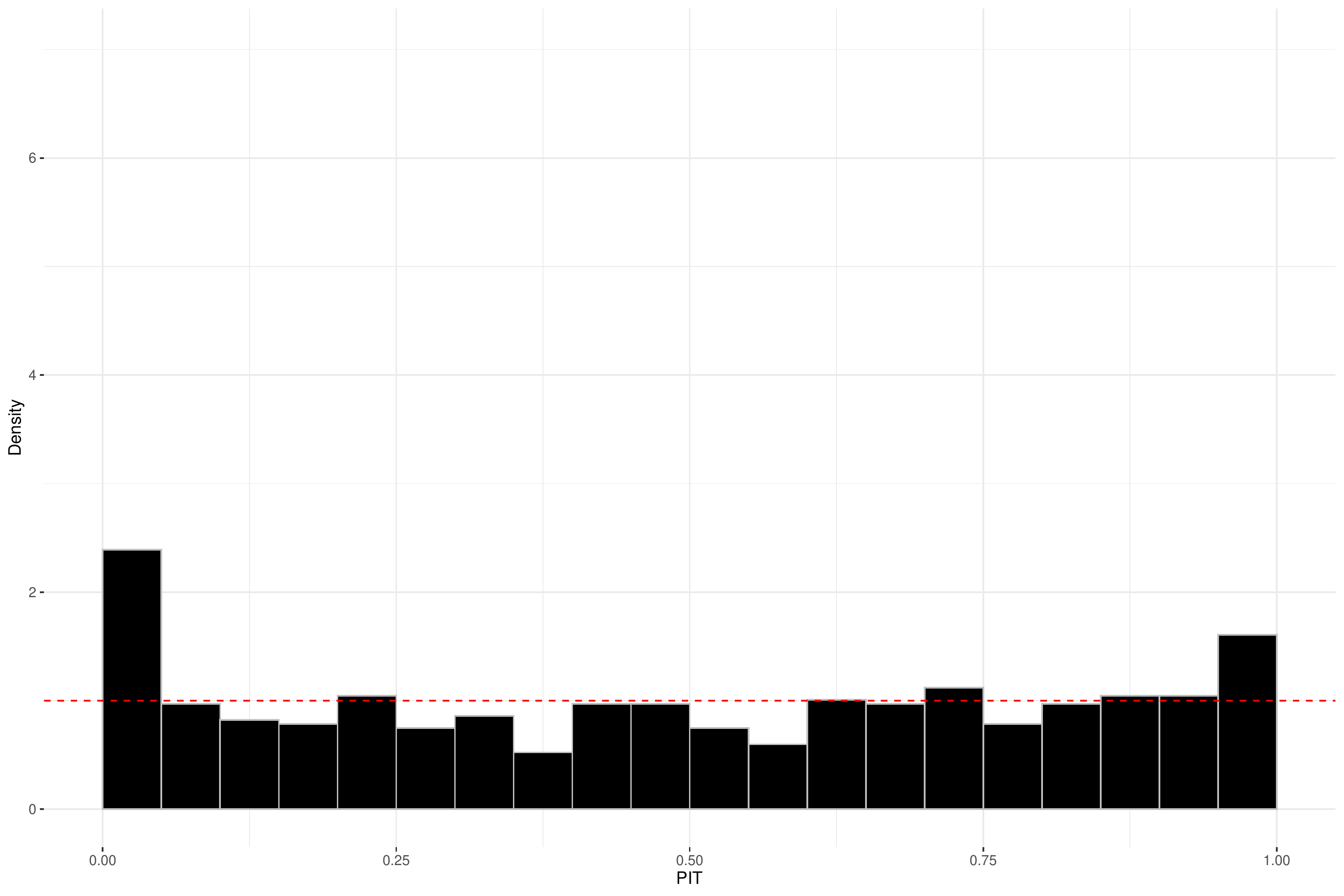}
		\caption{Temperature}
	\end{subfigure}
	\begin{subfigure}[b]{0.49\textwidth}
		\includegraphics[width=\textwidth]{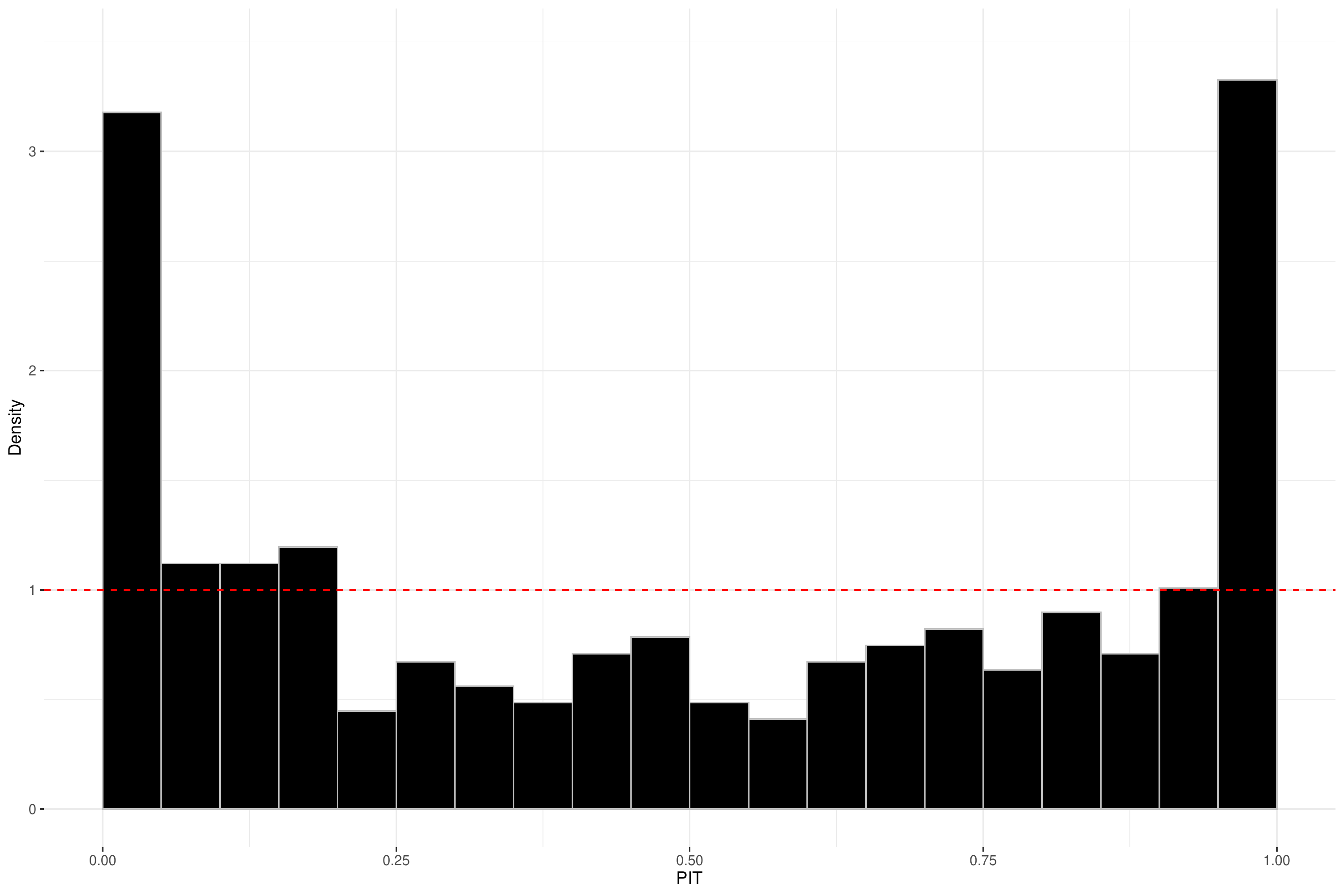}
		\caption{Wind U10}
	\end{subfigure}
	\begin{subfigure}[b]{0.49\textwidth}
		\includegraphics[width=\textwidth]{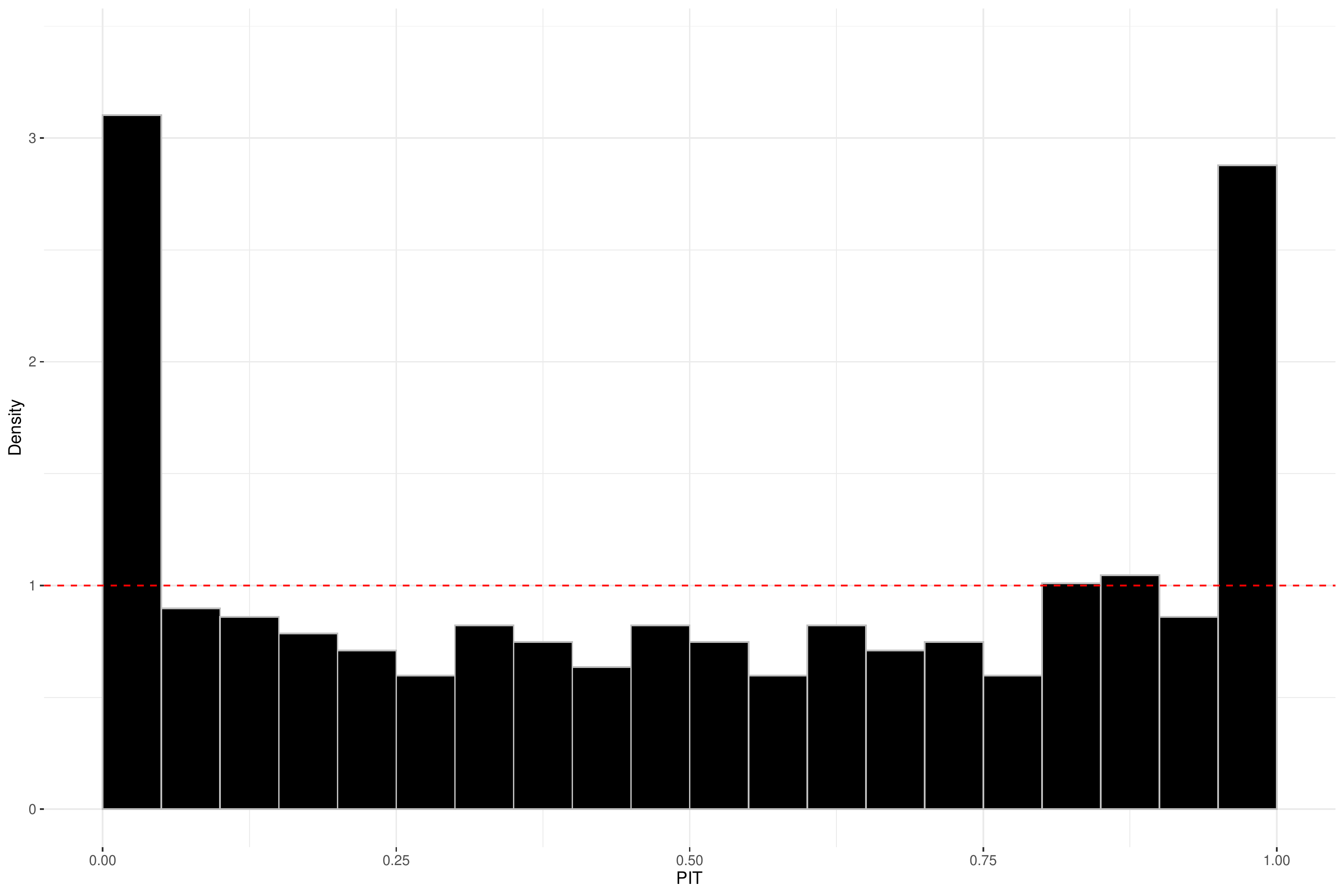}
		\caption{Wind V10}
	\end{subfigure}
	\caption{PIT Histograms showing the calibration of weather ensembles using BMA for the onshore benchmark data set with a forecast horizon of 12h. In comparison to EMOS, the ensembles are poorly calibrated.}
	\label{fig:bma_bad}
\end{figure}

Figure ~\ref{fig:bma_bad} exemplarily shows the calibration performance when using BMA for a forecast horizon of 12h for the onshore benchmark data set, the same scenario used in Figure ~\ref{fig:onshore_calibration_results}. Based on this figure, it is clear that the performance of BMA is inferior to \gls{emos}. Although temperature is only slightly underdispersed, all other variables are poorly calibrated with wind speed being overdispersed and both the u- and v-components of wind very underdispersed. The calibration performance of EMOS for the same scenario (see Figure~\ref{fig:onshore_calibration_results}) is far superior. We discover similar results for the other data sets and forecast horizons. We believe this is due to the computational complexity of BMA which caused convergence problems on our data set. Based on these initial results, we select \gls{emos} as the best performing post-processing method on our data set and only consider \gls{emos} for the evaluation of our developed post-processing strategies.

\subsection{Robustness of Results}\label{sec:appendix:robust}

To show the robust nature of our approach, we perform numerous further experiments. First, we implement both an alternative linear regression and a neural network as further forecasting methods, before also including the full results when considering different distributions for wind power and wind speed. Finally, we show that the selected probability distributions are a reasonable fit for our data and therefore a suitable choice for \gls{emos}.

\paragraph{Alternative Linear Regression Model}
To discuss the results of the alternative linear regression model proposed by Zhang and Wang~\cite{ZHANG20161074}, we firstly describe the linear regression used before showing the results of the different post-processing strategies.

The linear regression proposed by Zhang and Wang~\cite{ZHANG20161074} only considers wind speed and the hour of the day as regressors. The linear regression model can be described as
\begin{equation}
	\begin{split}
	y_{t+h} =  \beta_0 + \beta_1 S_{t+h} + \beta_2 \left(S_{t+h}\right)^2 &+ \beta_3 \left(S_{t+h}\right)^3 \beta_4 \cos\left(\frac{2\pi}{24}H_{t+h}\right) + \beta_5 \sin\left(\frac{2\pi}{24}H_{t+h}\right) \\
	& + \beta_6 \cos\left(\frac{4\pi}{24}H_{t+h}\right) + \beta_7 \sin\left(\frac{4\pi}{24}H_{t+h}\right) + \varepsilon_{t+h},
	\end{split}
\end{equation}
where $y_t$ is the dependent variable, which in this case is the wind power, $S_{t+h}$ is the forecasted wind speed for the considered forecast horizon, $H_{t+h}$ is the hour of the day for the considered forecast horizon, and $\beta_i, i = 0, \dots, 7$ are the regression coefficients. As observed by Zhang and Wang~\cite{ZHANG20161074}, the sine and cosine terms account for diurnal periodicities through a Fourier series considering the time of the day. We fit and apply this alternative linear regression model in the same way as the linear regression model described in \Cref{sec:forecastModels}.

\begin{figure}[t]
	\centering
	\includegraphics[width=\textwidth]{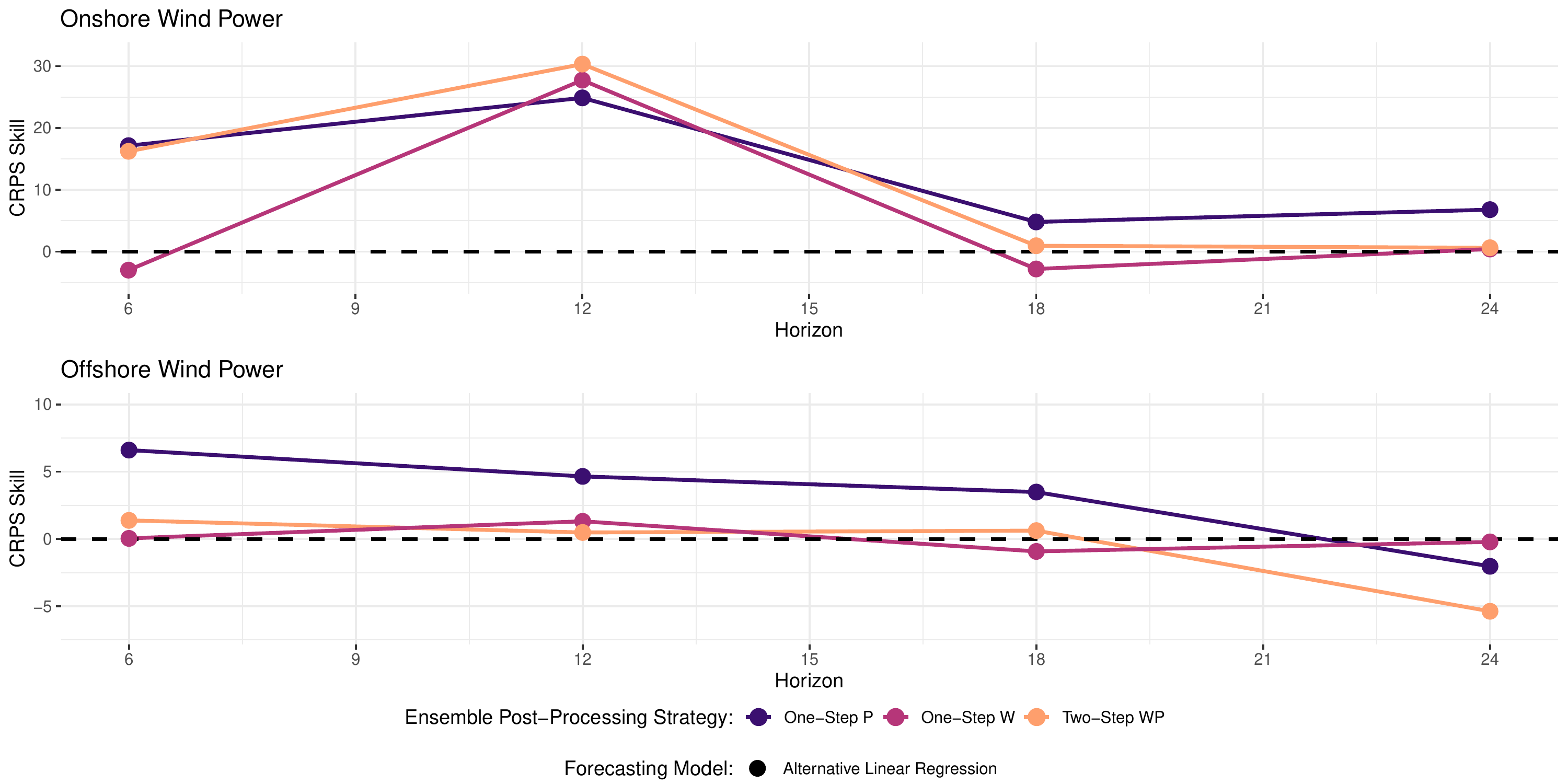}
	\caption{The CRPS skill score plotted against the forecast horizon on the test data for the alternative linear regression on the onshore benchmark (top figure) and the offshore benchmark (bottom figure). Positive values indicate an improvement over the \emph{Raw} strategy in percent. In general, post-processing leads to an improvement in forecasting performance.}
	\label{fig:alternative_linear_crpss_benchmark}
\end{figure}

\begin{figure}[t]
	\centering
	\includegraphics[width=\textwidth]{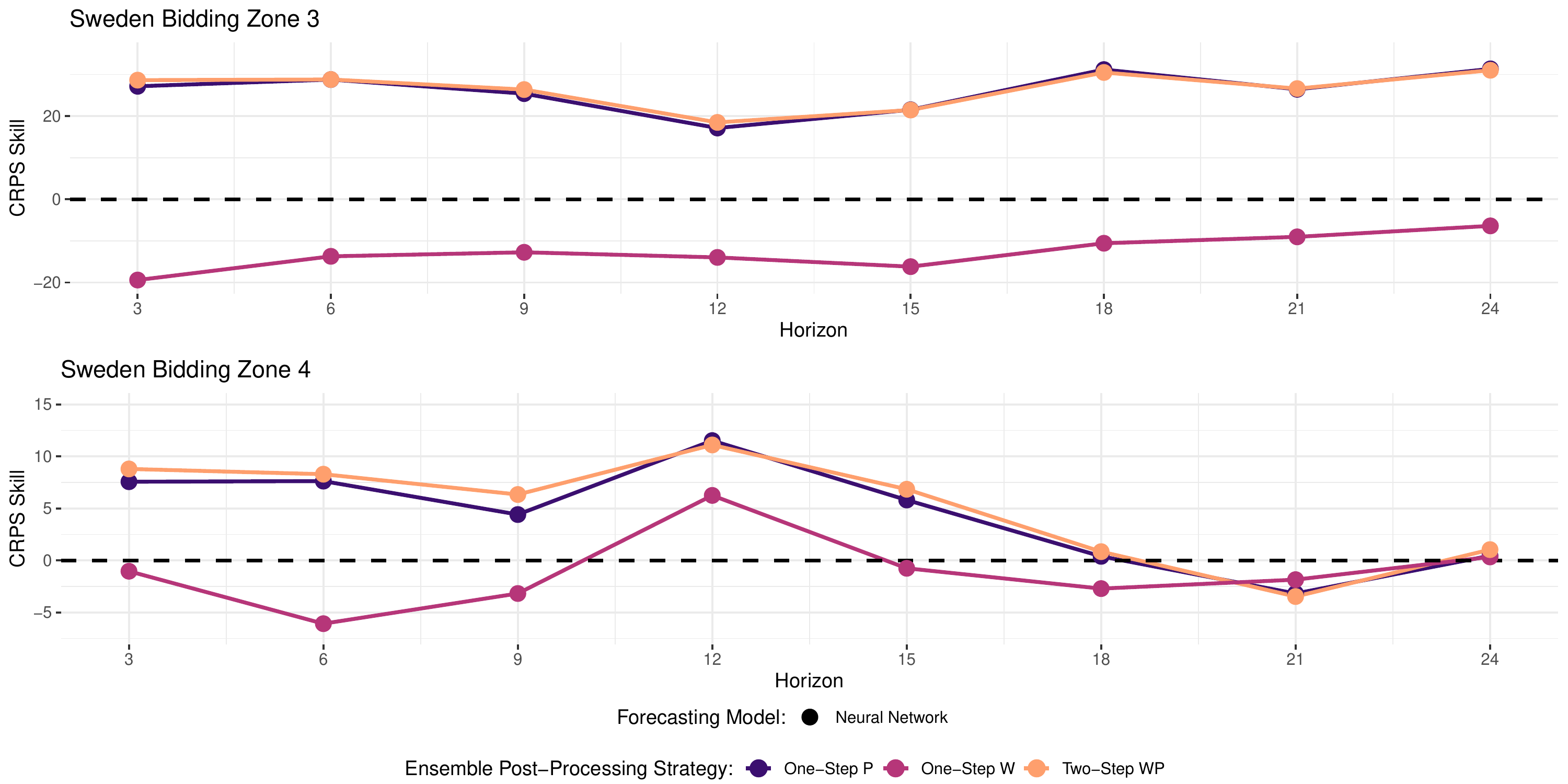}
	\caption{The CRPS skill score plotted against the forecast horizon on the test data for the alternative linear regression on both bidding zone 3 (top figure) and bidding zone 4 (bottom figure) in Sweden. Positive values indicate an improvement over the \emph{Raw} strategy in percent. Post-processing the final power ensemble, either directly or as part of a two-step process almost always clearly improves the forecast, whilst only post-processing the weather ensemble leads to worse or similar forecast performance.}
	\label{fig:sweden_crpss_alternative_linear}
\end{figure}

The CRPS skill score plotted against the forecast horizon for the different post-processing strategies is shown for the benchmark data sets in Figure ~\ref{fig:alternative_linear_crpss_benchmark} and for both bidding zones in Sweden in Figure ~\ref{fig:sweden_crpss_alternative_linear}. These results again highlight that post-processing the final wind power ensemble is the crucial step. The \emph{One-Step-P} and \emph{Two-Step-WP} strategies, which post-process the wind power ensemble, almost always improve forecasting performance on all data sets. The only exceptions are for a 24 hour forecast horizon in the offshore benchmark and a 21 hour forecast performance in bidding zone 4, which deliver forecast performance similar to the \emph{Raw} strategy. As with the linear regression model presented in \Cref{sec:forecastModels}, post-processing the weather ensembles does not generally lead to improvements. In general, the improvements achieved through post-processing with the alternative linear model, are similar to the improvements achieved with other forecasting models and therefore show the robust nature of our approach.

\paragraph{Neural Network as an Alternative Non-Linear Forecasting Model}
To discuss the results using neural networks as an additional forecasting model, we first describe the selected neural network architecture and then show the results of the different post-processing strategies. 

We select a simple feed forward neural network for our evaluation and test multiple configurations before choosing a configuration with two hidden layers of 10 and 7 neurons respectively. This network architecture is selected because it is the simplest we found that still returns accurate forecasts. Given this architecture, we train the neural network with the resilient backpropagation algorithm. The chosen activation function is a hyperbolic tangent and the input features are the same as those selected for the linear regression and random forest models (see Section ~\ref{sec:forecastModels}). The parameters (\ie weights) are fitted using the actual historical weather data and each ensemble member is passed through the network to get an ensemble of wind power forecasts.

\begin{figure}[t]
	\centering
	\includegraphics[width=\textwidth]{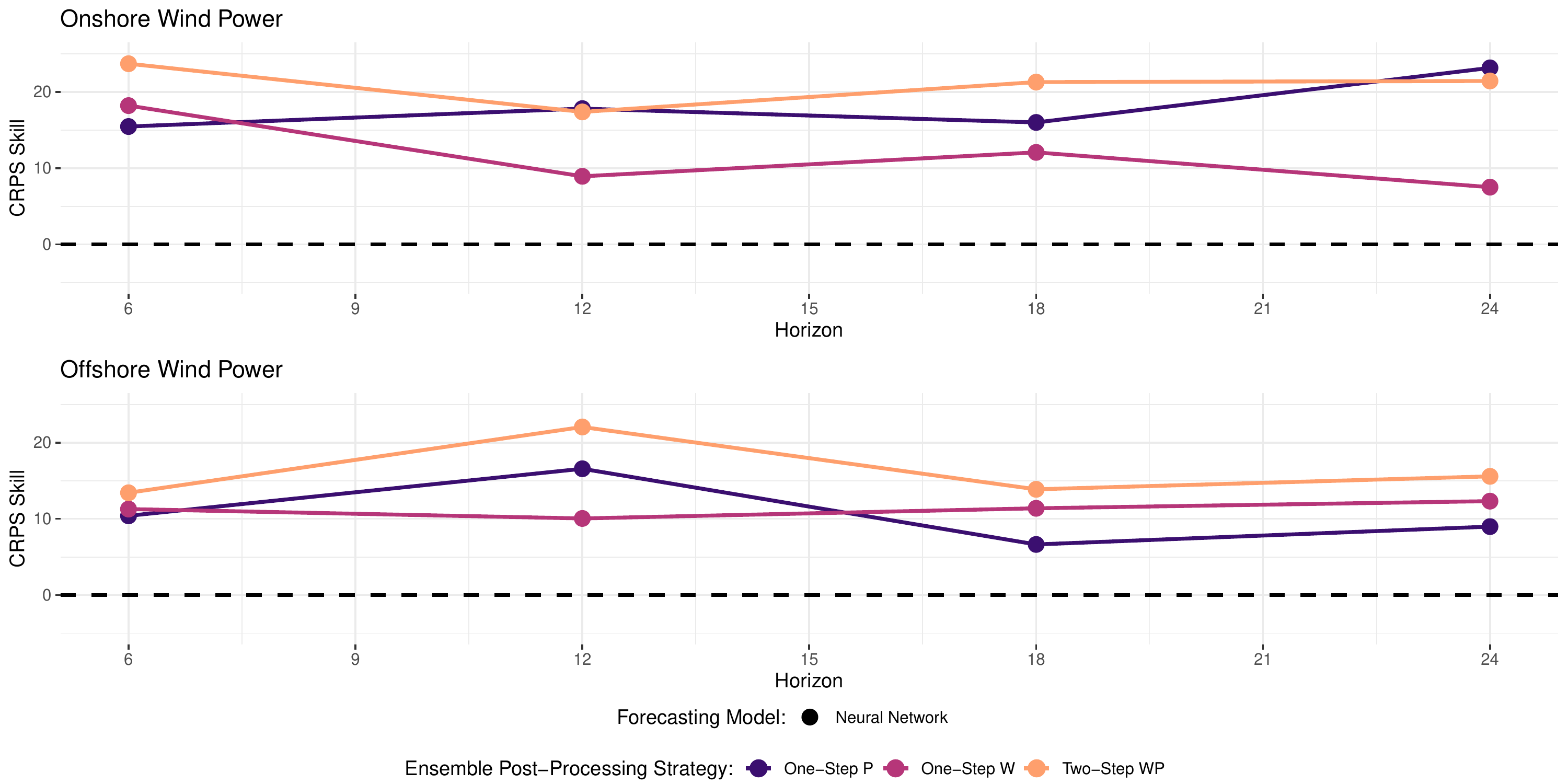}
	\caption{The CRPS skill score plotted against the forecast horizon on the test data for the neural networks on the onshore benchmark (top figure) and the offshore benchmark (bottom figure). Positive values indicate an improvement over the \emph{Raw} strategy in percent. In general post-processing leads to an improvement in forecasting performance.}
	\label{fig:neural_crps_benchmark}
\end{figure}

\begin{figure}[t]
	\centering
	\includegraphics[width=\textwidth]{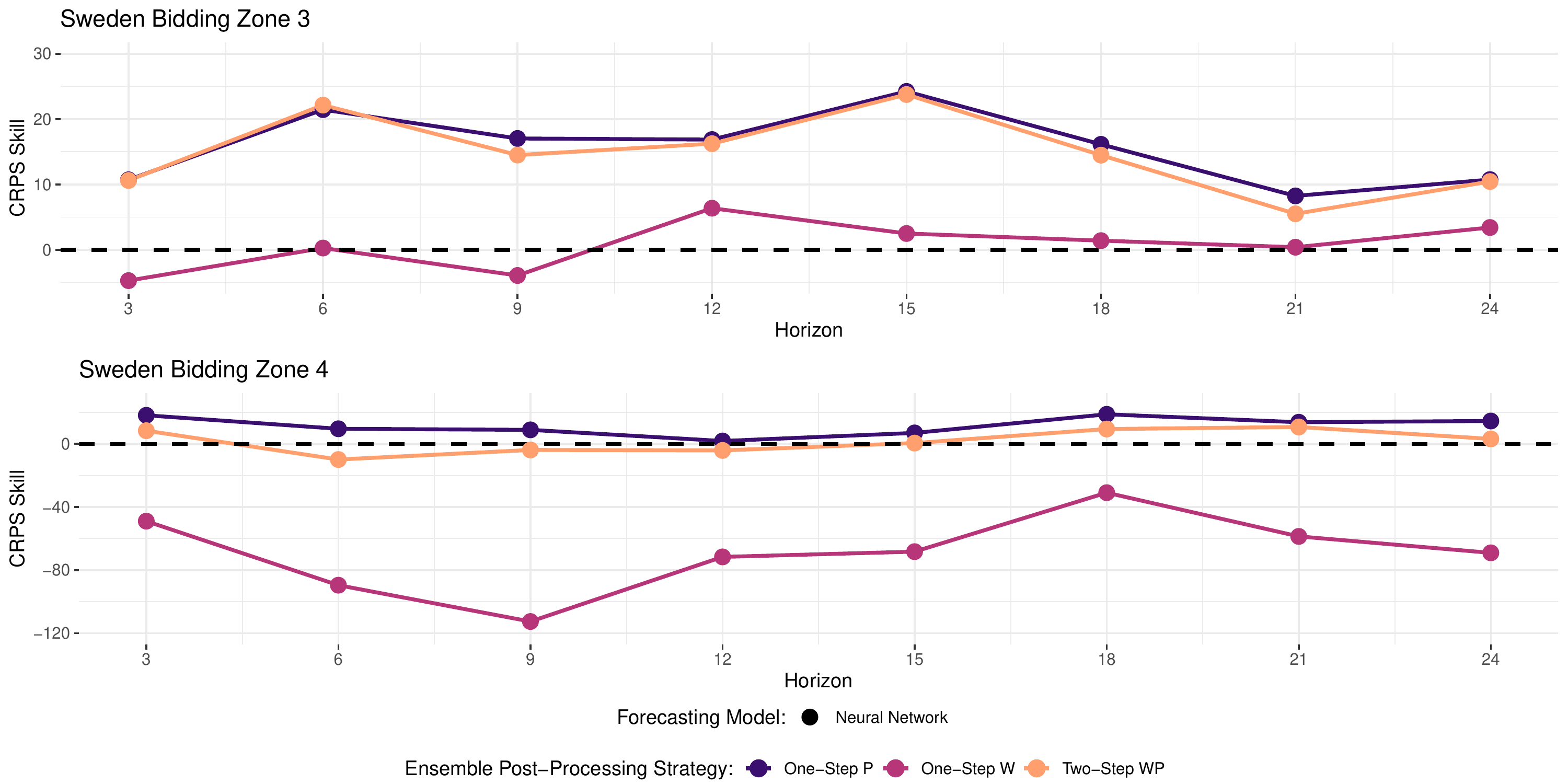}
	\caption{The CRPS skill score plotted against the forecast horizon on the test data for the neural networks on both bidding zone 3 (top figure) and bidding zone 4 (bottom figure) in Sweden. Positive values indicate an improvement over the \emph{Raw} strategy in percent. Post-processing the final power ensemble, either directly or as part of a two-step process clearly improves the forecast, whilst only post-processing the weather ensemble leads to worse or similar forecast performance.}
	\label{fig:sweden_crpss_nerual}
\end{figure}

The CRPS skill score plotted against the forecast horizon for the different post-processing strategies is shown for the benchmark data sets in Figure ~\ref{fig:neural_crps_benchmark} and for both bidding zones in Sweden in Figure ~\ref{fig:sweden_crpss_nerual}. With these results we again see, that post-processing the final wind power ensemble is the crucial step. Both the \emph{One-Step-P} and \emph{Two-Step-WP} strategies, which post-process the wind power ensemble, improve performance on the benchmark and the Swedish data sets. Only post-processing the weather ensembles, however, does not necessarily lead to improvements. Although \emph{One-Step-W} leads to improved performance on the benchmark data set, this strategy performs worse on the real world Swedish data sets.  Furthermore, we note that the improvements through the post-processing strategies are similar to the improvements achieved with other forecasting models.

\paragraph{Considering Different Distributions for Wind Speed and Wind Power}

From the literature, we identify both the truncated normal distribution and gamma distribution as possible candidates for modeling wind speed and wind power. To ensure robust results we consider both distributions and report the \gls{crps} values when using different combinations of the two distributions here. Following Gneiting~\cite{gneiting2014calibration}, we implement a gamma distribution using $\mathrm{Gamma}(\alpha, \beta) = F_{\alpha, \beta}$, with $\alpha = a + b_1x_1 + \ldots + b_Mx_M$ and $\beta = c + d m_{\mathrm{ENS}}$ with $m_{\mathrm{ENS}} = a + b \sum_{m=1}^M{x_m}$. The following tables (\Cref{tab:fullcrpsOnshore} - \Cref{tab:fullcrpsSweden4}) show that there is no noticeable performance difference when using different distributions. Additionally, we also report the results of models that do not include weather forecasts as input and are based solely on historical wind power values.

\begin{table}[ht]
	\centering
	\footnotesize
	\caption{Summary of mean CRPS on the test data for the onshore benchmark for all forecast horizons. The distribution used for post-processing the wind speed (W) and wind power (P) is shown in brackets.}
	\label{tab:fullcrpsOnshore}
	\begin{tabular}{r|rrrr}
		\toprule
		Strategy  & 6h & 12h & 18h & 24h \\ 
		\midrule
		Linear No Weather & 7.24 & 11.38 & 11.42 & 10.97 \\ 
		Linear Raw & 3.91 & 5.60 & 5.67 & 6.08 \\ 
		Linear \emph{One-Step-P} (Gamma) & 3.42 & 4.68 & 5.60 & 5.48 \\ 
		Linear \emph{One-Step-P} (T-Normal) & 3.53 & 4.88 & 5.71 & 5.76 \\ 
		Linear \emph{One-Step-W} (Gamma) & 3.80 & 5.29 & 5.66 & 5.96 \\ 
		Linear \emph{One-Step-W} (T-Normal) & 3.81 & 5.26 & 5.51 & 5.90 \\ 
		Linear \emph{Two-Step-WP} (W: Gamma, P: Gamma) & 3.52 & 4.96 & 5.63 & 5.78 \\ 
		Linear \emph{Two-Step-WP} (W: Gamma, P: T-Normal) & 3.42 & 4.96 & 5.61 & 5.72 \\ 
		Linear \emph{Two-Step-WP} (W: T-Normal, P: Gamma) & 3.50 & 4.87 & 5.64 & 5.84 \\ 
		Linear \emph{Two-Step-WP} (W: T-Normal, P: T-Normal) & 3.48 & 4.97 & 5.68 & 5.77 \\ 
		\midrule
		Alternative Linear No Weather & 13.45 & 15.73 & 12.45 & 11.46 \\ 
		Alternative Linear Raw & 4.59 & 5.79 & 6.21 & 5.93 \\ 
		Alternative Linear \emph{One-Step-P} (Gamma) & 3.67 & 4.12 & 5.78 & 5.43 \\ 
		Alternative Linear \emph{One-Step-P} (T-Normal) & 3.81 & 4.35 & 5.91 & 5.53 \\ 
		Alternative Linear \emph{One-Step-W} (Gamma) & 4.73 & 4.32 & 6.31 & 5.82 \\ 
		Alternative Linear \emph{One-Step-W} (T-Normal) & 4.73 & 4.19 & 6.38 & 5.91  \\ 
		Alternative Linear \emph{Two-Step-WP} (W: Gamma, P: Gamma) & 4.01 & 4.22 & 5.86 & 5.52  \\ 
		Alternative Linear \emph{Two-Step-WP} (W: Gamma, P: T-Normal) & 3.99 & 4.22 & 5.99 & 5.63 \\ 
		Alternative Linear \emph{Two-Step-WP} (W: T-Normal, P: Gamma) & 3.98 & 3.93 & 5.89 & 5.56 \\ 
		Alternative Linear \emph{Two-Step-WP} (W: T-Normal, P: T-Normal) & 3.85 & 4.04 & 6.15 & 5.89 \\
		
		\midrule
		Random Forest No Weather & 9.52 & 12.18 & 10.41 & 10.31 \\ 
		Random Forest Raw & 3.98 & 5.34 & 5.68 & 5.84 \\ 
		Random Forest \emph{One-Step-P} (Gamma) & 3.31 & 4.21 & 5.67 & 5.23 \\ 
		Random Forest \emph{One-Step-P} (T-Normal) & 3.61 & 4.63 & 5.84 & 5.38 \\ 
		Random Forest \emph{One-Step-W} (Gamma) & 3.85 & 4.44 & 5.82 & 5.83 \\ 
		Random Forest \emph{One-Step-W} (T-Normal) & 3.97 & 4.53 & 5.78 & 5.83 \\ 
		Random Forest \emph{Two-Step-WP} (W: Gamma, P: Gamma) & 3.34 & 4.08 & 5.73 & 5.45 \\ 
		Random Forest \emph{Two-Step-WP} (W: Gamma, P: T-Normal) & 3.57 & 4.30 & 5.86 & 5.52 \\ 
		Random Forest \emph{Two-Step-WP} (W: T-Normal, P: Gamma) & 3.41 & 4.05 & 5.73 & 5.42 \\ 
		Random Forest \emph{Two-Step-WP} (W: T-Normal, P: T-Normal) & 3.67 & 4.34 & 5.84 & 5.70 \\ 
		\midrule
		Neural Network No Weather & 7.80 & 11.34 & 10.15 & 10.31 \\ 
		Neural Network Raw & 7.16 & 10.42 & 10.19 & 10.19 \\ 
		Neural Network \emph{One-Step-P} (Gamma) & 6.05 & 8.57 & 8.56 & 7.83 \\ 
		Neural Network \emph{One-Step-P} (T-Normal) & 6.19 & 9.12 & 8.56 & 8.02 \\ 
		Neural Network \emph{One-Step-W} (Gamma) & 5.85 & 9.49 & 8.96 & 9.43 \\ 
		Neural Network \emph{One-Step-W} (T-Normal) & 5.90 & 9.60 & 9.38 & 9.44\\ 
		Neural Network \emph{Two-Step-WP} (W: Gamma, P: Gamma) & 5.46 & 8.61 & 8.02 & 8.01 \\ 
		Neural Network \emph{Two-Step-WP} (W: Gamma, P: T-Normal) & 5.54 & 9.00 & 7.71 & 7.76 \\ 
		Neural Network \emph{Two-Step-WP} (W: T-Normal, P: Gamma) & 5.53 & 8.76 & 8.21 & 7.89 \\ 
		Neural Network \emph{Two-Step-WP} (W: T-Normal, P: T-Normal) & 5.44 & 8.99 & 7.71 & 7.65 \\ 
		\bottomrule
	\end{tabular}
\end{table}

\begin{table}[ht]
	\centering
	\footnotesize
	\caption{Summary of mean CRPS on the test data for the offshore benchmark for all forecast horizons. The distribution used for post-processing the wind speed (W) and wind power (P) is shown in brackets.}
	\label{tab:fullcrpsOffshore}
	\begin{tabular}{r|rrrr}
		\toprule
		Strategy & 6h & 12h & 18h & 24h \\ 
		\midrule
		Linear No Weather & 35.64 & 54.01 & 66.41 & 68.15 \\ 
		Linear Raw & 18.98 & 23.51 & 25.46 & 24.11 \\ 
		Linear \emph{One-Step-P} (Gamma) & 17.97 & 22.52 & 24.95 & 23.69 \\ 
		Linear \emph{One-Step-P} (T-Normal) & 17.92 & 22.23 & 24.85 & 23.44 \\ 
		Linear \emph{One-Step-W} (Gamma) & 18.33 & 23.54 & 24.15 & 23.81 \\ 
		Linear \emph{One-Step-W} (T-Normal) & 18.74 & 23.31 & 24.51 & 23.88 \\ 
		Linear \emph{Two-Step-WP} (W: Gamma, P: Gamma) & 17.64 & 22.98 & 23.84 & 24.76 \\ 
		Linear \emph{Two-Step-WP} (W: Gamma, P: T-Normal) & 17.21 & 22.36 & 24.03 & 23.38 \\ 
		Linear \emph{Two-Step-WP} (W: T-Normal, P: Gamma) & 18.19 & 22.67 & 24.10 & 25.10 \\ 
		Linear \emph{Two-Step-WP} (W: T-Normal, P: T-Normal) & 17.30 & 22.08 & 24.28 & 23.49 \\ 
		\midrule
		Alternative Linear No Weather & 69.48 & 70.90 & 72.09 & 69.61 \\ 
		Alternative Linear Raw & 21.25 & 24.25 & 26.25 & 24.60 \\ 
		Alternative Linear \emph{One-Step-P} (Gamma) & 19.94 & 23.48 & 25.86 & 24.68 \\ 
		Alternative Linear \emph{One-Step-P} (T-Normal) & 19.85 & 23.12 & 25.34 & 25.10 \\ 
		Alternative Linear \emph{One-Step-W} (Gamma) & 20.97 & 24.12 & 25.93 & 23.88 \\ 
		Alternative Linear \emph{One-Step-W} (T-Normal) & 21.24 & 23.93 & 26.49 & 24.65 \\ 
		Alternative Linear \emph{Two-Step-WP} (W: Gamma, P: Gamma) & 20.19 & 24.21 & 26.42 & 25.79 \\ 
		Alternative Linear \emph{Two-Step-WP} (W: Gamma, P: T-Normal) & 20.22 & 23.67 & 25.29 & 24.94 \\ 
		Alternative Linear \emph{Two-Step-WP} (W: T-Normal, P: Gamma) & 20.48 & 23.81 & 27.14 & 26.40 \\ 
		Alternative Linear \emph{Two-Step-WP} (W: T-Normal, P: T-Normal) & 20.96 & 24.13 & 26.09 & 25.92 \\
		\midrule
		Random Forest No Weather & 51.37 & 58.67 & 64.36 & 64.75 \\ 
		Random Forest Raw & 20.77 & 23.80 & 25.00 & 25.54 \\ 
		Random Forest \emph{One-Step-P} (Gamma) & 18.91 & 21.97 & 25.03 & 24.54 \\ 
		Random Forest \emph{One-Step-P} (T-Normal) & 19.47 & 22.15 & 24.24 & 23.83 \\ 
		Random Forest \emph{One-Step-W} (Gamma) & 21.26 & 23.20 & 25.07 & 24.79 \\ 
		Random Forest \emph{One-Step-W} (T-Normal) & 21.45 & 23.22 & 25.41 & 24.93 \\ 
		Random Forest \emph{Two-Step-WP} (W: Gamma, P: Gamma) & 20.31 & 21.94 & 25.23 & 25.26 \\ 
		Random Forest \emph{Two-Step-WP} (W: Gamma, P: T-Normal) & 20.24 & 22.27 & 24.49 & 23.64 \\ 
		Random Forest \emph{Two-Step-WP} (W: T-Normal, P: Gamma) & 20.33 & 22.51 & 25.67 & 25.13 \\ 
		Random Forest \emph{Two-Step-WP} (W: T-Normal, P: T-Normal) & 20.47 & 22.43 & 25.56 & 24.08 \\ 
		\midrule
		Neural Network No Weather & 37.99 & 54.47 & 64.61 & 68.55 \\ 
		Neural Network Raw & 42.61 & 55.57 & 49.78 & 47.01 \\ 
		Neural Network \emph{One-Step-P} (Gamma) & 38.18 & 46.36 & 46.47 & 42.79 \\ 
		Neural Network \emph{One-Step-P} (T-Normal) & 38.19 & 45.67 & 44.03 & 43.71 \\ 
		Neural Network \emph{One-Step-W} (Gamma) & 37.80 & 49.98 & 44.11 & 41.22 \\ 
		Neural Network \emph{One-Step-W} (T-Normal) & 38.58 & 51.33 & 45.52 & 42.19 \\ 
		Neural Network \emph{Two-Step-WP} (W: Gamma, P: Gamma) & 36.89 & 43.32 & 42.87 & 39.69 \\ 
		Neural Network \emph{Two-Step-WP} (W: Gamma, P: T-Normal) &  36.07 & 42.74 & 40.47 & 38.90 \\ 
		Neural Network \emph{Two-Step-WP} (W: T-Normal, P: Gamma) & 38.13 & 43.60 & 44.01 & 40.38 \\ 
		Neural Network \emph{Two-Step-WP} (W: T-Normal, P: T-Normal) & 38.12 & 43.20 & 42.24 & 39.70 \\ 
		\bottomrule
	\end{tabular}
\end{table}

\begin{table}[ht]
	\centering
	\scriptsize
	\caption{Summary of mean CRPS on the test data for the use case of bidding zone 3 in Sweden for all forecast horizons. The distribution used for post-processing the wind speed (W) and wind power (P) is shown in brackets.}
	\label{tab:fullcrpsSweden3}
	\begin{tabular}{r|rrrrrrrr}
		\toprule
		Strategy & 3h & 6h & 9h & 12h & 15h & 18h & 21h & 24h \\ 
		\midrule
		Linear No Weather & 68.25 & 125.24 & 169.95 & 206.38 & 228.61 & 244.21 & 255.13 & 273.54 \\ 
		Linear Raw & 57.84 & 78.46 & 81.03 & 79.75 & 86.20 & 89.05 & 89.06 & 93.33 \\ 
		Linear \emph{One-Step-P} (Gamma) & 45.87 & 55.41 & 62.07 & 67.34 & 70.53 & 64.36 & 67.97 & 67.45 \\ 
		Linear \emph{One-Step-P} (T-Normal) & 47.30 & 55.99 & 62.01 & 67.39 & 67.31 & 64.36 & 67.17 & 68.00 \\ 
		Linear \emph{One-Step-W} (Gamma) & 63.98 & 86.41 & 89.92 & 87.47 & 95.43 & 98.89 & 97.16 & 98.64 \\ 
		Linear \emph{One-Step-W} (T-Normal) & 64.51 & 86.61 & 90.05 & 87.27 & 96.89 & 99.55 & 97.95 & 98.99 \\ 
		Linear \emph{Two-Step-WP} (W: Gamma, P: Gamma) & 46.03 & 55.53 & 62.90 & 67.13 & 69.74 & 64.54 & 67.28 & 67.65 \\ 
		Linear \emph{Two-Step-WP} (W: Gamma, P: T-Normal) & 46.42 & 54.69 & 61.06 & 66.47 & 67.40 & 64.72 & 65.97 & 65.77 \\ 
		Linear \emph{Two-Step-WP} (W: T-Normal, P: Gamma) & 46.73 & 54.88 & 62.44 & 66.65 & 69.04 & 62.50 & 68.34 & 66.76 \\ 
		Linear \emph{Two-Step-WP} (W: T-Normal, P: T-Normal) & 47.08 & 55.32 & 62.04 & 65.56 & 66.86 & 63.70 & 67.05 & 65.83 \\ 
		\midrule
		Alternative Linear No Weather & 276.01 & 282.10 & 280.85 & 287.40 & 292.19 & 290.27 & 279.83 & 280.57 \\ 
		Alternative Linear Raw & 82.32 & 82.86 & 76.98 & 70.64 & 73.19 & 81.16 & 82.95 & 83.68 \\ 
		Alternative Linear \emph{One-Step-P} (Gamma) & 59.74 & 58.48 & 56.85 & 57.52 & 56.18 & 55.63 & 60.11 & 56.36 \\ 
		Alternative Linear \emph{One-Step-P} (T-Normal) & 59.95 & 59.01 & 57.42 & 58.53 & 57.43 & 55.85 & 61.02 & 57.44 \\ 
		Alternative Linear \emph{One-Step-W} (Gamma) & 97.98 & 93.85 & 86.63 & 80.77 & 84.45 & 90.63 & 91.10 & 89.64 \\ 
		Alternative Linear \emph{One-Step-W} (T-Normal) & 98.29 & 94.21 & 86.78 & 80.51 & 85.04 & 89.73 & 90.44 & 89.02 \\ 
		Alternative Linear \emph{Two-Step-WP} (W: Gamma, P: Gamma) & 57.43 & 56.73 & 56.79 & 55.95 & 56.55 & 55.19 & 60.48 & 58.29 \\ 
		Alternative Linear \emph{Two-Step-WP} (W: Gamma, P: T-Normal) & 58.22 & 58.27 & 56.81 & 56.76 & 56.73 & 56.47 & 60.51 & 57.52 \\ 
		Alternative Linear \emph{Two-Step-WP} (W: T-Normal, P: Gamma) & 57.62 & 57.29 & 55.46 & 55.96 & 55.87 & 55.07 & 60.19 & 56.55 \\ 
		Alternative Linear \emph{Two-Step-WP} (W: T-Normal, P: T-Normal) & 58.74 & 58.99 & 56.67 & 57.58 & 57.48 & 56.41 & 60.88 & 57.72 \\ 
		\midrule
		Random Forest No Weather & 170.90 & 192.13 & 211.51 & 235.68 & 252.47 & 252.25 & 251.61 & 260.93 \\ 
		Random Forest Raw & 60.71 & 71.31 & 69.72 & 67.74 & 73.39 & 82.01 & 83.67 & 88.09 \\ 
		Random Forest \emph{One-Step-P} (Gamma) & 43.54 & 48.30 & 52.78 & 55.29 & 54.46 & 52.92 & 57.36 & 57.11 \\ 
		Random Forest \emph{One-Step-P} (T-Normal) & 46.08 & 50.28 & 55.09 & 55.96 & 55.17 & 53.44 & 57.83 & 58.86 \\ 
		Random Forest \emph{One-Step-W} (Gamma) & 70.05 & 79.57 & 78.28 & 76.32 & 82.51 & 91.18 & 91.08 & 92.40 \\ 
		Random Forest \emph{One-Step-W} (T-Normal) & 69.96 & 79.95 & 78.21 & 76.50 & 82.77 & 90.38 & 91.47 & 91.15 \\ 
		Random Forest \emph{Two-Step-WP} (W: Gamma, P: Gamma) & 43.00 & 48.10 & 53.28 & 54.17 & 54.95 & 53.46 & 56.25 & 57.52 \\ 
		Random Forest \emph{Two-Step-WP} (W: Gamma, P: T-Normal) & 44.84 & 49.53 & 55.36 & 54.19 & 54.58 & 54.02 & 56.75 & 56.08 \\ 
		Random Forest \emph{Two-Step-WP} (W: T-Normal, P: Gamma) & 42.83 & 48.83 & 53.37 & 54.44 & 54.34 & 52.51 & 55.76 & 55.73 \\ 
		Random Forest \emph{Two-Step-WP} (W: T-Normal, P: T-Normal) & 45.20 & 49.92 & 55.48 & 54.27 & 55.36 & 53.95 & 57.34 & 57.24 \\ 
		\midrule
		Neural Network No Weather & 60.11 & 111.15 & 150.67 & 185.20 & 202.59 & 210.39 & 224.67 & 233.61 \\ 
		Neural Network Raw & 54.05 & 78.40 & 70.00 & 76.04 & 100.56 & 73.69 & 73.16 & 75.79 \\ 
		Neural Network \emph{One-Step-P} (Gamma) & 48.26 & 61.59 & 58.07 & 63.21 & 76.20 & 61.80 & 67.13 & 67.65 \\ 
		Neural Network \emph{One-Step-P} (T-Normal) & 48.38 & 64.54 & 59.89 & 66.47 & 78.67 & 64.66 & 69.14 & 68.11 \\ 
		Neural Network \emph{One-Step-W} (Gamma) & 56.59 & 78.18 & 72.74 & 71.19 & 98.05 & 72.66 & 72.87 & 73.20 \\ 
		Neural Network \emph{One-Step-W} (T-Normal) & 56.75 & 78.24 & 72.64 & 71.44 & 99.37 & 74.27 & 72.51 & 75.06 \\ 
		Neural Network \emph{Two-Step-WP} (W: Gamma, P: Gamma) & 48.31 & 61.04 & 59.85 & 63.69 & 76.67 & 63.02 & 69.13 & 67.87 \\ 
		Neural Network \emph{Two-Step-WP} (W: Gamma, P: T-Normal) & 48.56 & 64.67 & 60.48 & 64.27 & 77.11 & 63.77 & 68.66 & 66.32 \\ 
		Neural Network \emph{Two-Step-WP} (W: T-Normal, P: Gamma) & 48.09 & 62.30 & 60.30 & 63.38 & 75.97 & 63.69 & 68.40 & 67.69 \\ 
		Neural Network \emph{Two-Step-WP} (W: T-Normal, P: T-Normal) & 48.70 & 64.57 & 60.81 & 65.20 & 76.92 & 64.61 & 68.46 & 67.70 \\ 
		\bottomrule
	\end{tabular}
\end{table}

\begin{table}[ht]
	\centering
	\scriptsize
	\caption{Summary of mean CRPS on the test data for the use case of bidding zone 4 in Sweden for all forecast horizons. The distribution used for post-processing the wind speed (W) and wind power (P) is shown in brackets.}
	\label{tab:fullcrpsSweden4}
	\begin{tabular}{r|rrrrrrrr}
		\toprule
		Strategy & 3h & 6h & 9h & 12h & 15h & 18h & 21h & 24h \\ 
		\midrule
		Linear No Weather & 49.39 & 89.66 & 120.37 & 142.60 & 154.74 & 165.48 & 172.85 & 181.03 \\ 
		Linear Raw & 41.75 & 49.91 & 50.93 & 63.17 & 63.85 & 53.27 & 51.46 & 51.21 \\ 
		Linear \emph{One-Step-P} (Gamma) & 35.70 & 42.39 & 45.62 & 54.83 & 58.10 & 50.11 & 49.84 & 47.72 \\ 
		Linear \emph{One-Step-P} (T-Normal) & 36.17 & 41.60 & 43.15 & 49.68 & 53.51 & 47.86 & 45.92 & 43.69 \\ 
		Linear \emph{One-Step-W} (Gamma) & 41.78 & 51.18 & 52.17 & 57.57 & 61.95 & 52.73 & 51.68 & 49.12 \\ 
		Linear \emph{One-Step-W} (T-Normal) & 41.91 & 51.20 & 51.85 & 57.17 & 61.52 & 52.86 & 51.24 & 48.86 \\ 
		Linear \emph{Two-Step-WP} (W: Gamma, P: Gamma) & 35.44 & 42.47 & 45.94 & 54.84 & 58.20 & 50.14 & 49.22 & 47.74 \\ 
		Linear \emph{Two-Step-WP} (W: Gamma, P: T-Normal) & 36.08 & 40.73 & 42.56 & 49.52 & 53.96 & 48.05 & 45.62 & 43.49 \\ 
		Linear \emph{Two-Step-WP} (W: T-Normal, P: Gamma) & 35.31 & 42.24 & 45.74 & 54.04 & 57.33 & 49.50 & 49.64 & 46.95 \\ 
		Linear \emph{Two-Step-WP} (W: T-Normal, P: T-Normal) & 35.78 & 41.24 & 43.38 & 50.61 & 54.43 & 47.50 & 47.15 & 43.14 \\ 
		\midrule
		Alternative Linear No Weather & 184.17 & 189.06 & 196.37 & 202.03 & 197.75 & 190.11 & 186.76 & 187.90 \\ 
		Alternative Linear Raw & 42.92 & 44.07 & 42.82 & 54.38 & 55.16 & 44.77 & 40.33 & 38.30 \\ 
		Alternative Linear \emph{One-Step-P} (Gamma) & 39.76 & 40.06 & 40.17 & 48.42 & 50.34 & 43.93 & 41.02 & 37.65 \\ 
		Alternative Linear \emph{One-Step-P} (T-Normal) & 39.68 & 40.71 & 40.93 & 48.11 & 51.96 & 44.59 & 41.61 & 38.13 \\ 
		Alternative Linear \emph{One-Step-W} (Gamma) & 42.95 & 46.43 & 44.15 & 51.19 & 55.29 & 45.58 & 40.60 & 38.03 \\ 
		Alternative Linear \emph{One-Step-W} (T-Normal) & 43.37 & 46.75 & 44.18 & 50.98 & 55.58 & 45.98 & 41.08 & 38.17 \\ 
		Alternative Linear \emph{Two-Step-WP} (W: Gamma, P: Gamma) & 38.16 & 40.60 & 39.50 & 48.42 & 50.72 & 43.77 & 41.47 & 38.36 \\ 
		Alternative Linear \emph{Two-Step-WP} (W: Gamma, P: T-Normal) & 38.99 & 41.41 & 40.37 & 47.80 & 50.26 & 44.40 & 41.41 & 37.82 \\ 
		Alternative Linear \emph{Two-Step-WP} (W: T-Normal, P: Gamma) & 38.91 & 40.44 & 39.50 & 48.32 & 50.54 & 43.47 & 41.41 & 37.58 \\ 
		Alternative Linear \emph{Two-Step-WP} (W: T-Normal, P: T-Normal) & 39.15 & 40.42 & 40.11 & 48.34 & 51.39 & 44.39 & 41.73 & 37.90 \\ 
		\midrule
		Random Forest No Weather & 116.32 & 129.71 & 148.64 & 166.77 & 171.40 & 168.17 & 169.95 & 175.11 \\ 
		Random Forest Raw & 36.97 & 41.33 & 40.74 & 50.67 & 51.60 & 45.54 & 42.26 & 39.41 \\ 
		Random Forest \emph{One-Step-P} (Gamma) & 33.34 & 37.24 & 38.29 & 44.51 & 47.28 & 43.09 & 40.58 & 37.88 \\ 
		Random Forest \emph{One-Step-P} (T-Normal) & 34.33 & 38.25 & 39.30 & 44.60 & 48.02 & 44.28 & 41.72 & 38.18 \\ 
		Random Forest \emph{One-Step-W} (Gamma) & 36.63 & 42.63 & 41.89 & 46.57 & 51.59 & 45.69 & 43.07 & 38.57 \\ 
		Random Forest \emph{One-Step-W} (T-Normal) & 36.81 & 42.91 & 41.94 & 46.51 & 51.16 & 46.24 & 43.16 & 38.31 \\ 
		Random Forest \emph{Two-Step-WP} (W: Gamma, P: Gamma) & 32.47 & 37.03 & 37.82 & 43.25 & 48.01 & 43.12 & 41.11 & 37.42 \\ 
		Random Forest \emph{Two-Step-WP} (W: Gamma, P: T-Normal) & 33.71 & 37.91 & 38.76 & 43.33 & 47.90 & 44.73 & 41.72 & 37.32 \\ 
		Random Forest \emph{Two-Step-WP} (W: T-Normal, P: Gamma) & 32.44 & 37.31 & 37.87 & 43.74 & 46.81 & 43.13 & 40.73 & 37.25 \\ 
		Random Forest \emph{Two-Step-WP} (W: T-Normal, P: T-Normal) & 33.52 & 38.25 & 39.21 & 43.95 & 48.08 & 44.74 & 42.04 & 37.33 \\ 
		\midrule
		Neural Network No Weather & 49.75 & 90.51 & 121.89 & 150.22 & 157.56 & 163.37 & 168.79 & 174.76  \\ 
		Neural Network Raw & 47.52 & 48.68 & 49.41 & 58.64 & 58.55 & 61.21 & 55.49 & 51.71 \\ 
		Neural Network \emph{One-Step-P} (Gamma) & 38.92 & 44.02 & 45.03 & 57.58 & 54.49 & 49.75 & 47.89 & 44.22 \\ 
		Neural Network \emph{One-Step-P} (T-Normal) & 40.26 & 44.24 & 45.80 & 56.82 & 56.79 & 51.21 & 47.55 & 45.02 \\ 
		Neural Network \emph{One-Step-W} (Gamma) & 70.80 & 92.27 & 105.05 & 100.64 & 98.54 & 80.15 & 88.04 & 87.43 \\ 
		Neural Network \emph{One-Step-W} (T-Normal) & 51.76 & 68.62 & 77.67 & 78.28 & 74.47 & 61.77 & 65.74 & 65.53 \\ 
		Neural Network \emph{Two-Step-WP} (W: Gamma, P: Gamma) & 43.55 & 53.51 & 51.32 & 61.07 & 58.26 & 55.46 & 49.56 & 50.08 \\ 
		Neural Network \emph{Two-Step-WP} (W: Gamma, P: T-Normal) & 45.79 & 53.14 & 52.60 & 59.16 & 57.91 & 55.56 & 49.88 & 52.13 \\ 
		Neural Network \emph{Two-Step-WP} (W: T-Normal, P: Gamma) & 40.89 & 51.27 & 50.19 & 56.03 & 55.45 & 52.22 & 47.70 & 48.15 \\ 
		Neural Network \emph{Two-Step-WP} (W: T-Normal, P: T-Normal) & 43.45 & 50.78 & 49.57 & 56.19 & 54.98 & 52.43 & 48.17 & 49.43 \\ 
		\bottomrule
	\end{tabular}
\end{table}

\paragraph{Fitting Distributions}

\begin{figure}
	\centering
	\begin{subfigure}[b]{0.49\textwidth}
		\includegraphics[width=\textwidth]{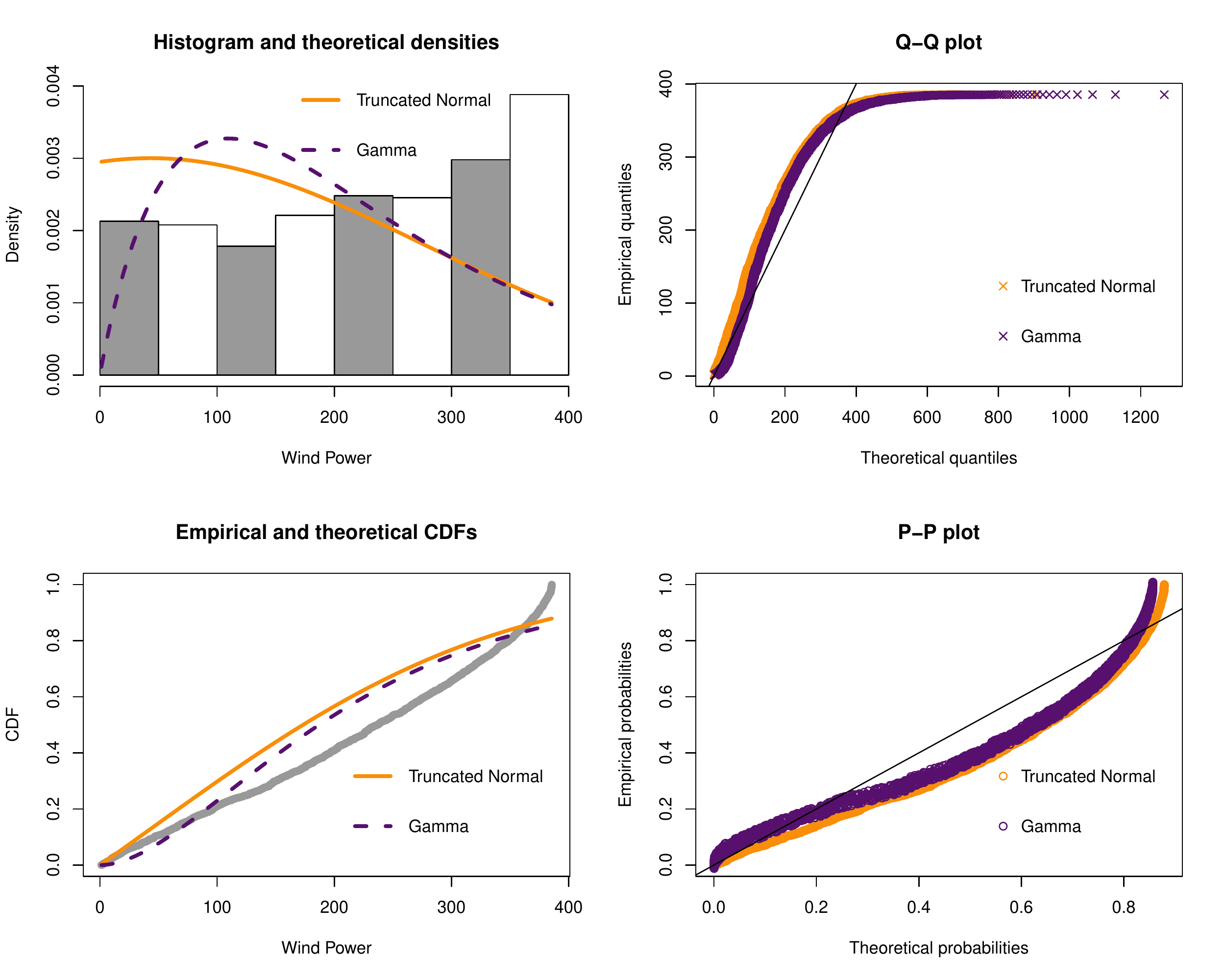}
		\caption{Wind Power}
	\end{subfigure}
	\begin{subfigure}[b]{0.49\textwidth}
		\includegraphics[width=\textwidth]{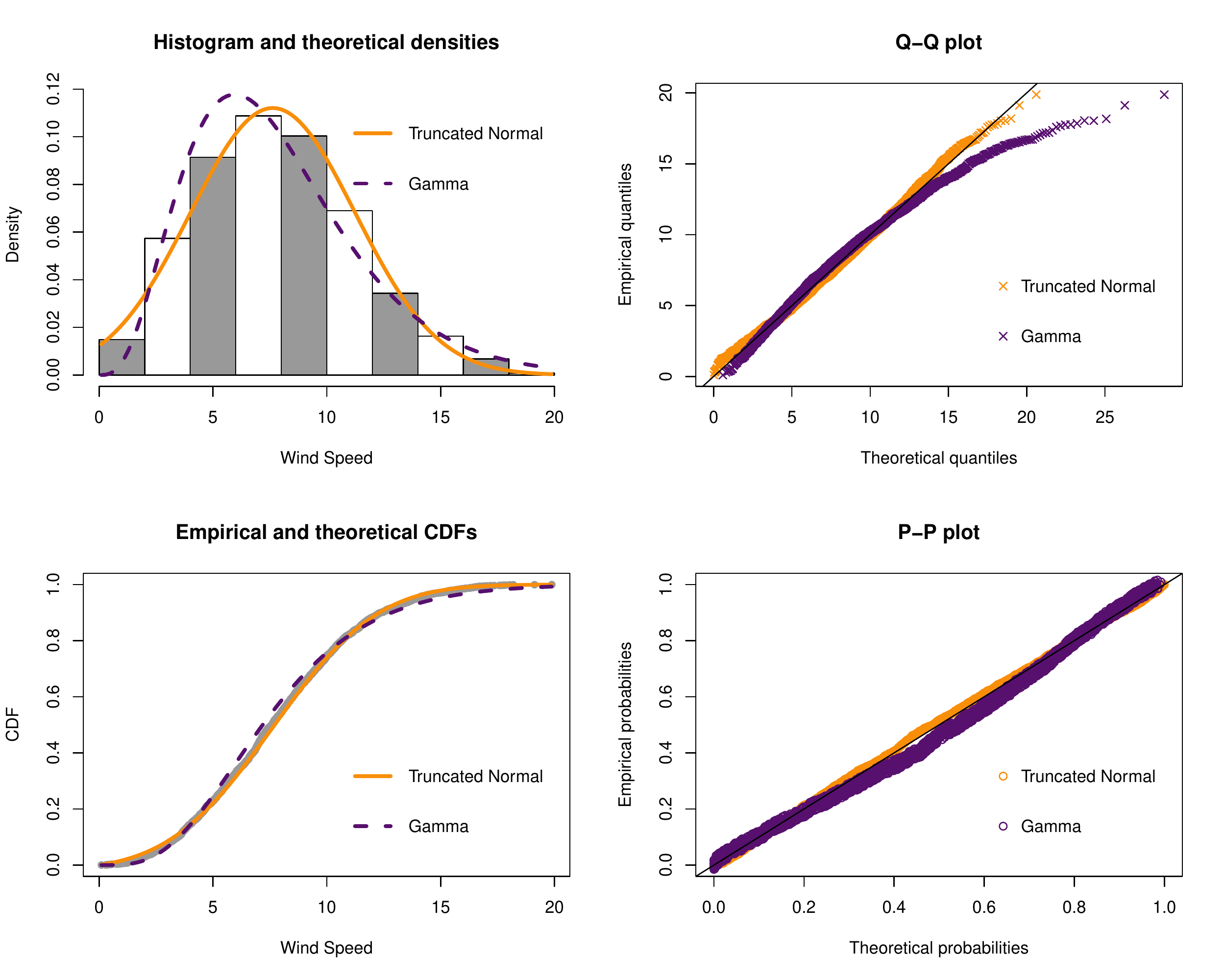}
		\caption{Wind Speed}
	\end{subfigure}
	\begin{subfigure}[b]{0.49\textwidth}
		\includegraphics[width=\textwidth]{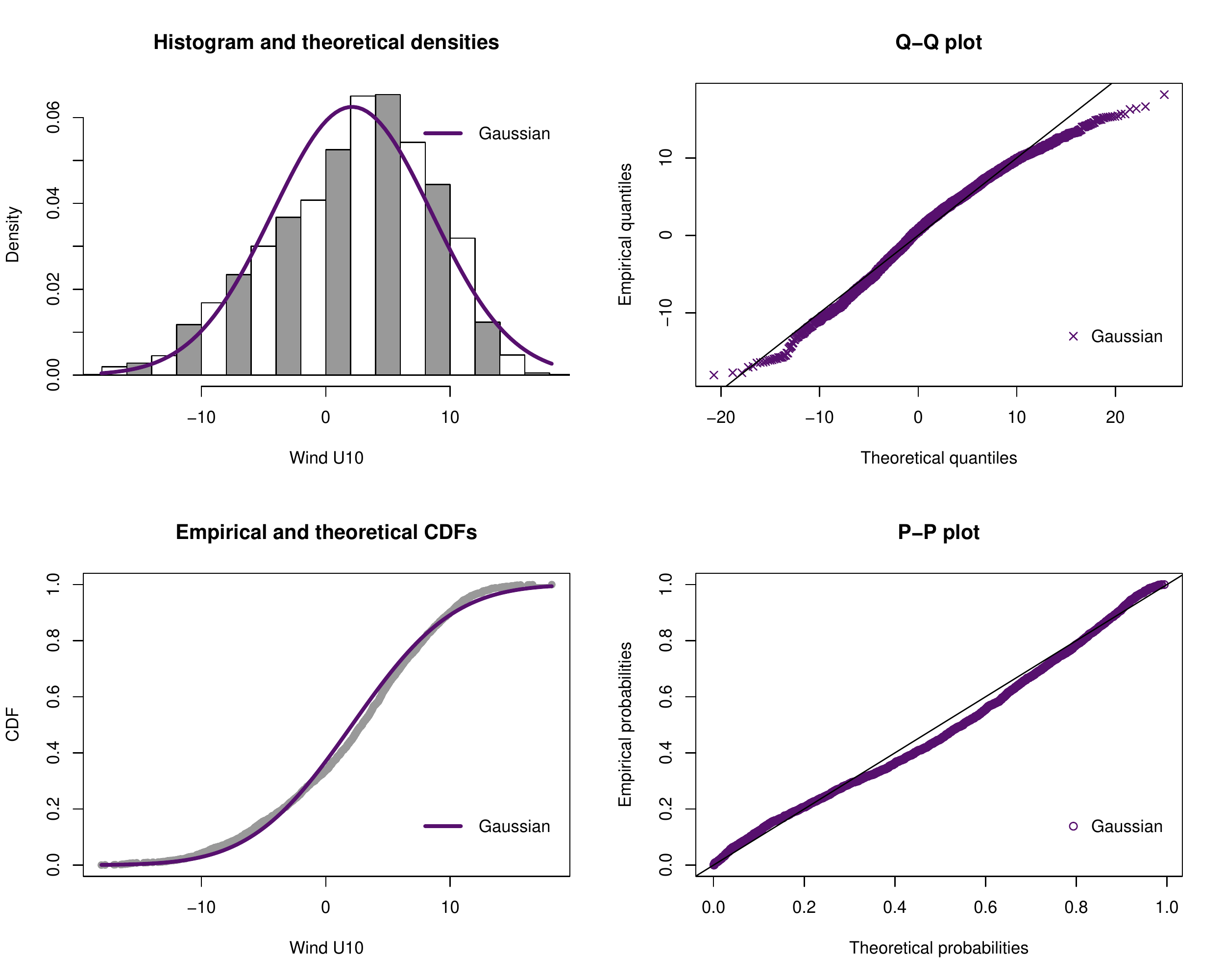}
		\caption{Wind U100}
	\end{subfigure}
	\begin{subfigure}[b]{0.49\textwidth}
		\includegraphics[width=\textwidth]{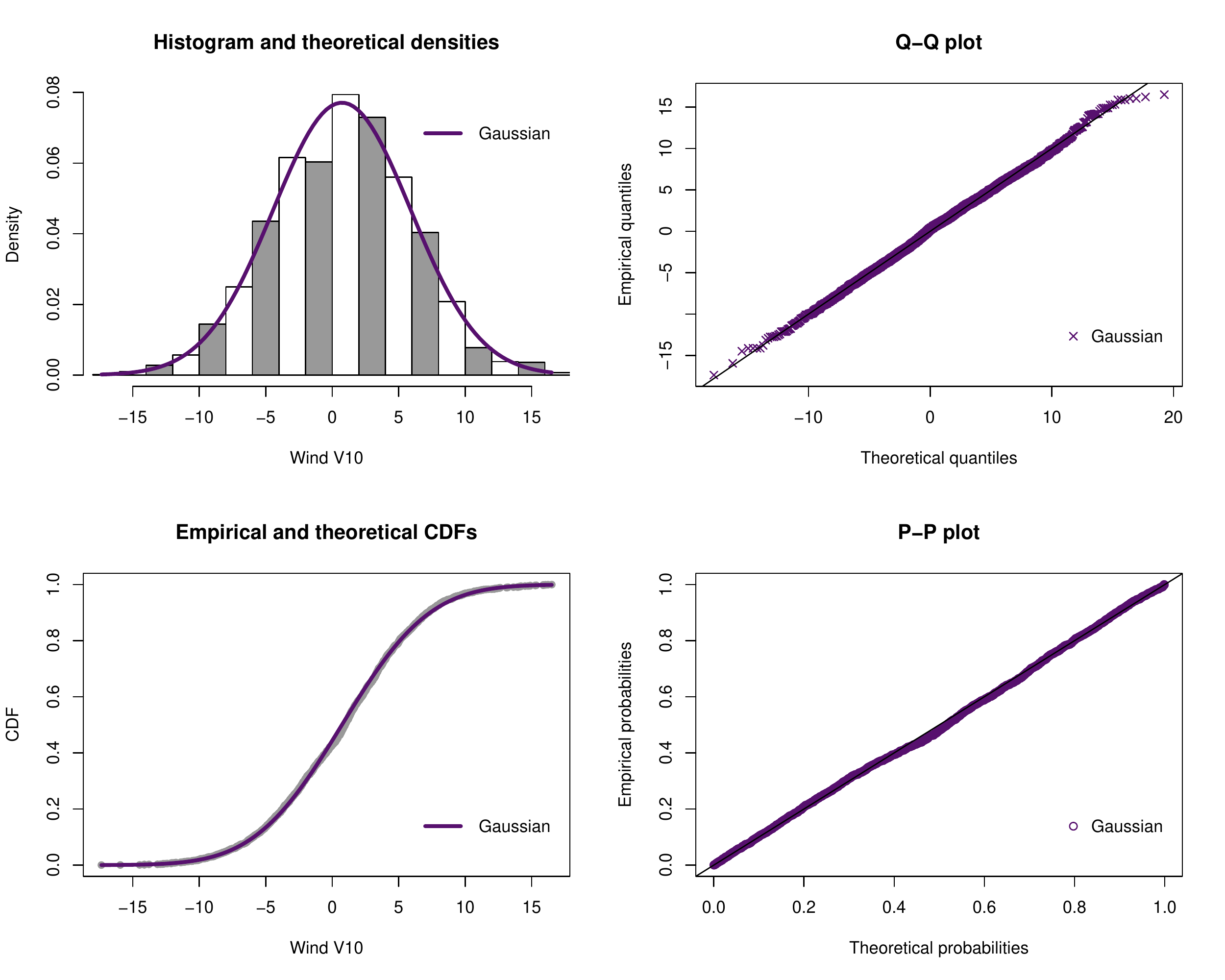}
		\caption{Wind V100}
	\end{subfigure}
	\begin{subfigure}[b]{0.49\textwidth}
		\includegraphics[width=\textwidth]{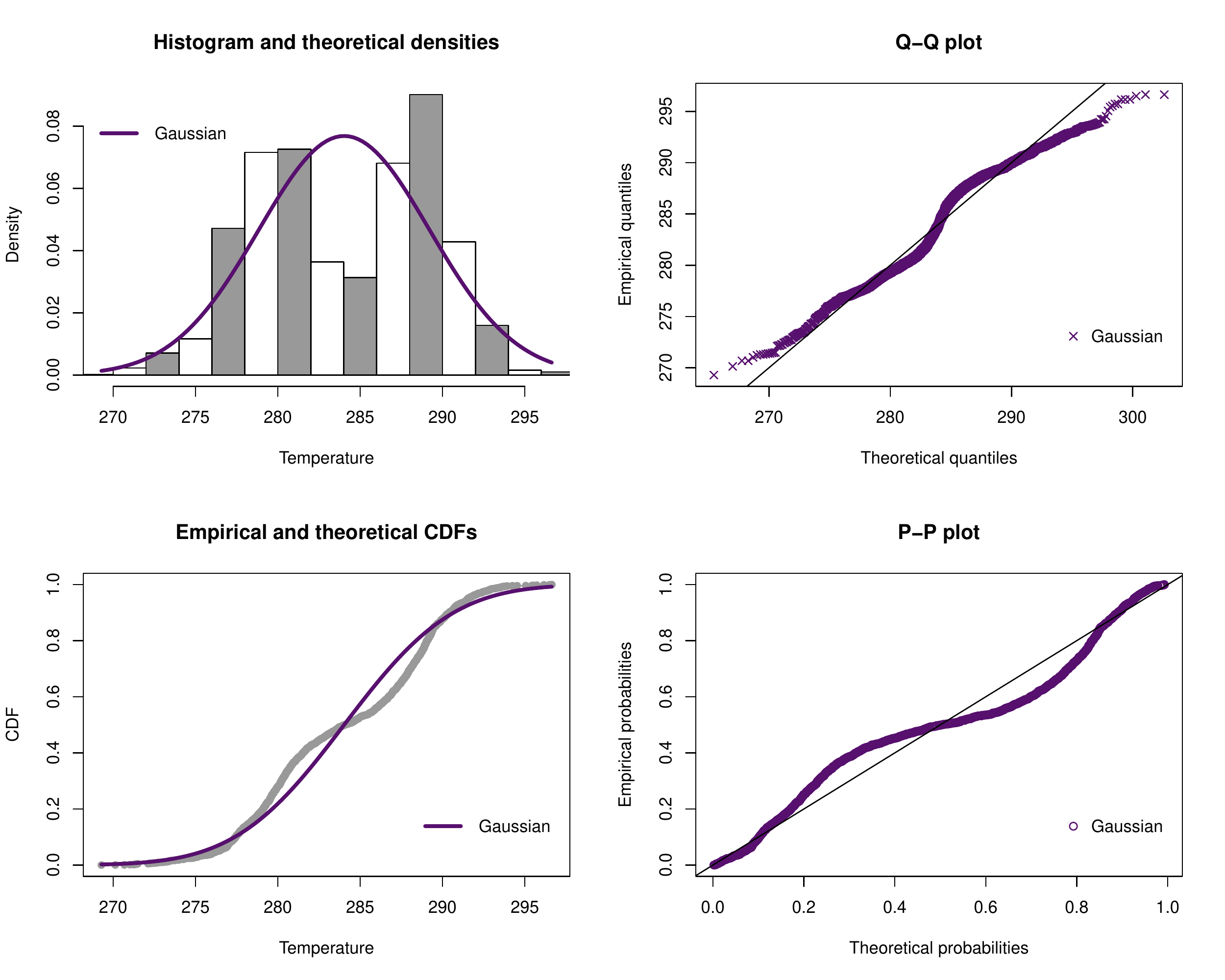}
		\caption{Temperature}
	\end{subfigure}
	\begin{subfigure}[b]{0.49\textwidth}
		\includegraphics[width=\textwidth]{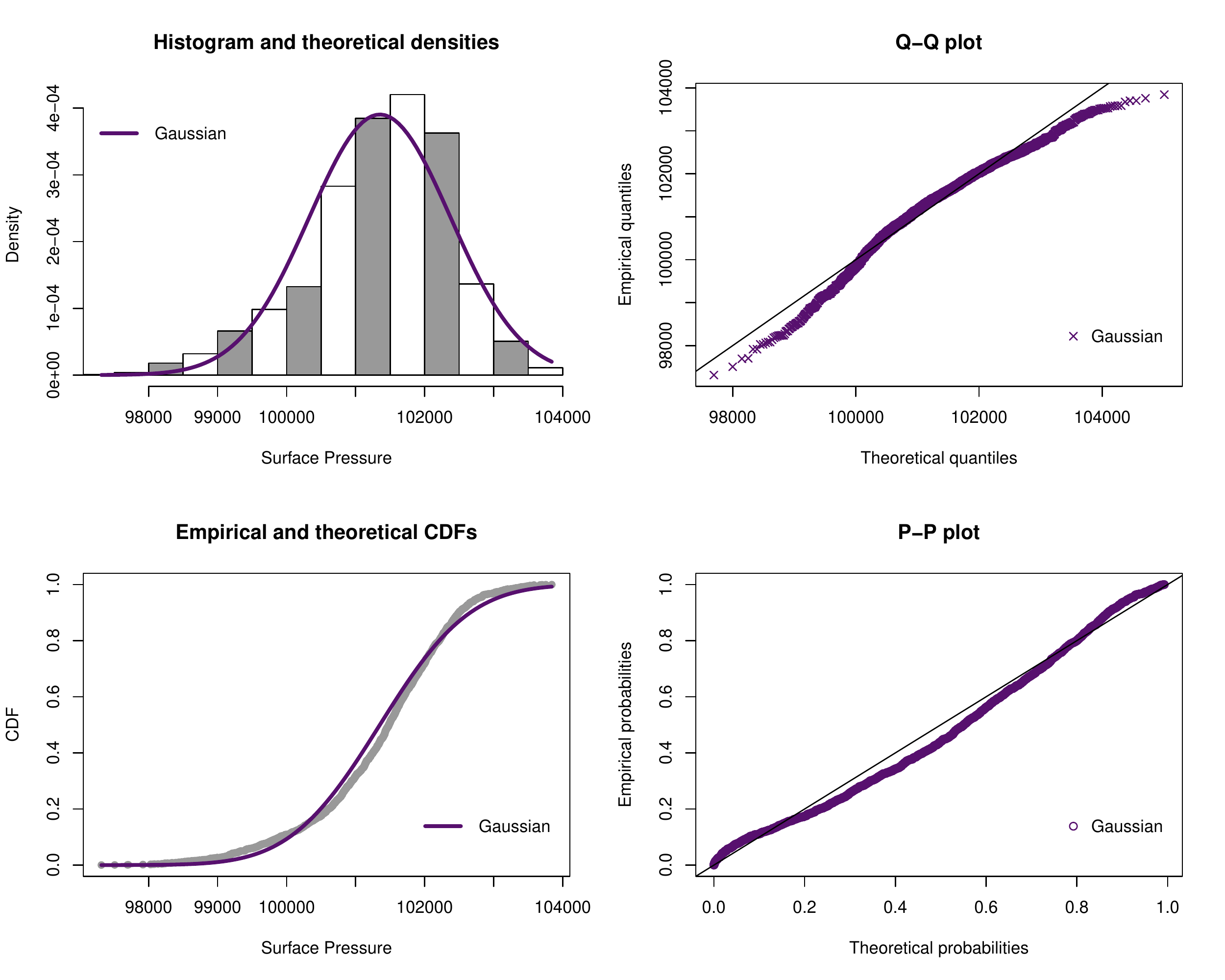}
		\caption{Surface Pressure}
	\end{subfigure}
	\caption{Fitted distributions for wind power and all weather variables for the offshore benchmark data set.}
	\label{fig:fit_offshore}
\end{figure}

\begin{figure}
	\centering
	\begin{subfigure}[b]{0.49\textwidth}
		\includegraphics[width=\textwidth]{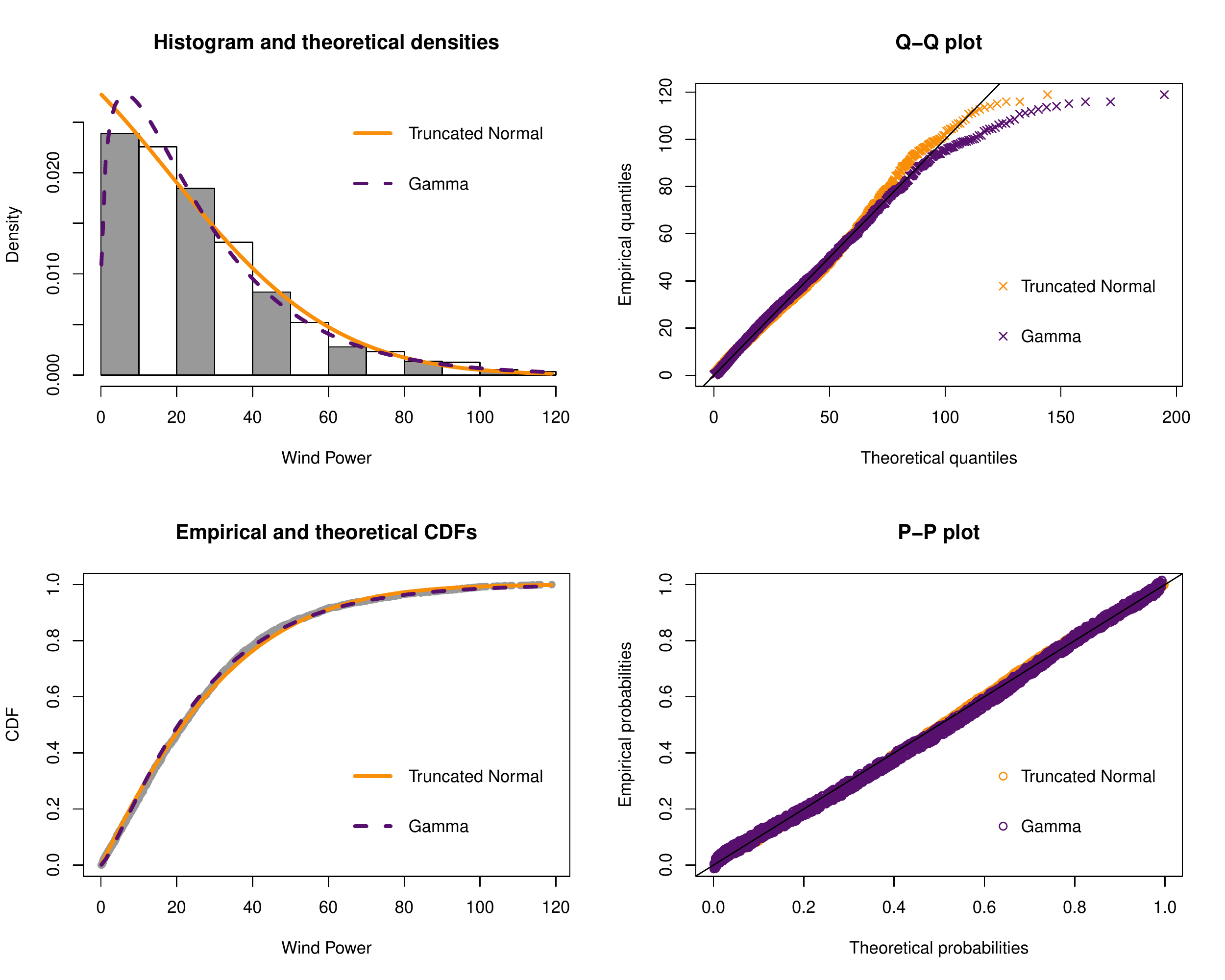}
		\caption{Wind Power}
	\end{subfigure}
	\begin{subfigure}[b]{0.49\textwidth}
		\includegraphics[width=\textwidth]{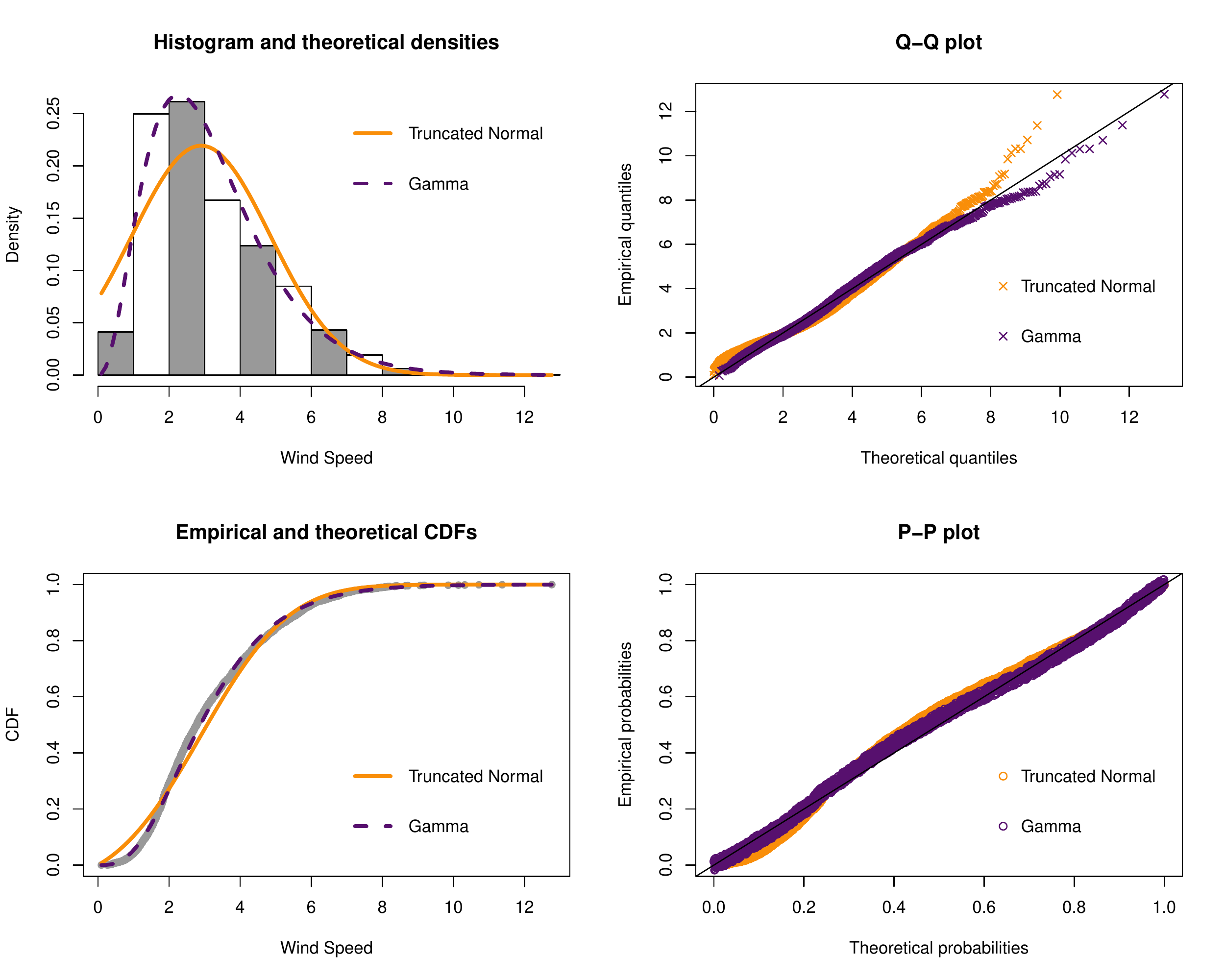}
		\caption{Wind Speed}
	\end{subfigure}
	\begin{subfigure}[b]{0.49\textwidth}
		\includegraphics[width=\textwidth]{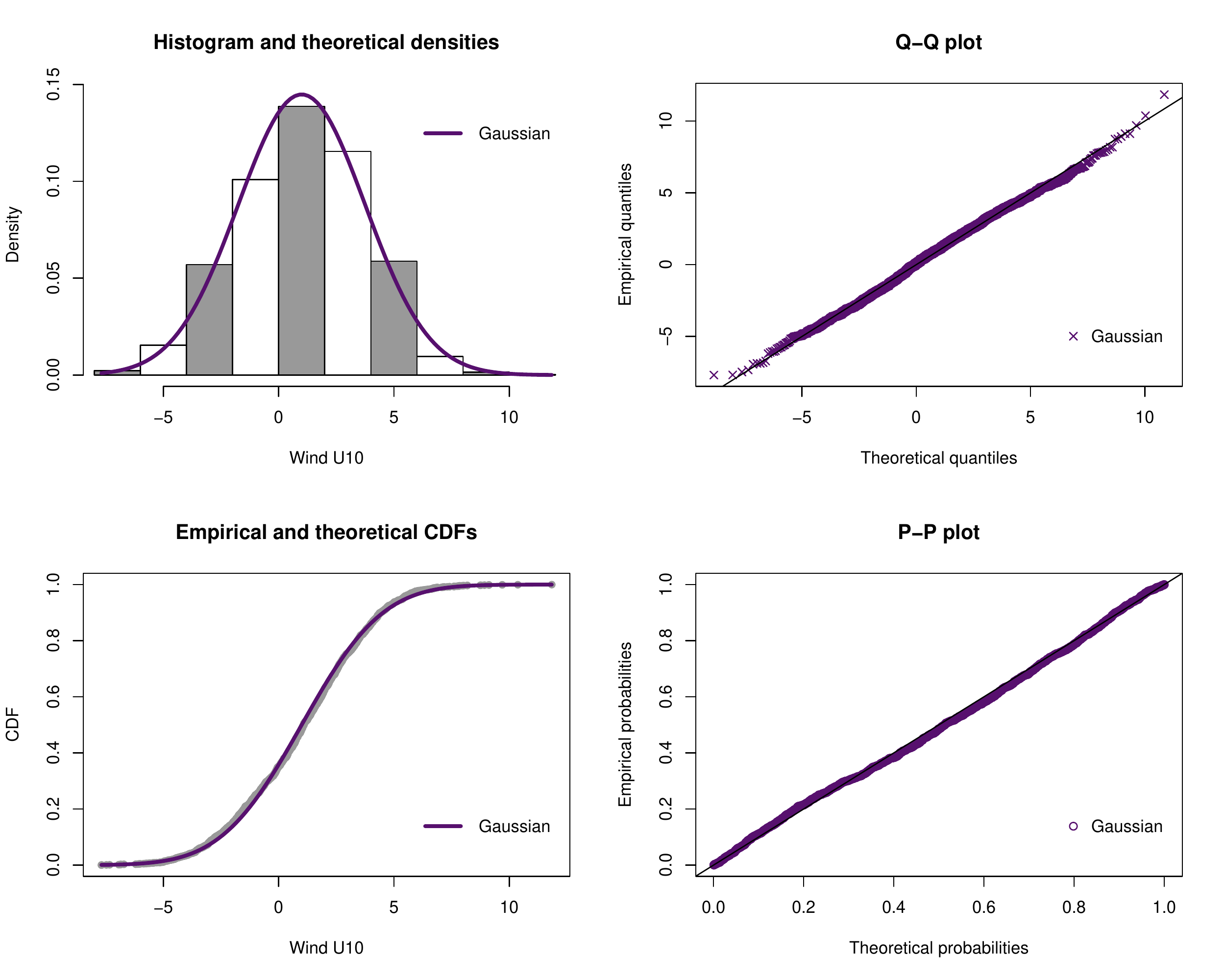}
		\caption{Wind U100}
	\end{subfigure}
	\begin{subfigure}[b]{0.49\textwidth}
		\includegraphics[width=\textwidth]{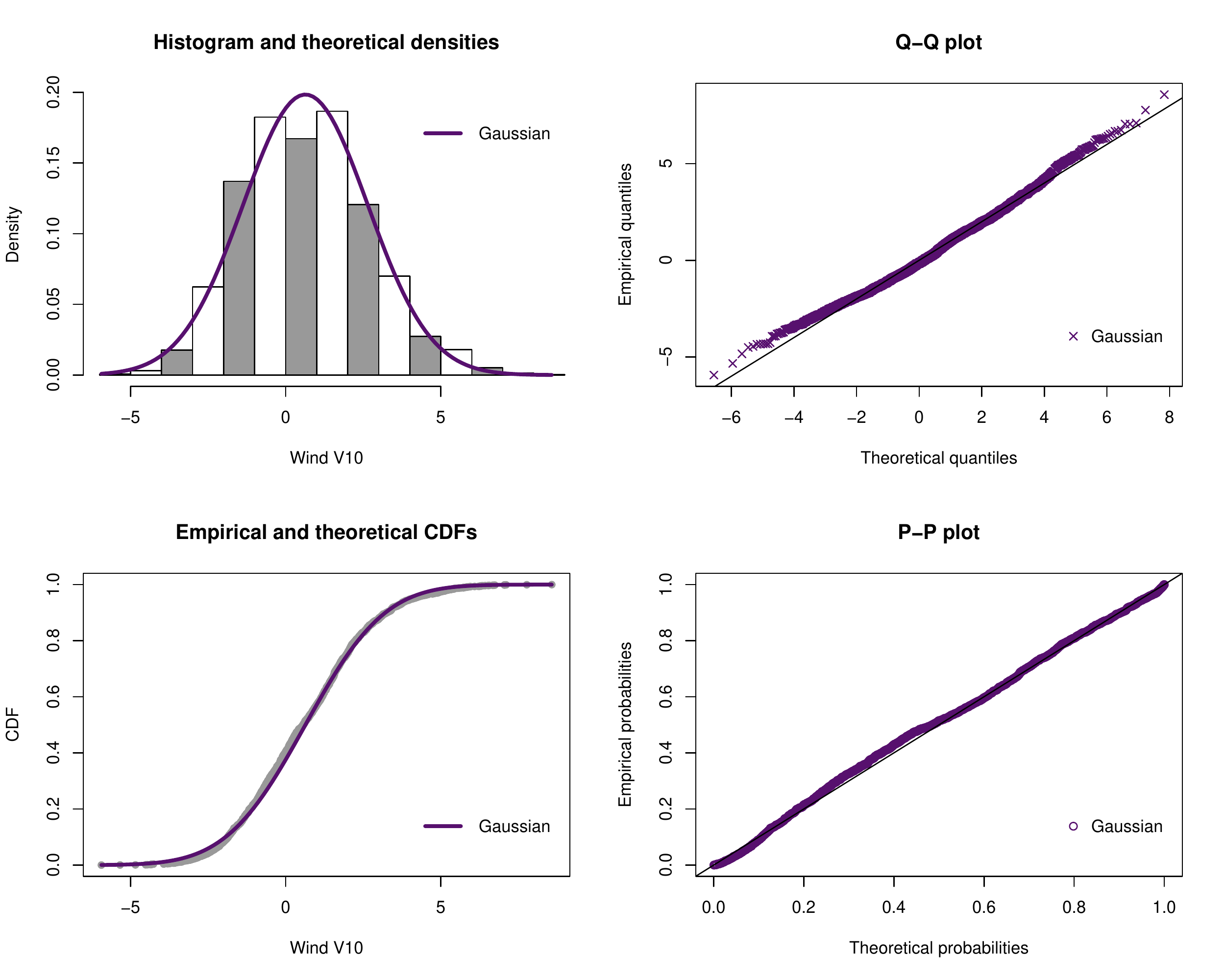}
		\caption{Wind V100}
	\end{subfigure}
	\begin{subfigure}[b]{0.49\textwidth}
		\includegraphics[width=\textwidth]{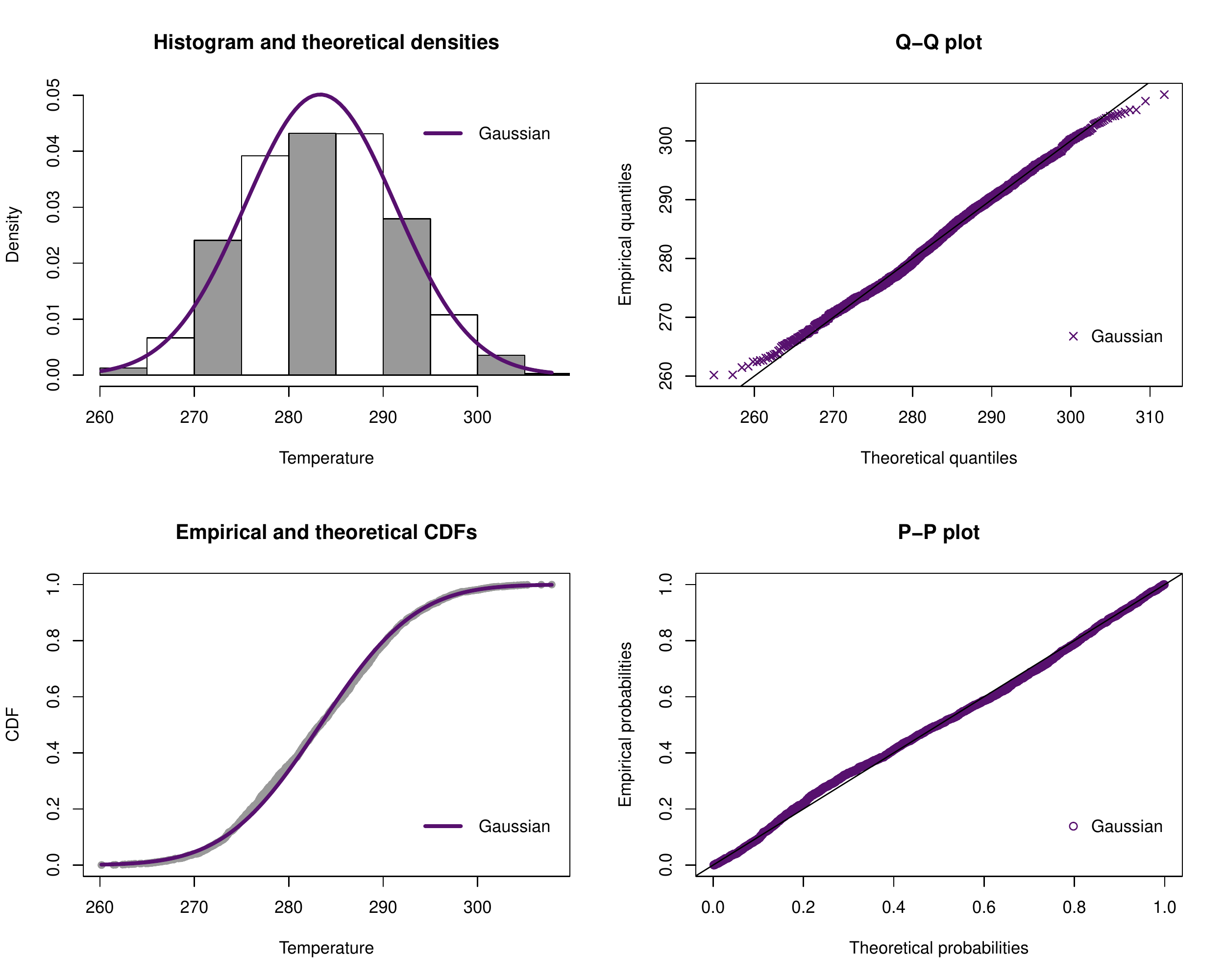}
		\caption{Temperature}
	\end{subfigure}
	\begin{subfigure}[b]{0.49\textwidth}
		\includegraphics[width=\textwidth]{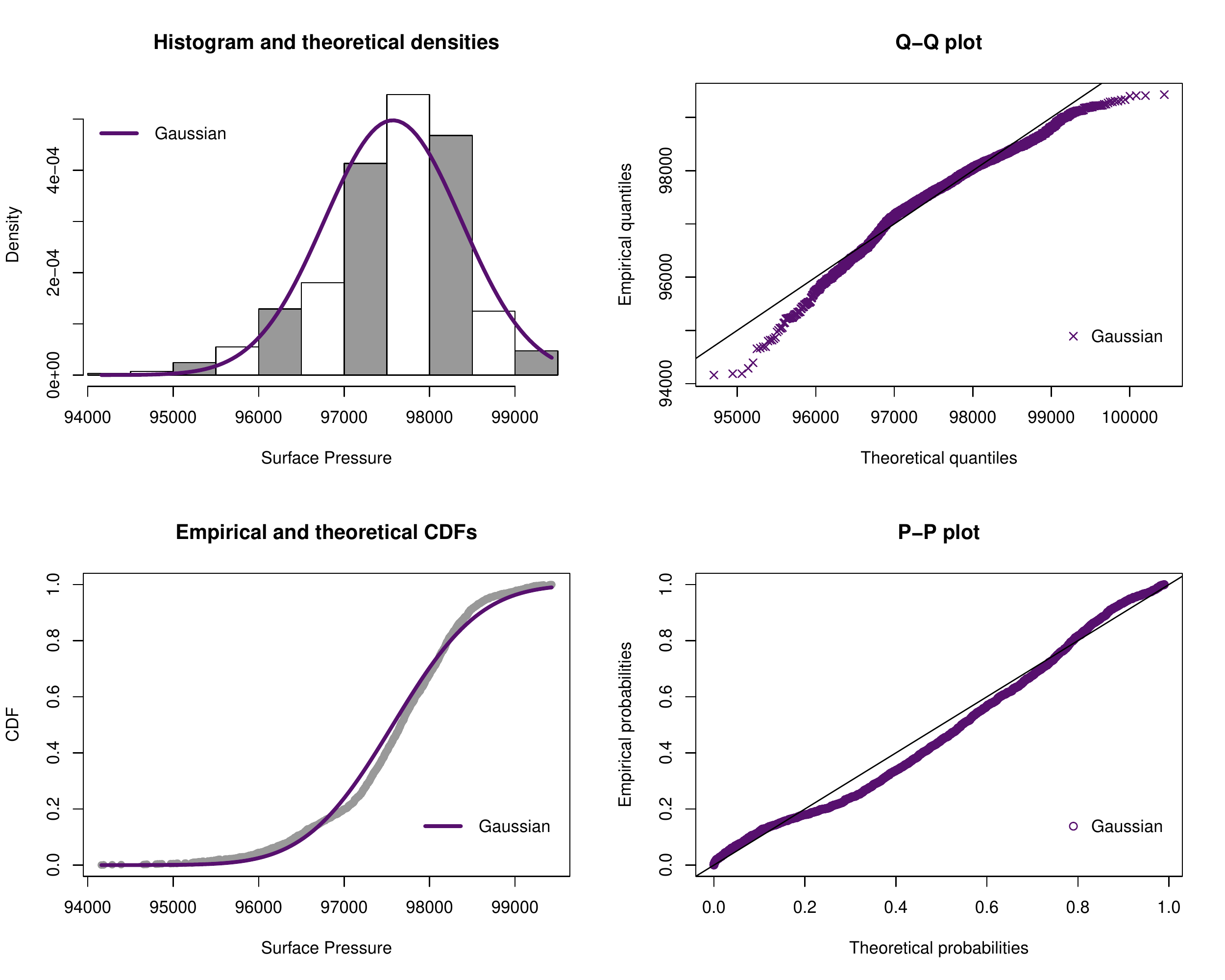}
		\caption{Surface Pressure}
	\end{subfigure}
	\caption{Fitted distributions for wind power and all weather variables for the onshore benchmark data set.}
	\label{fig:fit_onshore}
\end{figure}

\begin{figure}
	\centering
	\begin{subfigure}[b]{0.49\textwidth}
		\includegraphics[width=\textwidth]{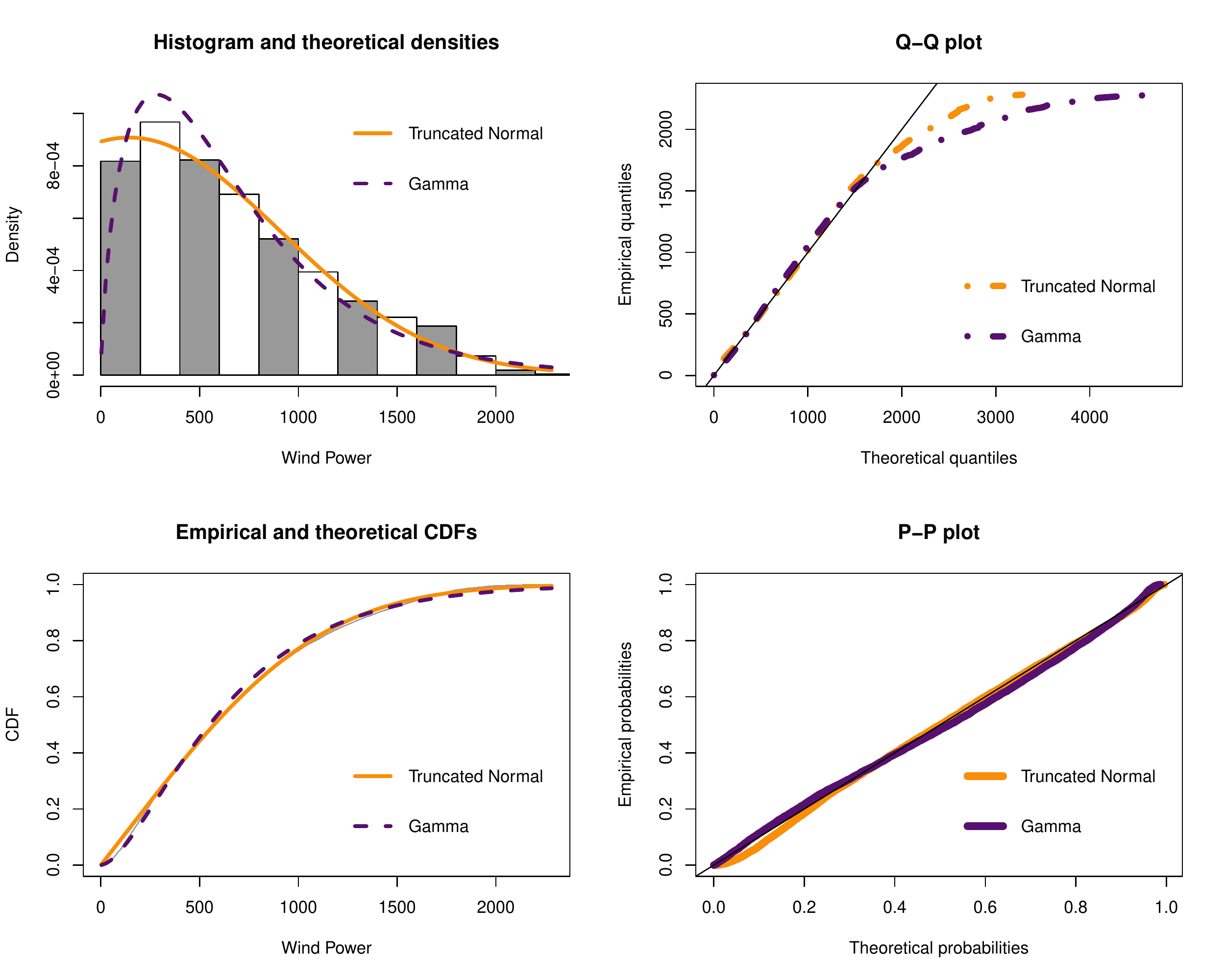}
		\caption{Wind Power}
	\end{subfigure}
	\begin{subfigure}[b]{0.49\textwidth}
		\includegraphics[width=\textwidth]{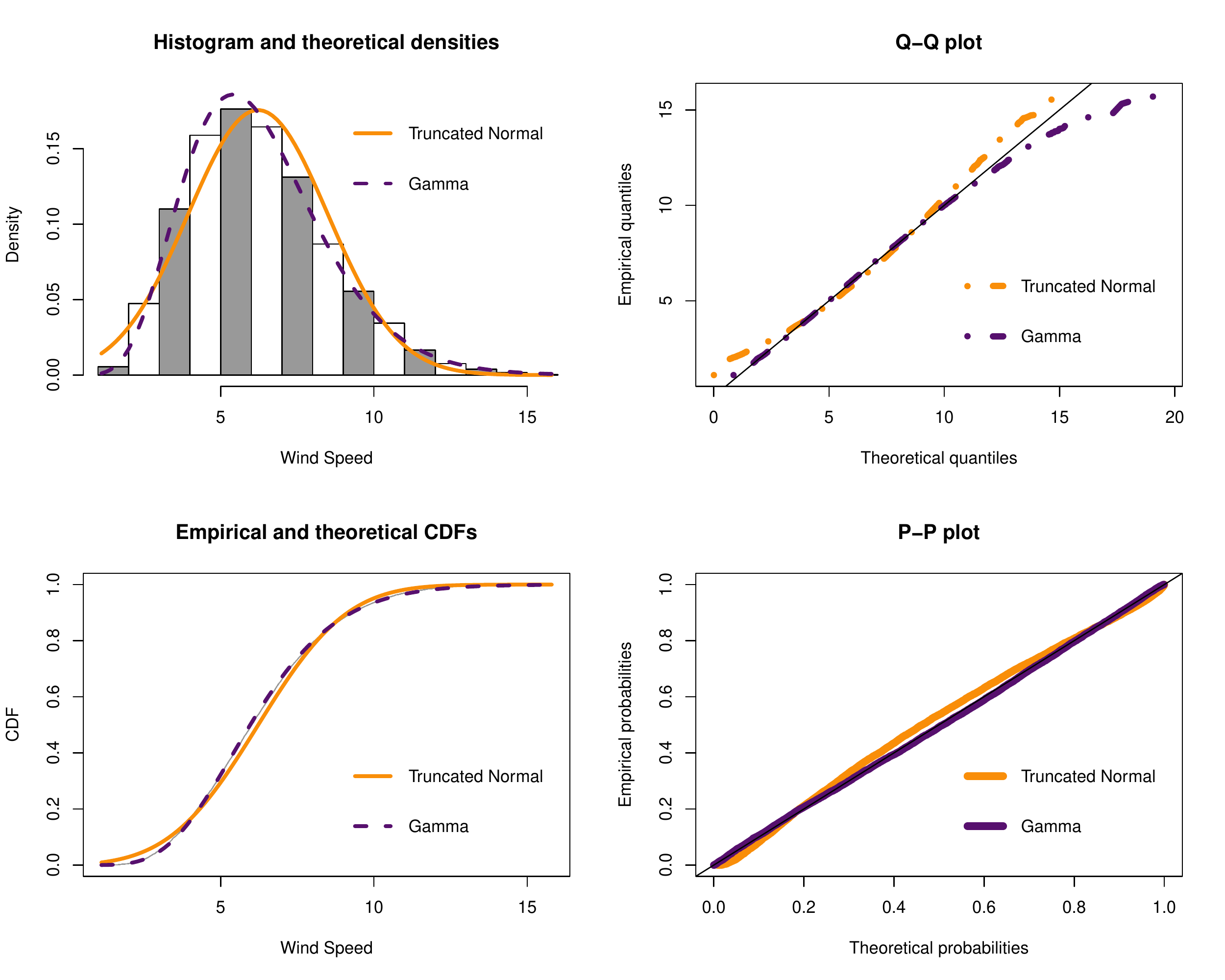}
		\caption{Wind Speed}
	\end{subfigure}
	\begin{subfigure}[b]{0.49\textwidth}
		\includegraphics[width=\textwidth]{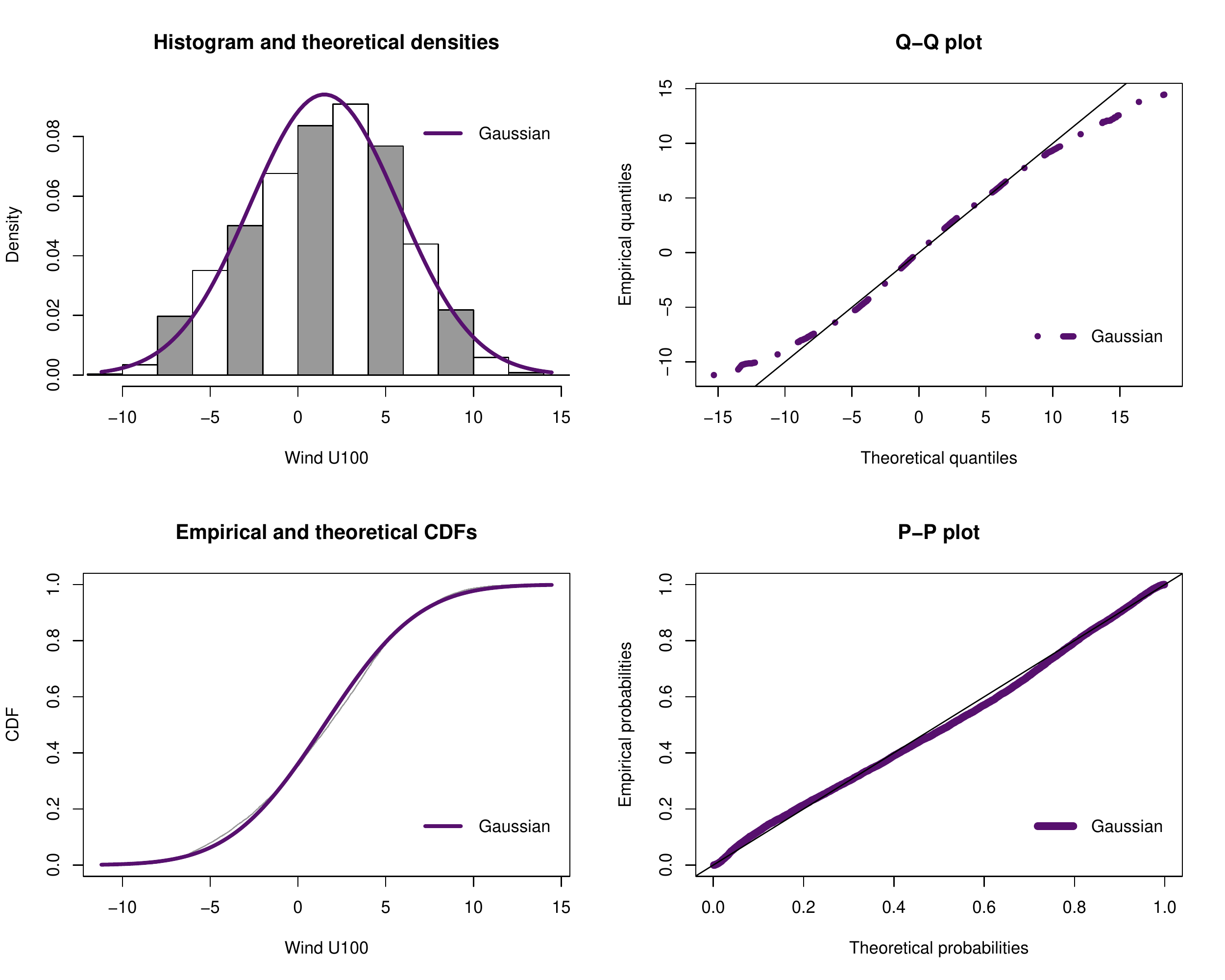}
		\caption{Wind U100}
	\end{subfigure}
	\begin{subfigure}[b]{0.49\textwidth}
		\includegraphics[width=\textwidth]{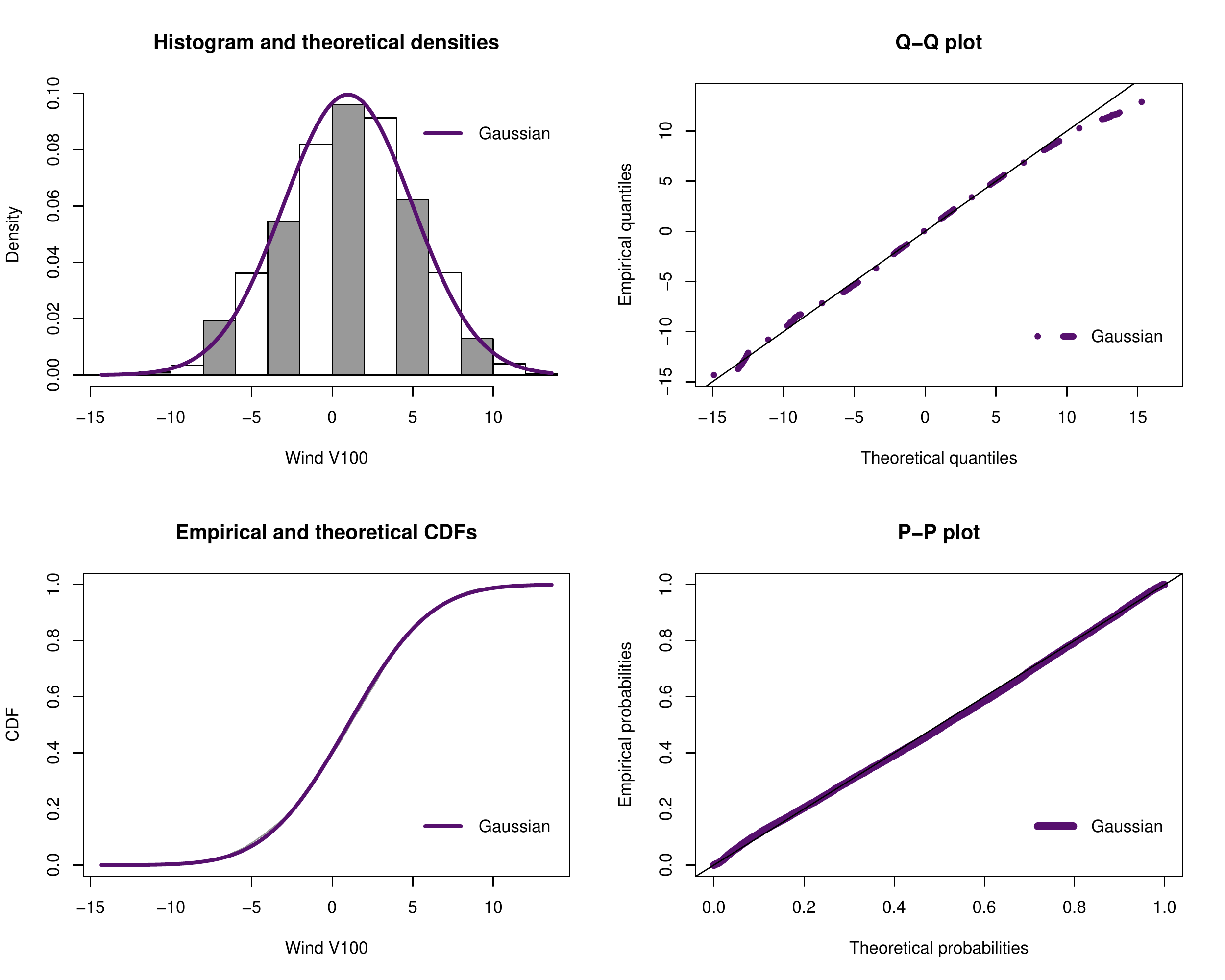}
		\caption{Wind V100}
	\end{subfigure}
	\begin{subfigure}[b]{0.49\textwidth}
		\includegraphics[width=\textwidth]{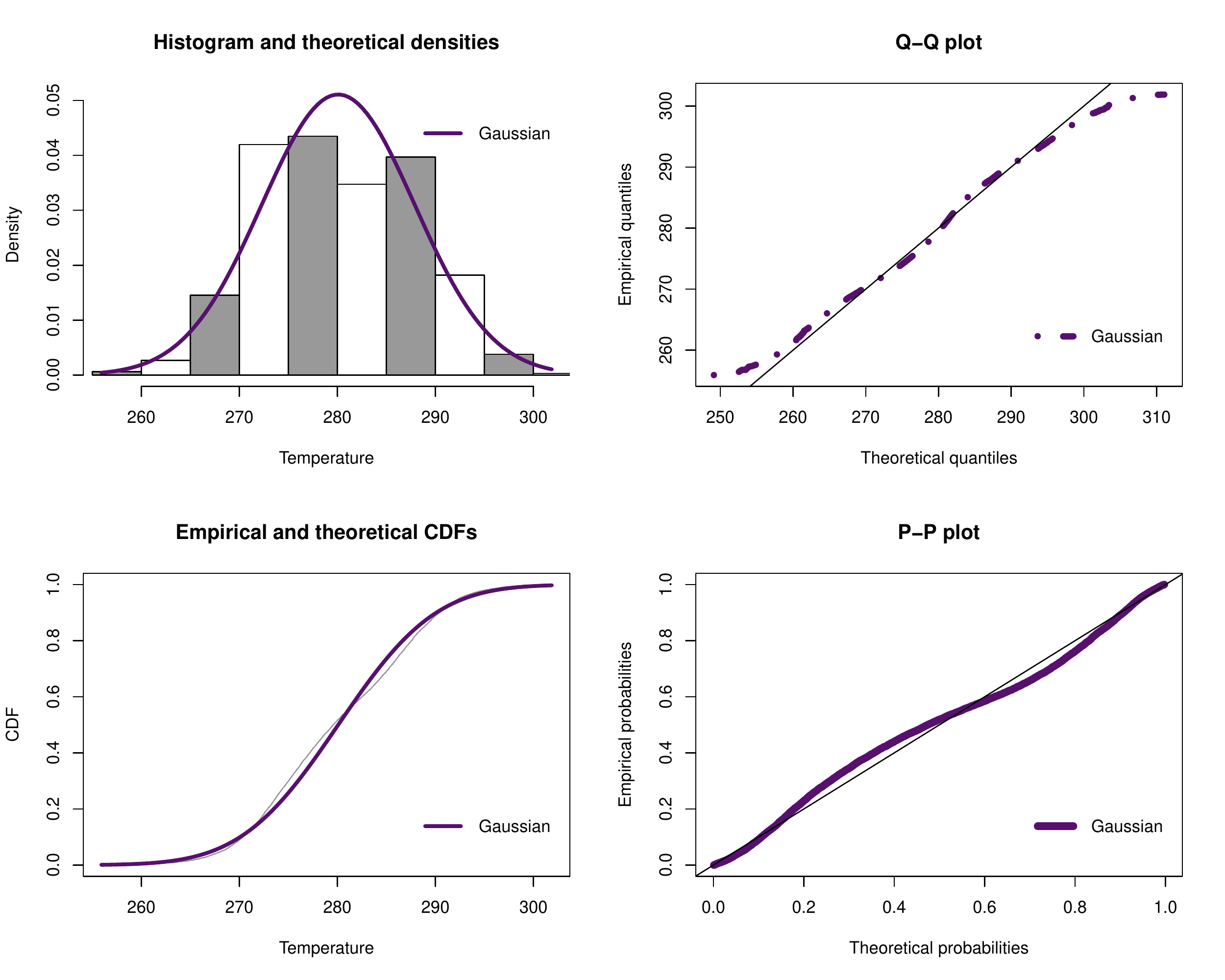}
		\caption{Temperature}
	\end{subfigure}
	\begin{subfigure}[b]{0.49\textwidth}
		\includegraphics[width=\textwidth]{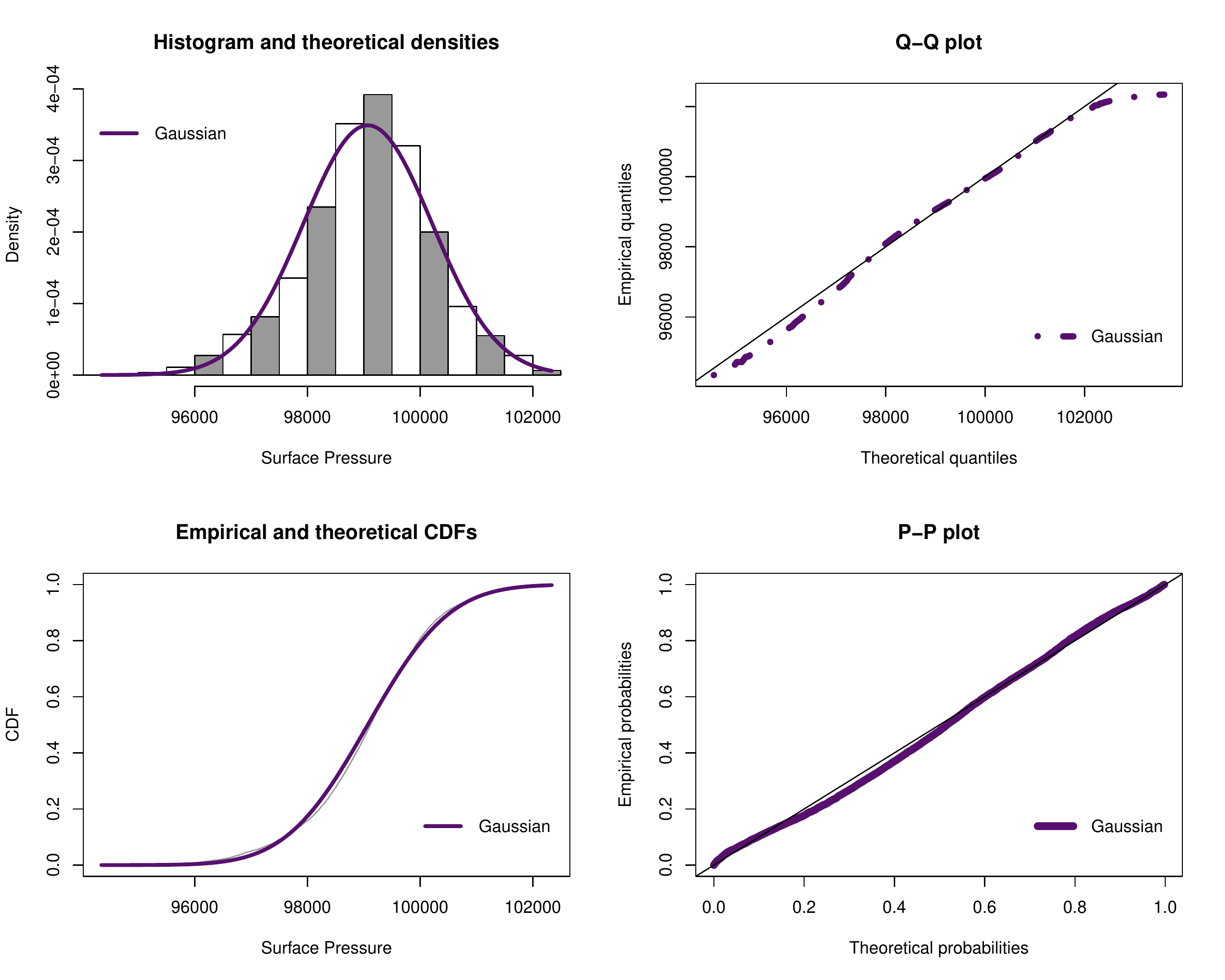}
		\caption{Surface Pressure}
	\end{subfigure}
	\caption{Fitted distributions for wind power and all weather variables for bidding zone 3 in Sweden.}
	\label{fig:fit_sweden_zone3}
\end{figure}

\begin{figure}
	\centering
	\begin{subfigure}[b]{0.49\textwidth}
		\includegraphics[width=\textwidth]{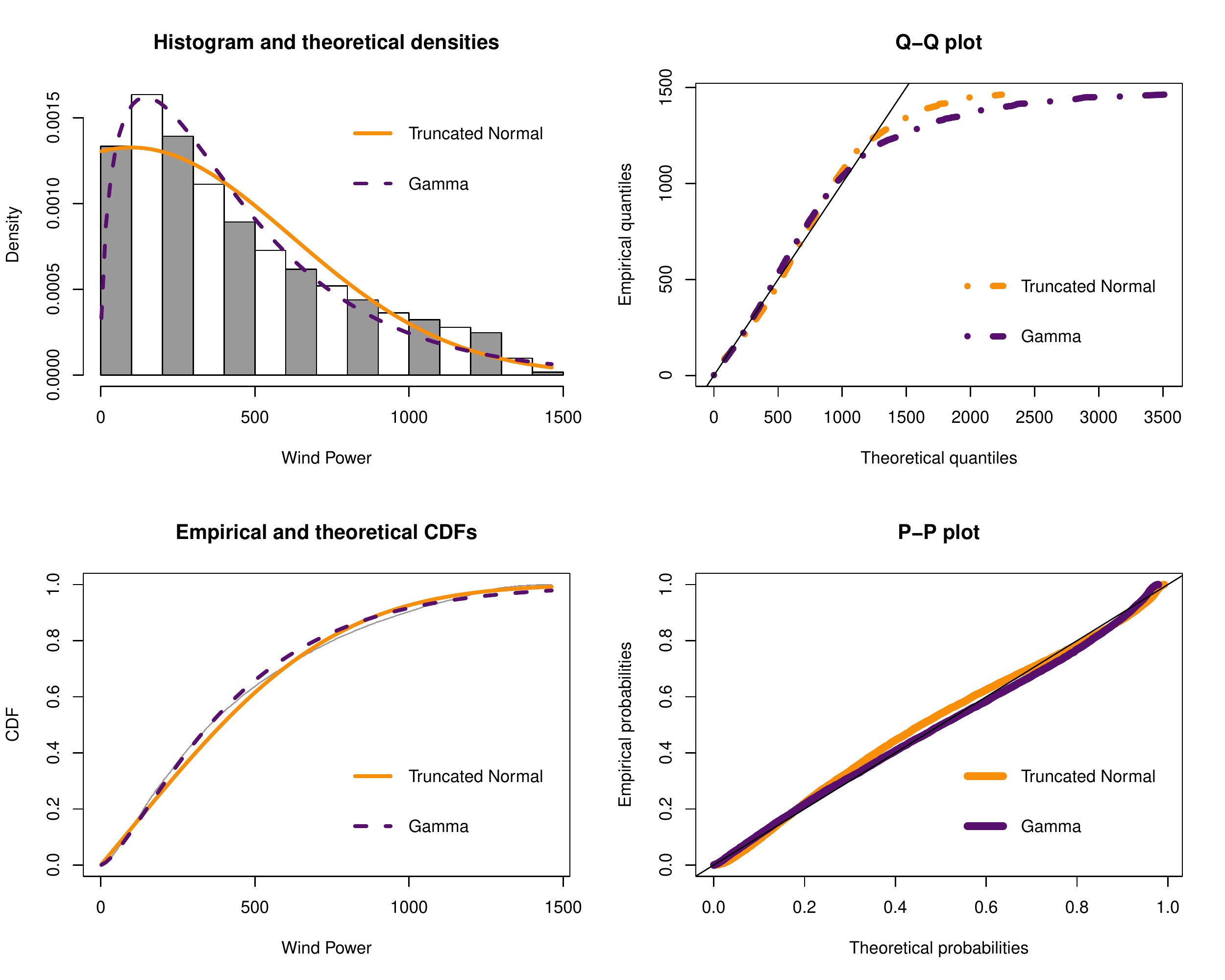}
		\caption{Wind Power}
	\end{subfigure}
	\begin{subfigure}[b]{0.49\textwidth}
		\includegraphics[width=\textwidth]{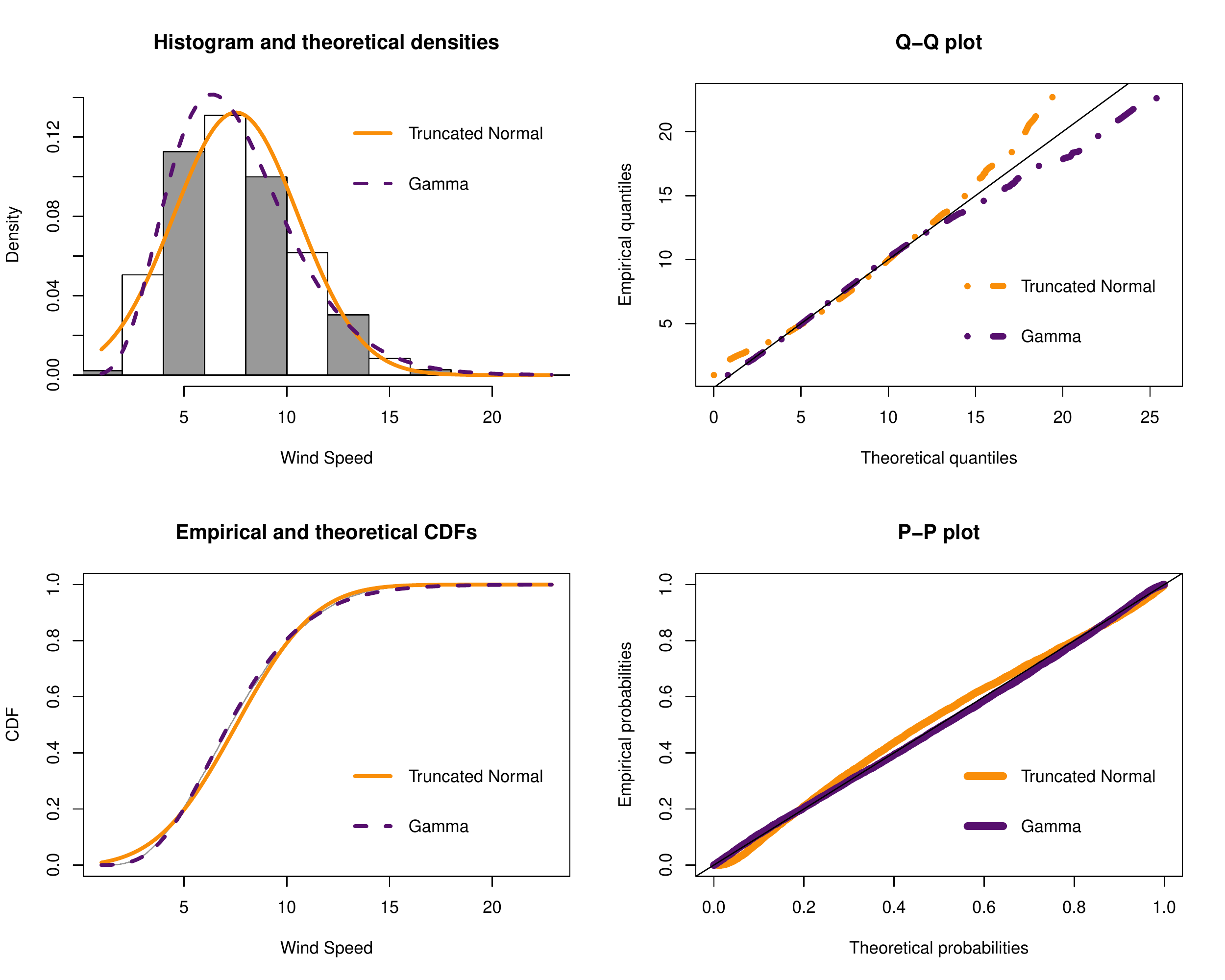}
		\caption{Wind Speed}
	\end{subfigure}
	\begin{subfigure}[b]{0.49\textwidth}
		\includegraphics[width=\textwidth]{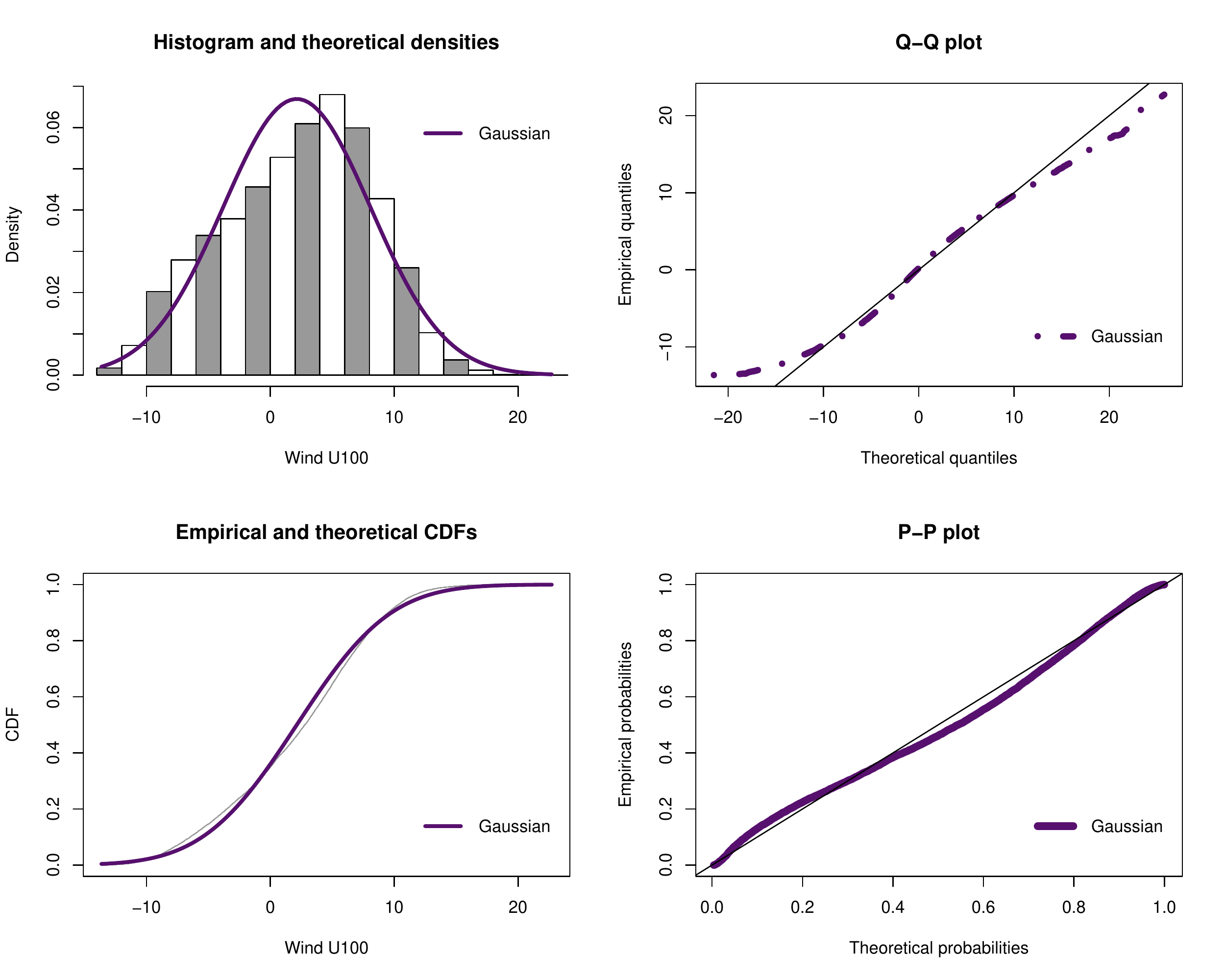}
		\caption{Wind U100}
	\end{subfigure}
	\begin{subfigure}[b]{0.49\textwidth}
		\includegraphics[width=\textwidth]{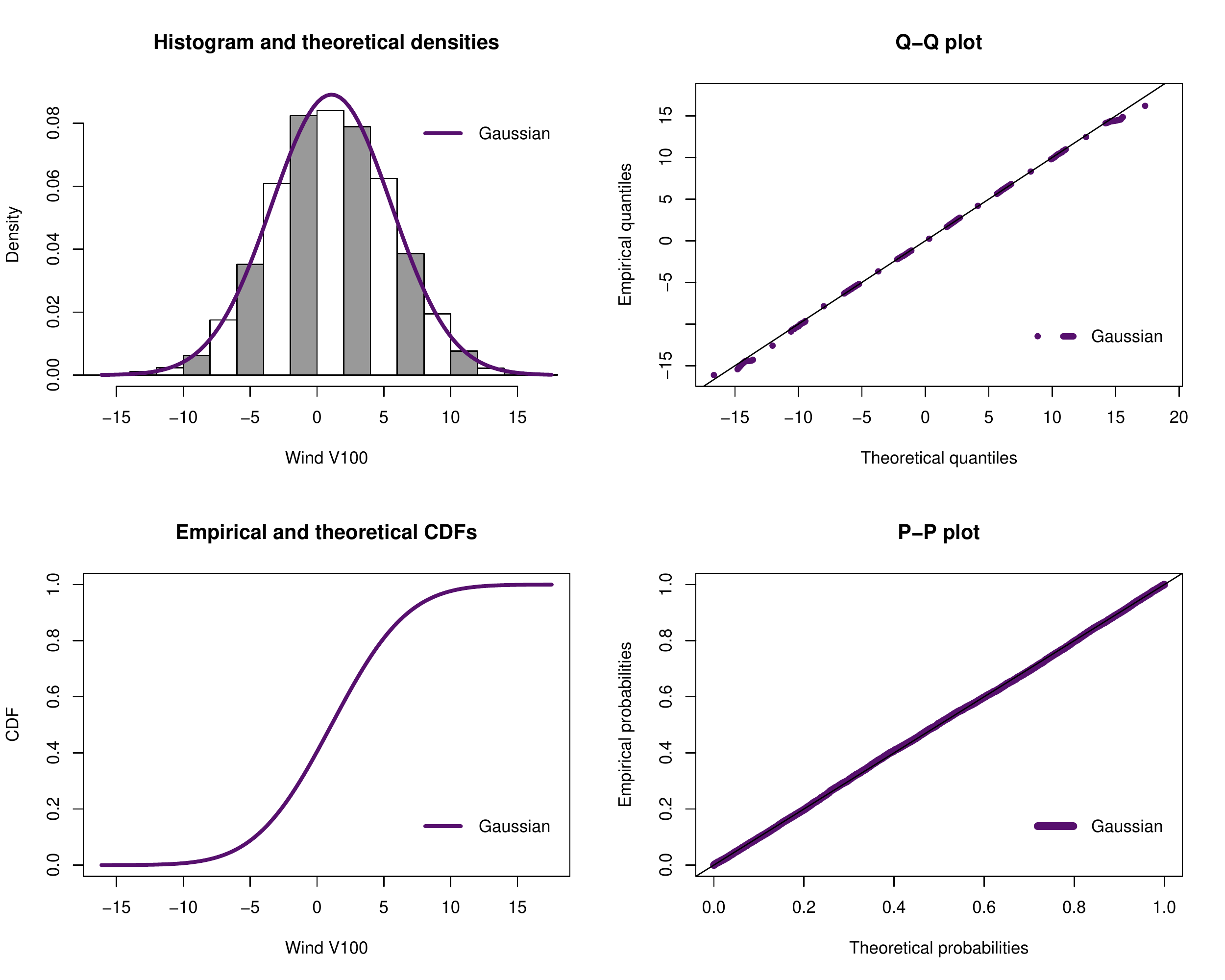}
		\caption{Wind V100}
	\end{subfigure}
	\begin{subfigure}[b]{0.49\textwidth}
		\includegraphics[width=\textwidth]{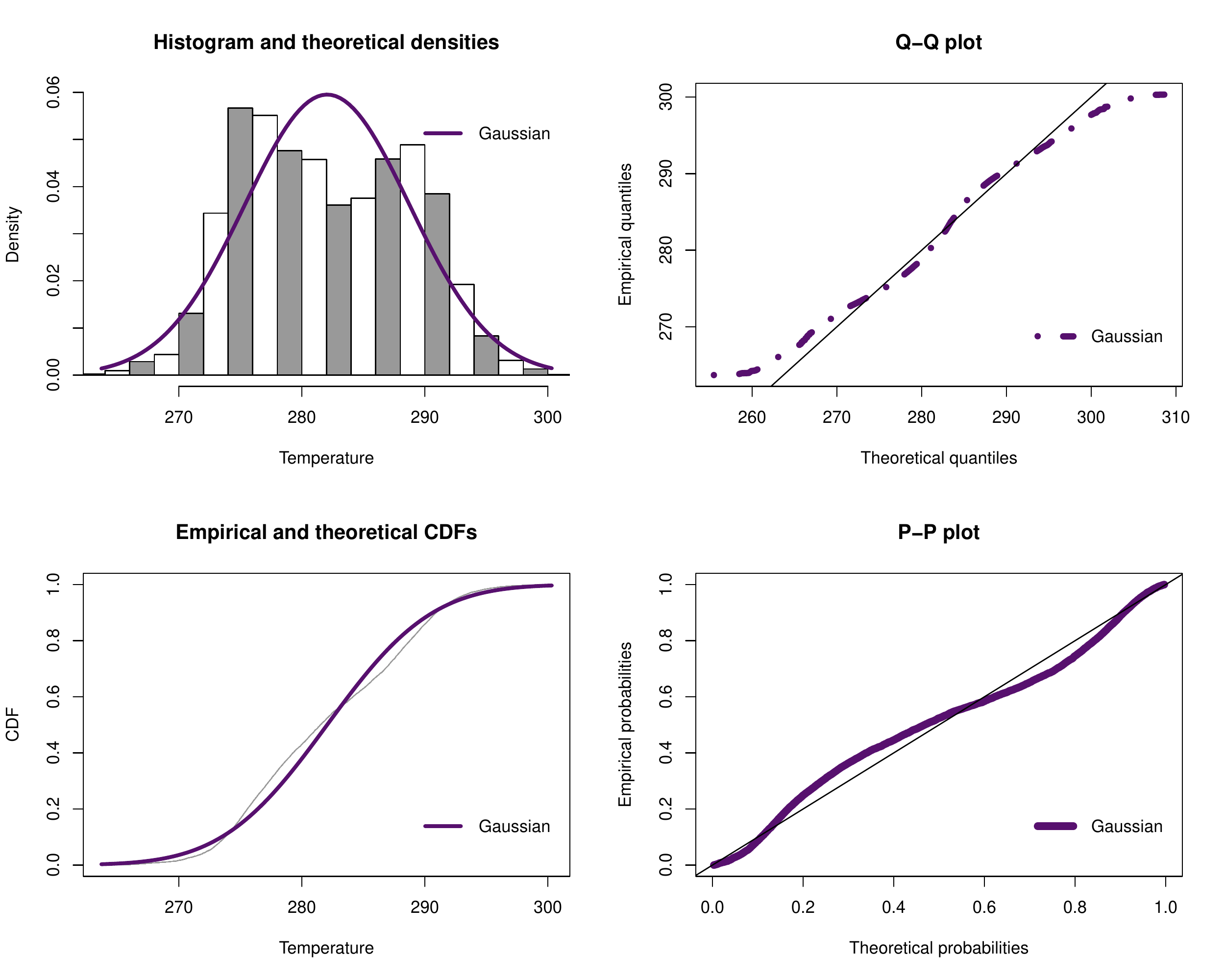}
		\caption{Temperature}
	\end{subfigure}
	\begin{subfigure}[b]{0.49\textwidth}
		\includegraphics[width=\textwidth]{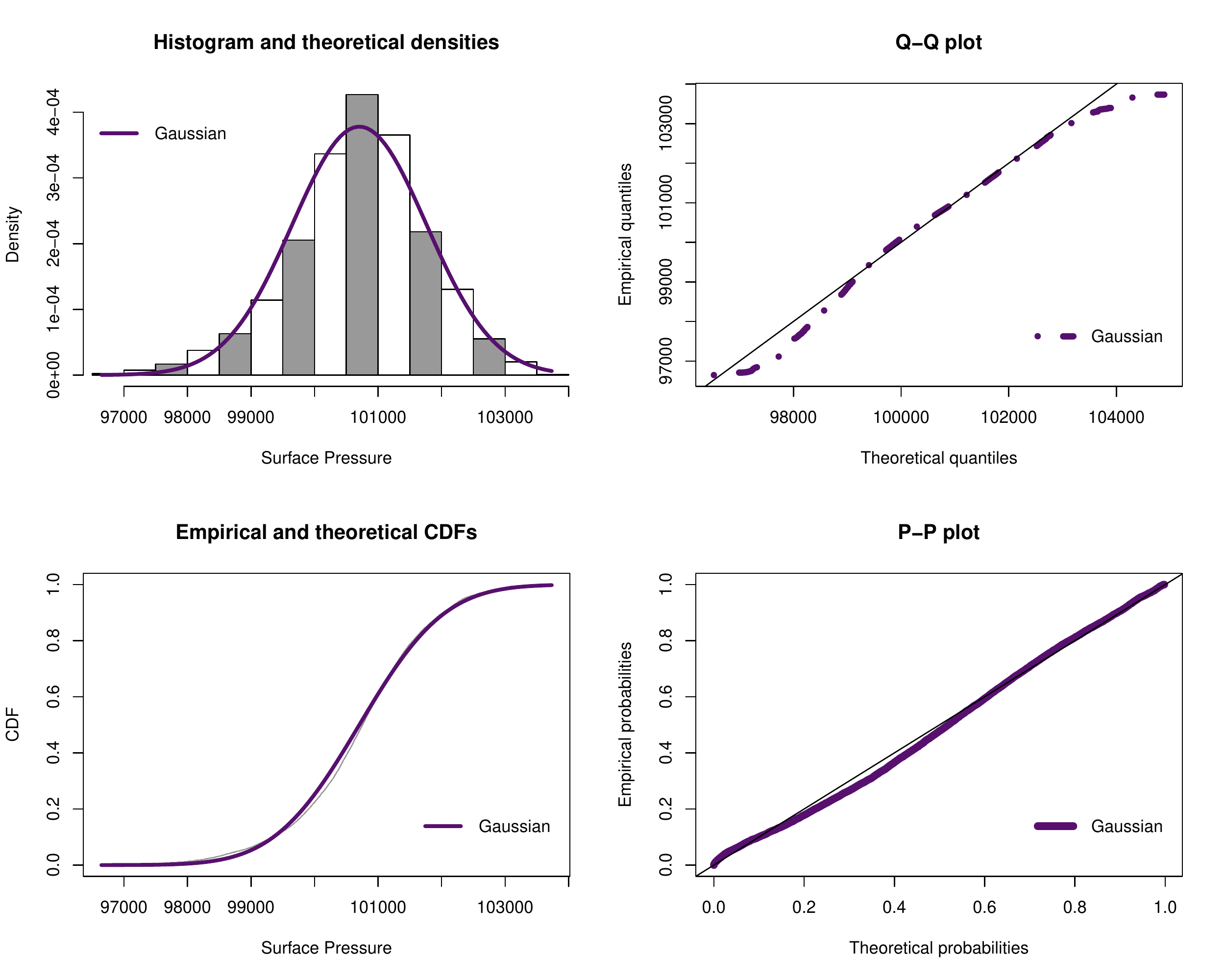}
		\caption{Surface Pressure}
	\end{subfigure}
	\caption{Fitted distributions for wind power and all weather variables for bidding zone 4 in Sweden.}
	\label{fig:fit_sweden_zone4}
\end{figure}

To verify that the choice of distributions for the EMOS calibration is valid, we fit the selected distributions for both the weather variables and resulting wind power generation to the data. \Cref{fig:fit_offshore} - \Cref{fig:fit_sweden_zone4} show the fit of the selected distributions: a truncated normal and gamma distribution for wind power and wind speed, and Gaussian distributions for all other weather variables.

It is noticeable, that although the fits are not perfect, the selected distributions are a good approximation for the data. Some of the empiric distributions are difficult to represent with parametric distributions and therefore these results further motivate non-parametric approaches, which we plan to consider in future work.

\end{document}